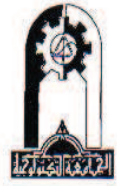

# *Structural, Morphological and Electrical Properties of Porous Silicon Prepared Under Laser Illumination*

This thesis is submitted to the Department of Applied Sciences, University of Technology
In a Partial Fulfillment of the Requirements for the Degree of Master of Science in Laser Physics

*By*
***Oday Arkan Abbas***

*Supervisor*
***Dr. Sabah Mohamed Ali***

**September /2007**            رمضان/1428

# *Acknowledgement*

It is my privilege to express my profound sense of gratitude and appreciation to my supervisor Dr. Sabah Mohamed Ali who supported me in every possible way with his experience, motivation, and his positive attitude. I feel obliged to express heartfelt thanks and gratitude to Dr. H. G. Bohn **(**Institute for Bio-and Nano-systems (IBN2)), Germany, for useful discussions and helpful references

I sincerely acknowledge the support offered by Dr. Alwan M. Alwan (University of Technology**)**, Iraq, during the research work and, also I'm very thankful to all people who are working in the Applied Sciences Department of the University of Technology, and I can't forget to thank my family who supported me with their kindness, patience and encouragement.

Oday

# *Supervisor's Certificate*

      I certify that this thesis entitled titled "***Structural, Morphological and Electrical Properties of Porous Silicon prepared under Laser Illumination***" was prepared by (***Oday Arkan Abbas***) under my supervision at the School of Applied Sciences of the University of Technology in a partial fulfillment of the requirements for the degree of Master of Science in Laser Physics.

Signature: 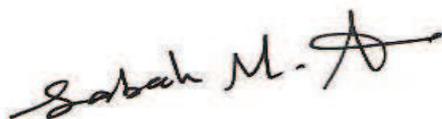

Name    : Assist. Prof. Dr. Sabah Mohamed Ali

Date     : 16 / 9 / 2007

# *Committee Certificate*

We certify that we have read this thesis entitled "**Structural, Morphological and Electrical Properties of Porous Silicon prepared under Laser Illumination**" and, as an examining committee, examined the student (**Oday Arkan Abbas**) in its content and that in our opinion; it is adequate for partial fulfillment of the requirements for the degree of Master of Science in Laser Physics.

Signature: 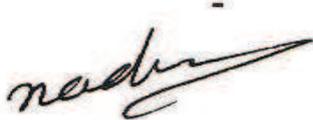   Signature: 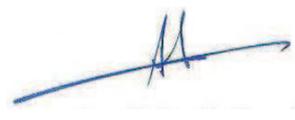

Name: Dr. Nadir Fadhil Habubi          Name: Dr. Ali H. Al-Hamdani
Title: Prof. Dr.                        Title: Assist. Prof. Dr.
Address: Al-Mustansiriyah University    Address: University of Technology
Data: 24/ 10 / 2007                     Data: 24 / 10 / 2007
(Chairman)                              (Member)

Signature: 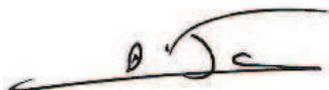   Signature: 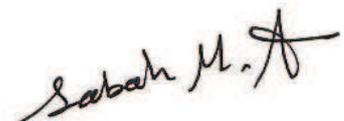

Name: Dr. Adawiya J. Haider             Name: Dr. Sabah M. Ali Ridha
Title: Assist. Prof. Dr.                Title: Assist. Prof. Dr.
Address: University of Technology       Address: University of Technology
Data: 24 / 10 / 2007                    Data: 24 / 10 / 2007
(Member)                                (Supervisor)

Approved by the School of applied sciences, University of Technology

Signature: 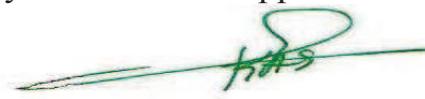
Name: Dr. Kassim S. Kassim
Title: Head of Applied Sciences
Data: 5 / 11 / 2007

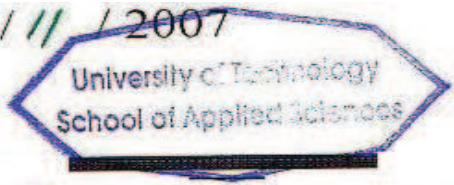

# *ABSTRACT*

Porous silicon (PSi) layers has been prepared in this work via photoelectrochemical (PEC) etching process of an n-type silicon wafers of two resistivities (3.5 Ω.cm and 0.02 Ω.cm) in hydrofluoric (HF) acid of 24.5% concentration at different etching times (5 – 25 min). The irradiation has been achieved using laser beam of 2W power and 810 nm wavelength.

We have studied the morphological and structural properties of PSi layers using the techniques of X-ray Diffraction (XRD) and Scanning Electron Microscopy (SEM) and Gravimetric method. The X-ray Diffraction data shows that the structure aspect of PSi layers remains crystalline as well as the decreasing of diffraction angle ($\theta_B$) of X-ray from PSi layers (29 – 26 degree) and increasing of the lattice parameter values of PSi structures with increasing of etching times from 5-25 min., and the resistivity of silicon substrates from 0.02-3.5 Ω.cm. The nanocrystallite size is decreasing from (20.72 – 5.13 nm) with increasing of etching times, and the resistivity of silicon substrates.

The SEM images shows that the values of pore width and PSi layer thickness increases from (0.5 – 6.25 μm) and (6.7 – 47 μm) respectively with increasing of etching times and silicon substrates resistivities, while the values of the thickness of walls between pores has been varied from (1.25 – 0.03 μm) with increasing of etching times and silicon substrates resistivities. The pore shape of pores has been varied from Cylindrical to Rectangular and to Starful with varied of etching conditions. Further the measured specific surface area of PSi layers has been increased from (7.43 – 235.35 $m^2/cm^3$) with increasing of etching times and silicon substrates resistivities.

The measured values of the porosity of PSi layers using gravimetric method shows that the porosity increases from (35 – 73%) with increasing of etching times and silicon substrates resistivities.

The electrical properties of prepared Al/PSi/n-Si/Al sandwich structures have been studied in this work using conventional electrical method which is represented by current-voltage characteristics under dark and illumination respectively. The obtained values of rectification ratio (8.7 – 67.2%), ideality factor (10 – 29), the heights of potential barriers (0.536 – 0.643 eV) and appearance of single or double saturation regions depending on porosity and layer thickness of the entered PSi layers in these devices and they are demonstrating that the formed (n-type Si/PSi) heterojunctions in these devices is responsible on electrical behavior of these devices as double-Schottky-diode. The obtained values photocurrent of these devices increases from (0.0002 – 19.83 mA/cm$^2$) with increasing of applied reverse bias voltage and decreasing of entered PSi layers thickness in these devices.

The PSi layers resistivity has been increased from (4565.8 – 32133.7 Ω.cm) with increasing porosity of PSi layer from (31 – 73 %).

The Spectral Sensitivity and Quantum Efficiency measurements of Al/PSi/n-Si/Al sandwich structures at 5 (V) reverse bias voltage, which include PSi layers prepared on 0.02 and 3.5 Ω.cm silicon substrates resistivities at 5 and 25 min. etching times shows that Al/PSi/n-Si/Al devices are very sensitive in visible region and the values of spectral sensitivity and quantum efficiency increases from 0.26 to 4.6 A/W and from 71% to 98% respectively with decreasing of etching times and silicon substrates resistivities.

# List of Symbols & Abbreviations

| Symbol & abbreviation | Meaning | Unit |
|---|---|---|
| a | Lattice constant | nm |
| $A_c$ | Contact area | $Cm^2$ |
| $A^{**}$ | Effective Richardson's constant | $A/K^2.cm^2$ |
| B | Full width half maximum of diffraction peak | radians |
| $C_{PSi}$ | Capacitance of PSi layer | F |
| $D_P$ | The hole diffusion constant | $cm^2/s$ |
| d | Layer thickness of porous silicon | μm |
| E | Electric field | V |
| $E_g$ | Energy gap of silicon wafer | eV |
| $E_g^*$ | Energy gap of porous silicon layer | eV |
| $\Delta E$ | The change in the energy gap | eV |
| G | Nanocrystallite size | nm |
| K | Boltzmann's constant | J/K |
| J | Current density | $mA/cm^2$ |
| $J_S$ | Saturation current density | $mA/cm^2$ |
| $J_p$ | Hole-current density | $mA/cm^2$ |
| FWHM | Full width half maximum | radians |
| I | Electric Current | A |
| $I_{ph}$ | Photocurrent | A |
| $I_b$ | Total reverse current | A |
| $I_d$ | Dark current | A |
| H | The distance between the silicon substrate and the spiral boat | cm |
| h | Height of pore | m |
| hν | Photon energy | eV |
| $M_1$ | Weight of silicon sample before etching process | g |
| $M_3$ | Weight of silicon sample before removing porous silicon layer | g |
| $M_2$ | Weight of silicon sample after etching process | g |
| n | Ideality factor | - |
| $N_D$ | Donor concentration | $cm^{-3}$ |
| P | The holes concentration | $cm^{-3}$ |
| $P_{in}$ | Power of incident light | W |
| ΔP | The magnitude of the change in holes concentration | $cm^{-3}$ |
| $Q_E(\lambda)$ | Spectral Quantum efficiency | % |
| R | Resistance | Ω |
| r | Radius of the pore | m |
| $R_s$ | Series resistance | Ω |
| $R_\lambda$ | Spectral sensitivity | A/W |
| S | Porous silicon area | $Cm^2$ |
| T | Absolute temperature | K |
| t | Etching time | Minute |
| ν | Etching rate | μm/min |
| W | Space charge region | μm |
| ε | Dielectric permittivity of the silicon | $F.cm^{-1}$ |
| $ε_∘$ | Permittivity of free space | F/cm |
| $ε_{psi}$ | Permittivity of porous silicon layer | F/cm |
| γ | Porosity of porous silicon layer | % |
| ρ | Resistivity | Ω.cm |
| Φ | Potential drop | V |
| $Φ_{Bn}$ | Barrier height | eV |
| $θ_B$ | Diffraction angle | radians |
| λ | Wavelength | nm |
| ΔP | The magnitude of the change in holes concentration | $cm^{-3}$ |
| $J_{Psi}$ | Critical current density | $mA/cm^2$ |
| SCR | Space-charge region | μm |
| $μ_P$ | Hole mobility | $Cm^2/V.s$ |
| PEC | Photoelectrochemical | - |
| EC | Electrochemical | - |
| HF | Hydrofluoric acid | - |
| FIPOS | Full isolation by porous oxidized silicon | - |
| c-Si | Crystalline Silicon | - |
| PSi | Porous Silicon | - |
| PL | Photoluminescence | - |
| SEM | Scanning electron microscopy | - |
| XRD | X-ray diffraction | - |
| TEM | Transmission electron microscopy | - |
| HRSEM | High resolution scanning electron microscopy | - |
| FTIR | Fourier transform infrared spectroscopy | - |
| LED | Light emitting diode | - |
| AFM | Atomic force microscopy | - |
| STM | Scanning tunneling microscopy | - |
| LED | Light emitting diode | - |
| AFM | Atomic force microscopy | - |
| PSi | Porous Silicon | - |
| PL | Photoluminescence | - |
| SEM | Scanning electron microscopy | - |
| XRD | X-ray diffraction | - |
| TEM | Transmission electron microscopy | - |
| HRSEM | High resolution scanning electron microscopy | - |
| FTIR | Fourier transform infrared spectroscopy | - |
| LED | Light emitting diode | - |
| AFM | Atomic force microscopy | - |
| PSi | Porous Silicon | - |
| PL | Photoluminescence | - |
| SEM | Scanning electron microscopy | - |

# *Contents*

# CHAPTER ONE

# Introduction

## 1-1 *Introduction*

Why nano? The discovery of novel materials, processes and phenomena at the nanoscale, as well as the development of new experimental and theoretical techniques for research provide fresh opportunities for the development of innovative nanosystems and nanostructured materials. Nanosystems are expected to find various unique applications. Nanostructured materials can be made with unique nanostructures and properties. This field is expected to open new venues in science and technology [1].

Crystalline silicon (c-Si) is well studied for production of electronic components, but it is useless for fabrication of light-emitting devices, because it possesses indirect nature of the fundamental bandgap [2]. *Canham* [3] has reported in 1990, observation of bright photoluminescence (PL) from porous silicon (PSi) even at room temperature. Therefore, PSi material has become a popular material among scientists and technologists, and has been applied in various fields during the past two decades. Porous silicon can be considered as crystalline silicon, it has a network of voids in its bulk. The voids in the silicon bulk result in a sponge-like structure of pores and channels with a skeleton of nanocrystals [4,5].

The interest in the electrical properties of PSi began prior to the discovery of its efficient luminescence [6,7]. The material was used for electrical isolation and sensing applications [8]. *Gaburro et al* [9] have reported that the electrical resistivity of PSi layer is five orders of magnitude higher than in intrinsic silicon. The electrical behavior of PSi layer is not emphasized [4], because it behaves like a Schottky junction with the metal electrode, or as a heterojunction with silicon substrate [10,11].

Therefore to integrate PSi layer into electronic circuits or to develop PSi based devices, the electrical properties of this material must be studied thoroughly [12].

Photoelectrochemical (PEC) etching process of III-V semiconductors has been used to fabricate unique structures in electronic and photonic devices, such as integral lenses on light-emitting diodes, gratings in laser devices [13]. For n-type silicon, where holes are the minority carriers, the electrochemical dissolution of the silicon depends strongly on the electron-hole generation by illumination or it needs high currents especially for lightly doped n-type. If the etching process is prepared under illumination, PSi layer is formed at low potential and the resulting material consists of two parts [4]. The top surface layer is nanoporous (pore diameter of about 3 nm) and its thickness lies in the 0.2-1 μm range depending on the formation conditions. The underlying part is macroporous. [9,12].

*Vincent* [14] has illustrated that photoluminescent PSi can be produced by PEC etching process. One of the driving forces behind the PEC approach is the perceived ability to form rectifying junctions in simpler ways compared to the relatively sophisticated techniques required in solid state processing [15].

## 1-2 *Historical Review*

In 1956, *Arthur Uhlir* at Bell Telephone laboratories was working on electrochemical (EC) etching of silicon using hydrofluoric acid (HF) solution. This was done with intent of polishing and shaping microstructures in silicon. However, these films were of relatively little interest at that time [16].

*Watanabe et al* in 1971 have demonstrated the first application of PSi material in electronics, the so-called full isolation by porous oxidized silicon process (FIPOS) [17]. *Canham* [18] in 1986 has observed photoluminescence (PL) from etched silicon at room temperature and studied the effect of chemical pretreatment on intensity of photoluminescence (PL). The more noticeable interest shown from the start of the last decade came with the demonstration of room temperature of PL from PSi material by *Canham* in 1990 [3]. The first model to explain the formation mechanism of PSi layer was reported in 1991 by *Lehmann et al* [19]. *Anderson et al* [7,8] in 1990 and 1991 have reported the first study for electrical properties of PSi layer and the utilizing of this material as vapor sensing. The important structural characteristics of PSi layer were investigated by *Lehmann et al* [20,21] in 1990 and 1993. *Ben-Chorin* [10,22] in 1995 has placed the important models to clarify the transport mechanisms of charge carrier. *Lee et al* [23] have reported in 2004, the effect of illumination on properties of PSi layer.

*B. G. Rasheed* and its groups in 2001, 2005 and 2006 have reported by depending on its experimental study that the PSi layers can be prepared using Laser-induced etching process by using diode laser and these layers have advantage and dramatic structural and optical properties.

*Alwan M. Alwan* in 2006 has reported that the PSi layers can be prepared by Photochemical etching using halogen lamp as illumination source, and demonstrated that the electrical behavior of Al/PSi/n-Si/Al sandwich structure like schottky diode and these electrical properties depends on structural properties of entered PSi layers in these devices. The related topics represent which are still a progressive research area as shown in figure (1-1) with a wide range of novel ideas and applications [3]

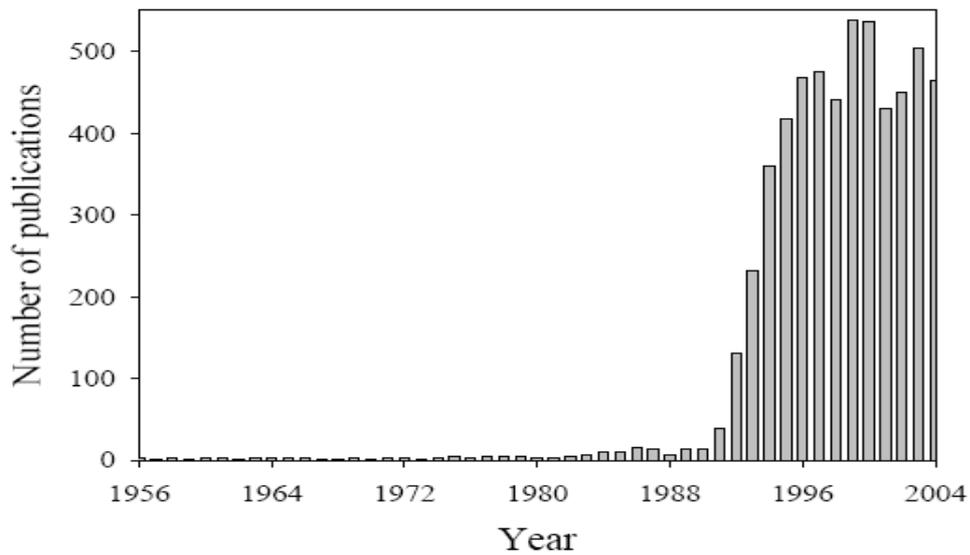

Figure (1-1); Number of publications in refereed journals which dealing with PSi material [3].

## 1-3 *Aims of Work*

The aims of this work can be summarized in the following points:

**1.** Preparation of PSi layers, which contain silicon nanocrystals via PEC etching process using diode laser.

**2.** Studying of Morphological properties of PSi layers.

**3.** Studying of Structural properties of PSi layers.

**4.** Threshing of the electrical features of Al/PSi/n-Si/Al sandwich structures.

**5.** Studying of the most effective parameters of this method on these properties.

# CHAPTER TWO
# Theoretical background

## 2-1 *Formation of Porous Silicon*

In order to understand the PSi layer formation it is necessary to deal with the dissolution chemistry of silicon anodically biased in HF acid. A silicon (Si) surface is known to be virtually inert against attack of HF acid [19]. If the silicon is under anodic bias, the reaction begins with the generation of holes, either thermally generated or photogenerated. The holes drift under the applied bias to the interface between silicon and solution, where they can react, and then formation PSi layer is observed as long as the reaction is limited by the charge supply of the electrode. This condition is fulfilled for current density (J) below the critical current density ($J_{PSi}$) [13,19,24,25].

If a hole ($h^+$) reaches the surface, nucleophilic attack on Si-H bonds by fluoride ions ($F^-$) can occur and a Si-F bond is established (step 1 in figure (2-1)). Due to the polarizing influence of the bonded F, another $F^-$ ion can attack and bond under generation of an $H_2$ molecule and injection of one electron into the bulk silicon (step 2). Because of the polarization induced by the Si-F groups, the electron density of the Si-Si back bonds is lowered and these weakened-bonds will now be attacked by HF (steps 4 and 5) in a way that the silicon surface atoms remain bonded to hydrogen (step 5) [19].

When a silicon atom becomes removed from an atomically flat surface by this reaction, an atomic size dip remains. This change in surface geometry will change the electric field distribution in such a way that holes transfer occurs at this location preferentially [19,24]. Therefore pores of about 1 μm will establish in a few minutes on polished n-type silicon surface by this process. If the walls between pores are depleted of holes they will be protected against dissolution, where the depletion layer acts as a barrier to transfer the holes into walls [20,19].

*Lehmann and Gosele* [19] have described the anodic reaction during pore formation as follows:

$$Si + 6HF \longrightarrow H_2SiF_6 + H_2 + 2H^+ + 2e^- \quad \ldots\ldots\ldots\ldots\ldots\ldots (2\text{-}1)$$

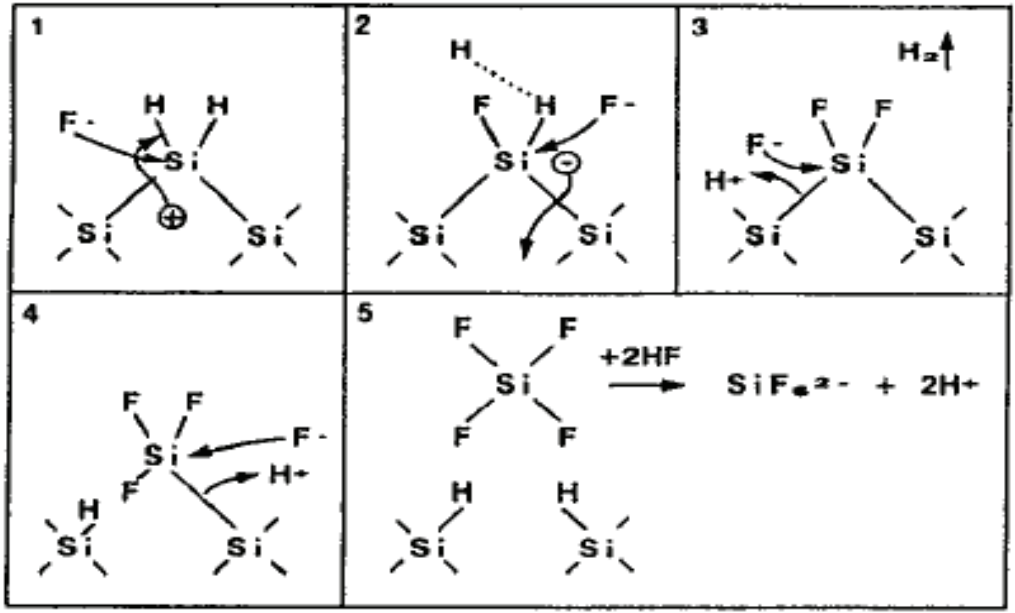

Figure (2-1); Proposed dissolution mechanism of silicon electrode in HF acid associated with PSi layer formation [19].

## 2-2 *Preparation Methods of Porous Silicon*

There are four important methods for fabrication of PSi layer, they are (stain, electrochemical, photochemical and photoelectrochemical) etching processes. Each one of these methods possesses properties which differ from other and the characteristics of formed PSi layer varies from one way to another. The joint feature of all methods is the simple components compared with those employed in developed and complicated techniques in solid state science.

## 2-2-1 *Electrochemical Etching*

Electrochemical dissolution of monocrystalline silicon in HF acid solution can be used to fabricate a silicon nanostructure known as PSi which consists of a network of pores separated by very small silicon columns [1].

The silicon wafer in this method serves as the anode and the cathode is made of platinum or any HF-resistance and conducting material. Since the entire silicon wafer serves as the anode, PSi is formed on any wafer surface in contact with the HF solution. The advantage of this method is its simplicity and ability to anodize silicon-on-insulator structures. Its drawback is the non-uniformity in both porosity and thickness of the resulting PSi layer [4,17]. In general, porosity between 10% to 80% and pore sizes of 50 nm down to less than 1 nm are easily obtained by varying the electrochemical parameters [20].

## 2-2-2 *Photochemical Etching*

In this technique a photon source such as lasers and intensive light is used to supply the required holes in the irradiated area of silicon wafer to initiate the etching process. Photochemical etching has been used to produce porous silicon (PSi) and nanocrystallites of silicon, and has advantages that can be summarized as follows: (1) Simple process (without external potential). (2) Controllable processes that have more accurate processing parameters. This way has some drawbacks such as nonuniformty of illumination distribution and also the etching process in this method is considered slow [26,27,28]. In over-view, the prepared PSi structure by photochemical etching process has different layer thickness and surface morphology according to wavelength and power of the utilized light [29,30].

## 2-2-3 *Stain Etching*

The stain etching is another technique which can be used to produce films similar in nature to anodically prepared PSi material [31,32]. The preparation of PSi films by stain etching process utilizes nitric and HF acids during normal etching. The most common etchant for stain etching is HF: $HNO_3$:$H_2O$ [30]. The implicit simplicity of fabrication steps are making the stain etching process an important option in PSi formation, but at the same time the stain etching process is considered very slow and the surface of formed PSi layer is very rough. Generally, it remains especially attractive when very thin PSi films of uniform depth ($10^1 - 10^2$ nm) are required [4].

## 2-2-4 *Photoelectrochemical Etching*

This technique is deemed an ultra important method in industry of PSi material because it is suitable for etching of n-type and p-type silicon in HF solution. This fact based on that the photoelectrochemical (PEC) etching process is collects between two ways (electrochemical and photochemical) etching processes. *Juhasz and Thonissen* [33,34] have reported that the illumination of n- as well as p-type PSi during formation causes dramatic changes in the structure of layers. The n-type PSi obtained under illumination by PEC etching process consists of layers of nanoporous silicon layer which covers a macroporous silicon layer with pores in the micron size range [35].

## 2-2-4-1 *Theoretical Fundament*

In PEC etching process the semiconductor is immersed in conductive electrolyte. The silicon is biased as anode, when the Fermi level of the silicon at the silicon-electrolyte interface is within the bandgap as shown in figure (2-2), which shows the case of n-type silicon in contact with electrolyte [13]. The density of charge carriers in the silicon is usually less than the density of carriers (ions) in the electrolyte, so that a majority of the potential difference across the interface occurs within the silicon [36]. The charge density in the silicon space-charge region (SCR) is $qN_D$ where $N_D$ (cm$^{-3}$) is the donor concentration. Therefore,

$$\nabla E = \frac{qN_D}{\varepsilon} \quad\quad\quad (2-2)$$

where $\varepsilon$ (F.cm$^{-1}$) is the dielectric permittivity of the silicon, $E$ (V) is the electric field, and $q$ (C) is the magnitude of the electronic charge [13,27]. The width of (SCR) $W$ (μm) is related to the potential drop $\Phi$ (V) as [13]:

$$W = \sqrt{\frac{2\varepsilon\Phi}{qN_D}} \quad\quad\quad (2-3)$$

Electron-hole pairs can be photogenerated by irradiating the silicon with light whose energy is greater than the bandgap [33]. The electrons and holes created within the (SCR) are transported by (1) migration under the influence of the electric field and (2) diffusion due to the gradient in the charge carrier concentration [36]. The holes concentration, $P$ (cm$^{-3}$) is determined from the transport relation for the hole-current density, $J_P$ (mA/cm$^2$) [13]:

$$J_P = q\mu_P PE - qD_P \Delta P \quad\quad\quad (2-4)$$

where $D_P$ (cm²/s) is the hole diffusion constant, $\mu_P$ (cm²/V.s) is the hole mobility and $\Delta P$ (cm⁻³) is the magnitude of the change in holes concentration.

Photogenerated holes which reach the silicon-electrolyte interface can react with solution species and the etching process is initiated as reported in section (2-1). The energy gap and electric field act as a kinetic barrier for electrons to reach the silicon-solution interface [13,17].

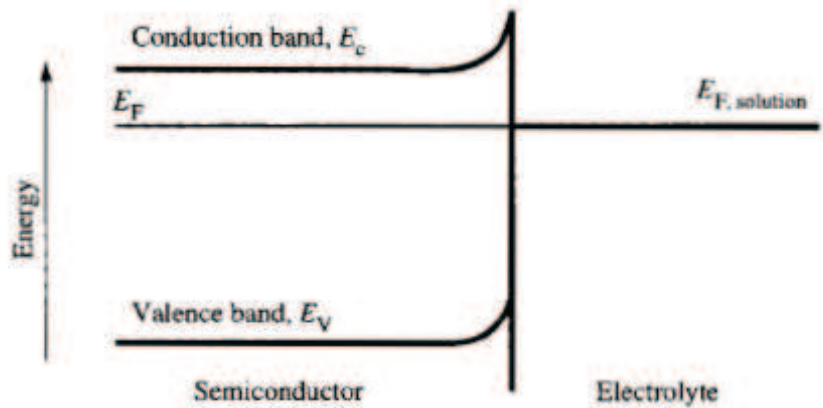

Figure (2-2), Energy-band structure of n-type silicon in contact with electrolyte [13].

## 2-2-4-2 *Properties of PEC Etching*

The properties of PEC etching process can be summarized in the following points:

1- *Light-intensity dependence*: The rate of etching depends on the rate of minority-carrier photogeneration (holes), and in many cases a linear relation exists between the etch rate and the light intensity [23,29].

2- *Spatial selectivity*: By modulation of the light intensity, one can anisotropically etch silicon [27].

3- *Bandgap selectivity*: Structures having materials with different bandgaps the narrow-gap materials could be selectively-etched by using light with a spectrum that is absorbed by the narrower-gap materials but not the wider [13].

## 2-3 *Etching Parameters*

All properties of PSi layer such as porosity, layer thickness, pore width and nanocrystallite size, are strongly dependent on the etching conditions. These conditions include illumination, etching time, doping density and current density [4,37].

### 2-3-1 *Illumination Power and Wavelength*

The illumination of the samples during or after etching process is known to be a further free etching parameter which can be used to modify the PSi layer morphology, where the illuminated sample during formation of PSi layer shows a greatly increased amount of small (< 12.5 $\text{A}°$) nanocrystals whereas the amount of larger crystals is reduced [27,34,38].

According to *Noguchi* model [26] the etching rate is a function of used illumination distribution, and also *Lim* [27] has confirmed that the etching rate relying on the intensity of illumination as shown in figure (2-3) as well as that the wavelength will limit the generating position of charge carrier.

*Vincent* [14] has reported that the short wavelength ensures that electron-hole pairs are generated close to the fiber/solution interface. Therefore a photoetching process can be much more effective at etching the thinner regions, while the charge carriers are generated deeply in the silicon bulk to result primarily in excitation of the substrate as shown in figure (2-4). In addition the type of utilized illumination source (coherent and incoherent) possesses very important effect on etching process and then on structural properties of PSi layer [23,33].

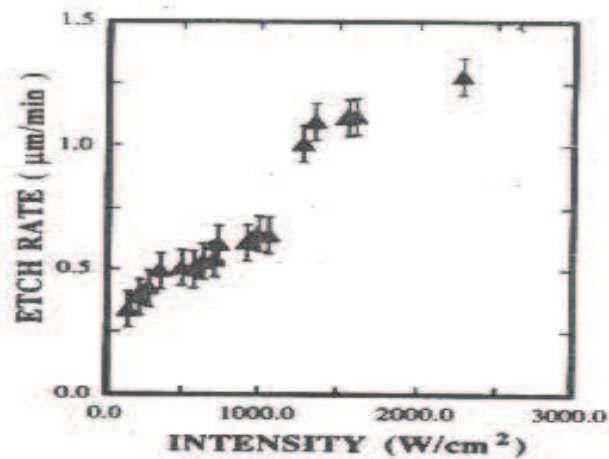

Figure (2-3); Effect of laser beam intensity on etching rate of n-type Si [27].

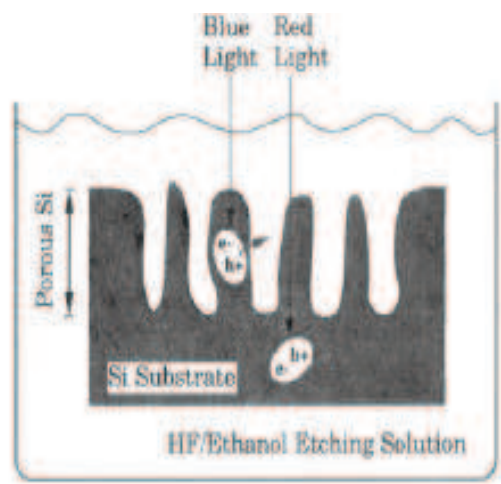

Figure (2-4); The generation of charge carriers by different wavelengths [14].

## 2-3-2 *Etching Time*

One of most important etching parameters is etching time, since this parameter has strong effect nearly on all characteristics of PSi layer, where the increase in the etching time causes an increase in the PSi layer thickness, as more silicon fragments dissolve when the etching time increases [20,29].

Further *Badel* [24] has illustrated from their practical observations of the constant current density and varied etching time; the pore width can get enlarged at the beginning of the pore. The surface morphology of PSi layer is dependent on etching time, since the PSi surface roughness increases gradually with increasing of time as shown in figure (2-5) [12].

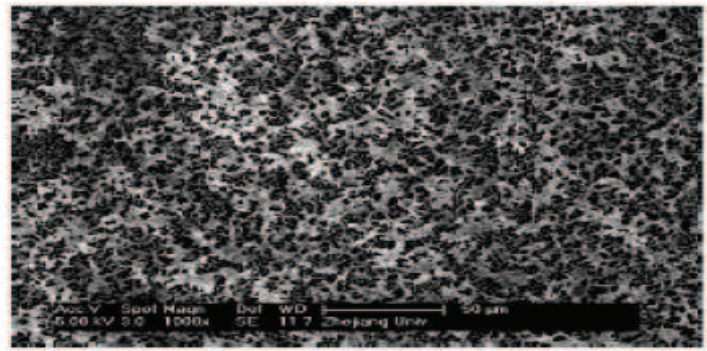
(a)

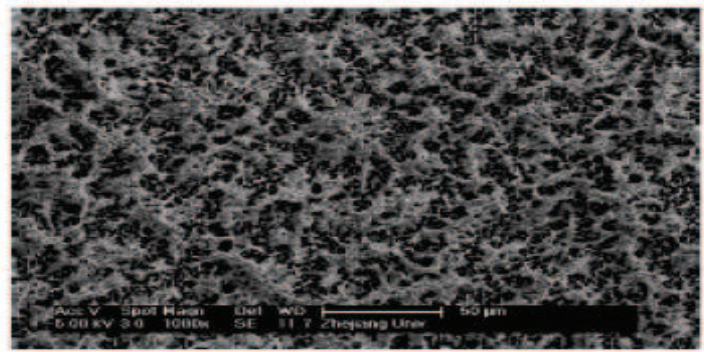
(b)

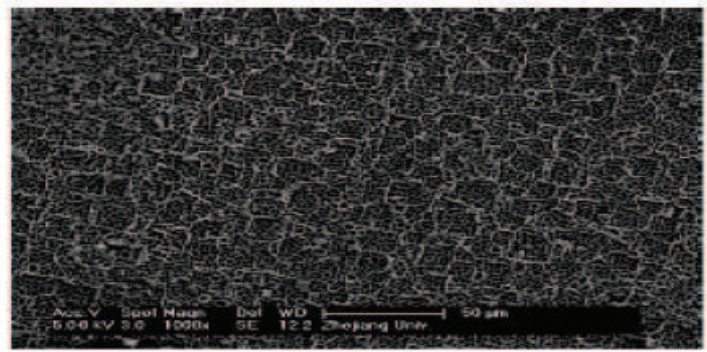
(c)

Figure (2-5); Scanning electron microscopy (SEM) images of formed PSi layer at different etching times (a) 90 min., (b) 60 min. and (c) 30 min. [12].

## 2-3-3 *Doping Density*

The scale of pore width and space charge region width of PSi material are dependent on doping density of silicon substrate.

*Lehmann and Pang* [20,21,37] have cleared that the various resistivity of substrate will likely change the depletion layer at the pore front which may change the resulting shape and diameter of pore as shown in figure (2-6,a,b).As well as the effect of doping density on pore shape and width, this parameter has influence on etching rate, according to *Dimitrov et al* [11,13] the etching process is faster in high doping sample compared with the low doping sample as shown in figure (2-8).

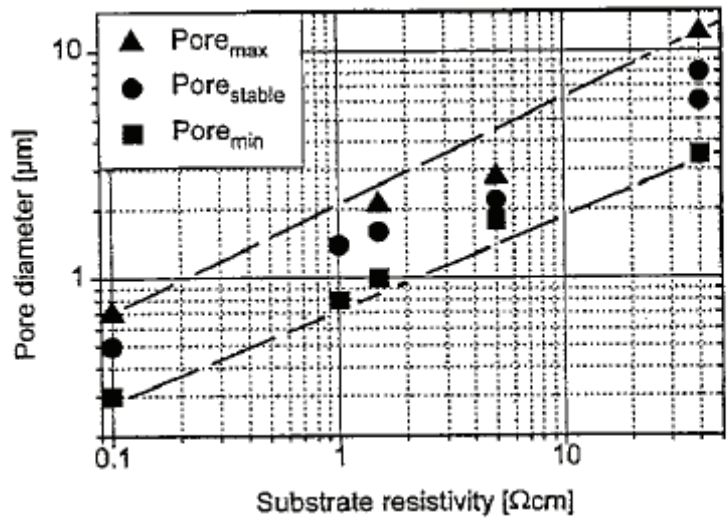

Figure (2-6, a); The pore width as a function of n-type doping density [37].

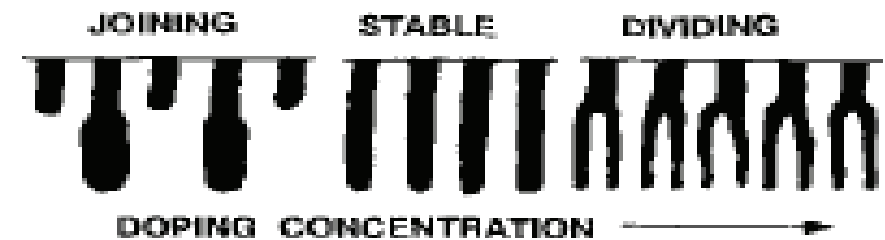

Figure (2-6, b); The pore shape as a function of n-type doping density [20].

## 2-3-4 *Current Density*

The porosity and pore width of a PSi layer is partly determined by current density. *Feng and Kohl* [13,17] have reported that the maintaining a constant current density leads to constant porosity for PSi layer, but if the current density changes during the etching process the porosity of the PSi layer and the pore shape will vary accordingly as shown in figure (2-7).

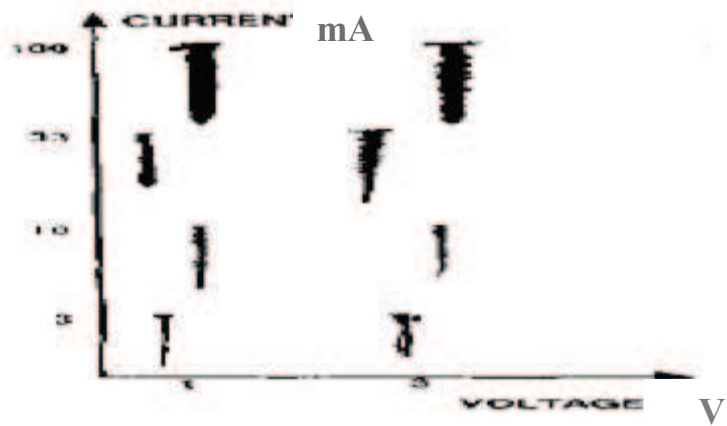

Figure (2-7); The pore shape as a function of current density [20].

The growth rate is governed by the current density according to figure (2-8). It is obvious from the graph that the etching rate increases with excess of current density [17].

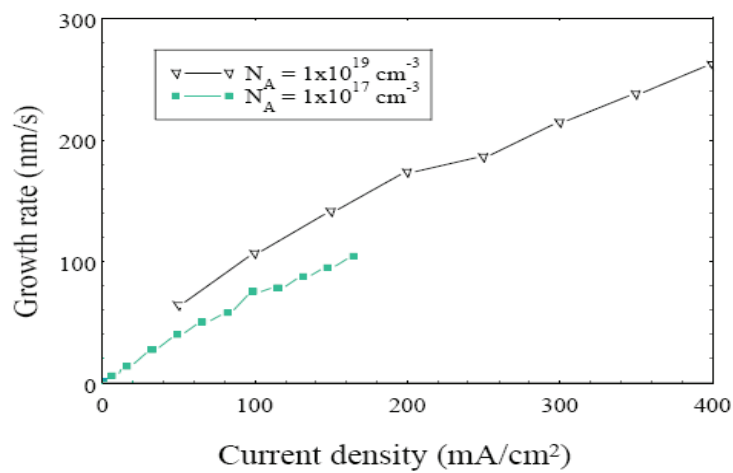

Figure (2-8); Growth rate of PSi layer as a function of current density [17].

## 2-4 *Properties of Porous Silicon*

Porous silicon (PSi) layer has unique and dramatic properties, which made this material is used in wide range of fields and applications.

### 2-4-1 *Structural Properties*

The morphological characteristics of PSi layer are considered the more important features because all other properties for this material such as electrical, optical, electronic…etc are dependent on structural properties [39,40]. Several techniques were employed for examining and studying the structural features of PSi material such as scanning electron microscopy (SEM), X-ray diffraction (XRD) and transmission electron microscopy (TEM) [4,21].

#### 2-4-1-1 *Surface Morphology*

Direct imaging of PSi layer by using high resolution scanning electron microscopy (HRSEM) technique gives perfect means for studying the surface structure of PSi layer. *Seejon et al* [23,24] have reported that the prepared PSi layer via PEC etching process has rough surface and three regions as shown in figure (2-9) which illustrates SEM images of both porous and electropolished silicon regions, viewed from the top. Furthermore, an intermediate regime, also called transition regime and characterized by the formation of pillar-like structures can be observed, as well as the roughness of the surface is a function of porosity of the PSi layer [3,23].

The porosity of PSi layer can be defined as the fraction of voids inside the PSi layer [4] which can be determined via gravimetric method as shown in equation (2-5), and by microsturctural analysis [19].

$$\gamma = \frac{M_1 - M_2}{M_1 - M_3} \dotfill (2-5)$$

where $M_1$ (g) and $M_2$ (g) are the weights of silicon sample before and after etching process respectively. $M_3$ (g) is the weight of the silicon sample after removing of PSi layer [17].

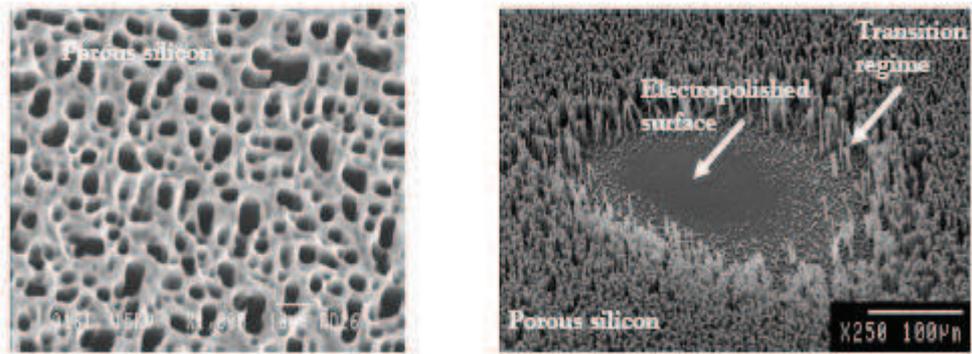

Figure (2-9); SEM images of PSi layer [24].

The porosity of PSi layer is very sensitive for etching parameters such as illumination, etching time, doping density and current density as shown previously in section (2-3). According to microsturctural analysis the porosity increases gradually from 25% to 80% or more [4]. Figure (2-10,a) shows an idealized plan view of the formed PSi layer of relatively low porosity (25%).

*Canham* [3] has suggested dependence on his experimental observations that the mesoporous silicon of low porosity can be adequately approximated by bulk material containing an array of noninteracting cylindrical pores of fixed radius that runs perpendicular to the surface. In reality, of course there is pore branching and variations in pore size, separation and perhaps even shape.

Figure (2-10,b-d,c) shows the increasing of porosity of the PSi layer from 50% to 80% with largeness of pores.

Due to crystallographic effect (crystalline orientations), the pore shape may not be cylindrical, but tends toward the square or rectangular shape as shown in figure (2-10,d-f) [3,37].

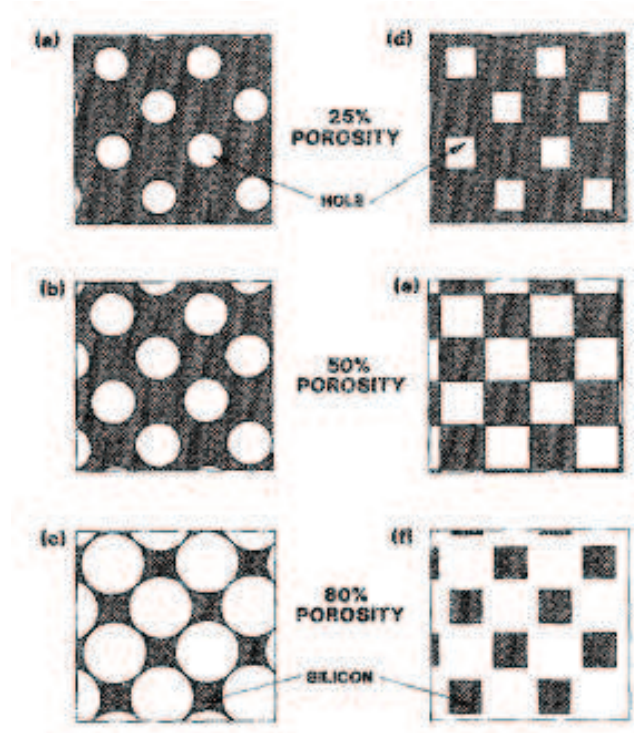

Figure (2-10); Idealized plan view of PSi layer shows the shape and width of pore and its relation with the porosity [3].

### 2-4-1-2 *Layer Thickness*

The thickness of PSi layer depends on etching parameters especially on the etching time, since the layer thickness increases with increasing of etching time [26]. Also *Choy* [28] has reported that the thickness of PSi layer is varied according to illumination distribution and doping density.

The most common technique to measure and study the layer thickness of PSi material is SEM measurement, where the measuring of this feature is considered very important to calculate and characterize the etching rates as shown in equation (2-6) [27,32].

$$v = \frac{d}{t} \quad \quad (2-6)$$

where $v$ (μm/min) is the etching rate, $d$ (μm) is the layer thickness and $t$ (min) is the etching time.

### 2-4-1-3 *Nanocrystallite Size and Lattice Constant*

The porous silicon material presents itself as crystalline silicon with structure contains crystals in nanoscale size. This characteristic has an important effect on each property of PSi layer; therefore it has been studied by many researchers. *Heuser and Pineik et al* [39,40] have demonstrated by achieving X-ray diffraction measurement for PSi layer that the structure of this material remains crystalline, but it has significant peak broadening, which is interpreted as a nanocrystallite size effect. The nanocrystallite size was calculated from the peak broadening as shown in figure (2-11), by using Scherer's formula as follows [41]:

$$G = \frac{0.9\lambda}{B\cos\theta_B} \quad\quad\quad (2-7)$$

where $G$ (nm) is the nanocrystallite size in PSi layer, $\lambda$ (nm) is the wavelength of employed radiation, $B$ (radians) is the full width half maximum (FWHM), $\theta_B$ (radians) is the diffraction angle and 0.9 is the value of shape factor [41].

The effect of nanocrystallitte size is interpreted as widening of bandgap of the PSi layer due to the quantum confinement phenomenon as shown in figure (2-12). *Lee* [42] has confirmed that the very small dimensions of PSi crystals lead to a heterojunction between silicon substrate and PSi layer because the latter has big bandgap (1.8 – 2.3 eV) compared with Si (1.1 eV).

The energy gap of PSi layer can be calculated dependence on the size of crystals as follows [9]:

$$E_g^* = E_g + \frac{88.34}{G^{(1.37)}} \quad \quad (2-8)$$

where $E_g^*$ (eV) is the energy gap of PSi layer, $E_g$ (eV) is the energy gap of bulk silicon and $G$ (nm) is nanocrystallite size.

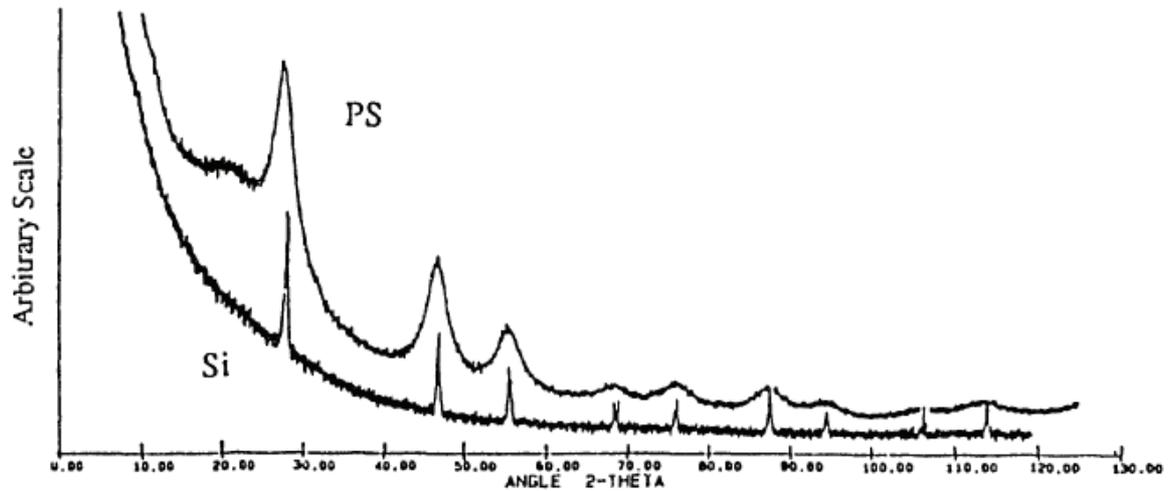

Figure (2-11); X-ray diffraction of silicon and PSi layer [39].

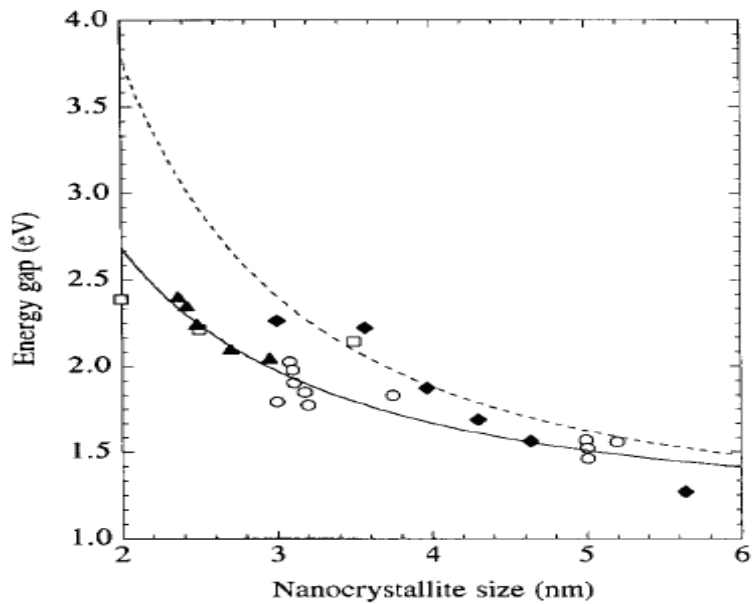

Figure (2-12); The relation between energy gap and nanocrystallite size of formed PSi layer under different etching conditions [42].

The lattice constant of PSi layer has been studied by a variety of researchers [40,42]. They found that the diffraction peak of PSi layer broadens and slightly shifts to a smaller diffraction angle and also confirm that the shifted peak is due to expansion in lattice constant and the latter as a consequence of the strain effect.

The lattice constant could be calculated using following equations [41]:

$$a = d(h^2 + k^2 + l^2)^{1/2} \quad \text{...............................................} (2-9)$$

Where the $a$ (Å) is the lattice constant and ($h,k,l$) represent Miller's indices.

$$n\lambda = 2d \sin\theta \quad \text{...............................................} (2-10)$$

where $n$ is the order of diffraction, $d$ (Å) is the distance between planes, $\lambda$ (Å) is the wavelength of incident radiation and $\theta$ is the diffraction angle.

## 2-4-2 *Electrical and Photoelectrical Properties*

The great interest in the PSi material is because this material possesses unusual physical properties, which gives new chances to develop many technological applications [43]. Within the last decade the PSi structures have been used not only to emit light efficiently, but also be capable of guiding, modulating and detecting light [44]. Consequently, the studies of electrical and photoelectrical properties of PSi-based devices have taken a large space of scientific publications [45]. *Ben-Chorin* [46] has reported that the progress towards silicon optoelectronics and photovoltaic devices using PSi layer requires good understanding of the electrical behavior of PSi layer.

## 2-4-2-1 J-V Characteristics of M/PSi/c-Si/M Sandwich Structure

The current-voltage characteristics of PSi-based devices have been studied by many groups of researchers [8,10,11,47,48,49,50,51]. They have shown that the behavior of current-voltage characteristics of the M/PSi/c-Si/M sandwich structure vary according to structural properties of PSi layer because it behaves as Schottky diode but in another case it behaves as heterojunction.

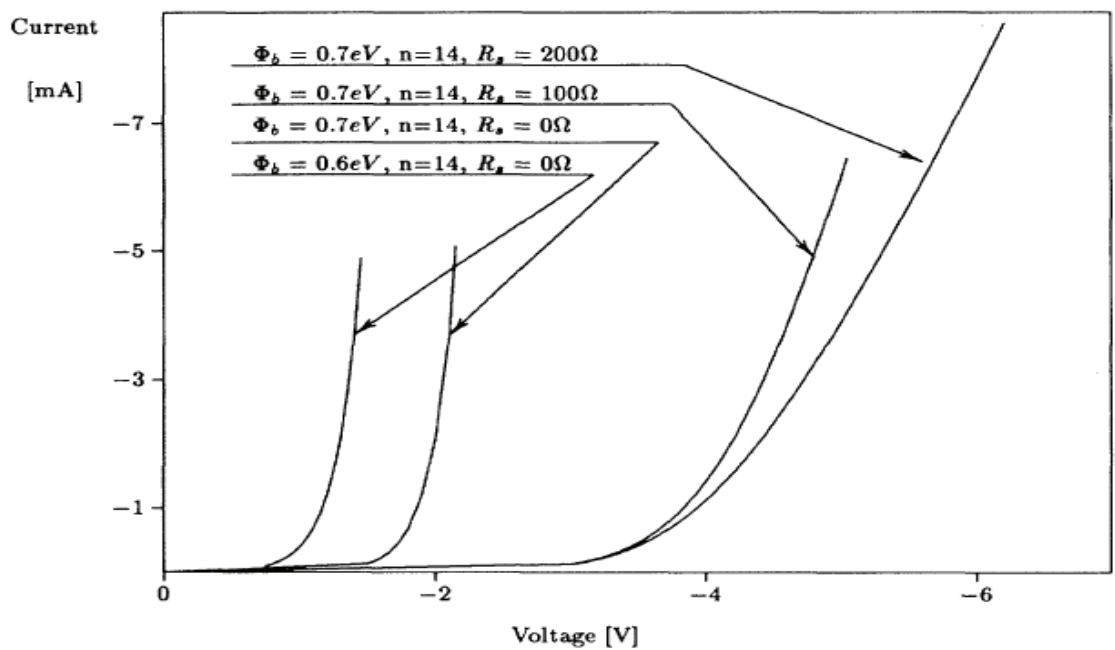

Figure (2-13); The current-voltage characteristics of M/PSi/c-Si/M sandwich structure [11].

Typical I-V characteristics are shown in figure (2-13). The forward part of the measured I-V curve is very similar to Schottky diode characteristics and may be analyzed by using the following equation which described by *Sze* [36].

$$J = J_S \left[ \exp\left(\frac{qV}{nKT}\right) - 1 \right] \quad \text{.................................} (2-11)$$

where $J$ (mA/cm$^2$) is the current density, $J_s$ (mA/cm$^2$) is the saturation current density, $T$ (K) is the absolute temperature, $K$ (J/K) is the Boltzmann's constant
$q$ (C) is the electronic charge, $V$ (V) is the applied voltage and $n$ is the ideality factor. Equation (2-11) depicts current transport across a metal-semiconductor contact. A hybrid theory based on both thermionic emission and diffusion of charge carriers yields an expression for the saturation-current density ($J_s$) [8]:

$$J_S = A^{**}T^2 \exp\left(-q\Phi_{Bn}/KT\right) \quad\quad (2-12)$$

where $A^{**}$ is the effective Richardson's constant, which equals 120 (A/K$^2$.cm$^2$) for n-type silicon and $\Phi_{Bn}$ (eV) is the barrier height. It is assumed in the derivation of this theory that the barrier height is much larger than thermal fluctuations ($\Phi_{Bn} > 3KT$) and that thermal-equilibrium fluxes are not altered by electrical bias [8]. The ideality factor ($n$) in equation (2-11) can deviate from unity for thermionic emission-diffusion because both effective Richardson's constant and Schottky barrier lowering are functions of the applied voltage bias. The ideality factor is given by [8,36]:

$$n = \frac{q}{KT}\frac{\partial V}{\partial(\ln J)} \quad\quad (2-13)$$

The barrier height ($\Phi_{Bn}$) can be calculated using the following equation [8]:

$$\Phi_{Bn} = \frac{KT}{q}\frac{A^{**}T^2}{J_S} \quad\quad (2-14)$$

*Dimitrov* [11] has reported that the obtained values of barrier height are very high in comparison to the barrier height at an Aluminum/silicon Schottky junction, as well as the large values of ideality factor. These considerations imply that the PSi/c-Si junction may be treated as a heterojunction between two semiconductors with forbidden gaps [45,52].

The current-voltage curves may take several shapes as shown in figure (2-14) depending on structural properties of PSi layer [45,53].

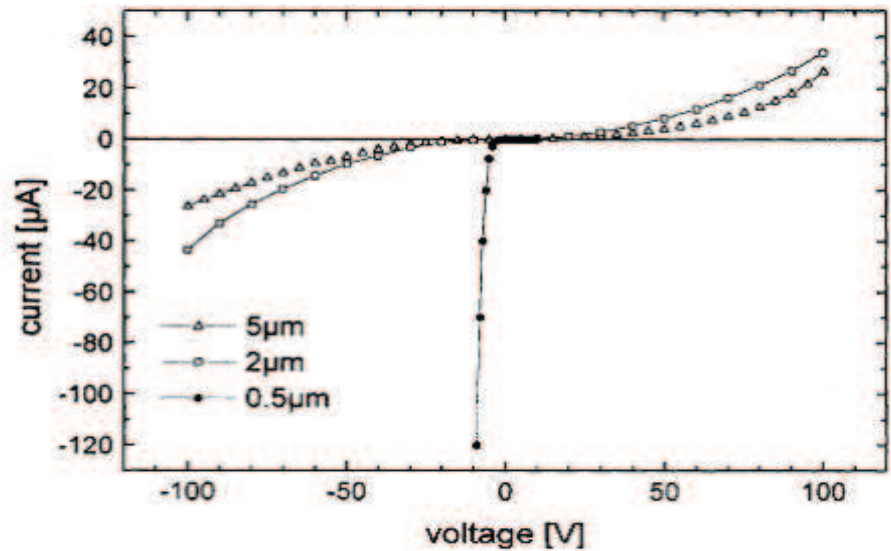

Figure (2-14); The current-voltage characteristics of M/PSi/c-Si/M sandwich structure with different thicknesses of PSi layer [10].

According to figure (2-14) there are obvious differences in electrical behavior of sandwich structure. While the thin sample exhibits diode like rectification, the thick PSi layers have nearly symmetric I-V features. *Ben-Chorin et al* [10,54] have suggested that the layered structure consist of the metal contact, the PSi layer and the doped silicon substrate should be understood qualitatively in terms of a series combination of a voltage-dependent resistance and a rectifying barrier. The latter could be the M/PSi contact [8-53] or the boundary layer between PSi and the substrate [55].

From the distinguished aspects of I-V characteristics is the saturation region (single or double) as shown in figure (2-15). *Monastyrskii et al* [11,55] have observed this aspect in their experimental results and have explained it according to theory of the isotype heterojunction presented by *Milnes and Feucht* [56].

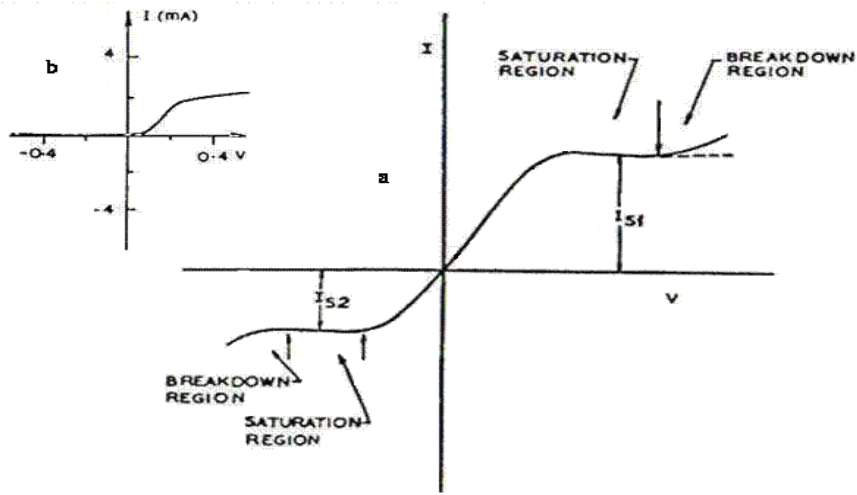

Figure (2-15); Schematic depicts the observed I-V characteristics of a n-n heterojunction (a) double saturation regions and (b) single saturation region [57].

## 2-4-2-2 *Resistivity and Conductivity of Porous Silicon*

The resistivity of PSi layer is a function of preparation conditions [4]. *Anderson and Canham* [8,58] have confirmed that the resistance which is a function of the resistivity of a PSi layer is very sensitive to the thickness and porosity of the layer as shown in figures (2-16) and (2-17).

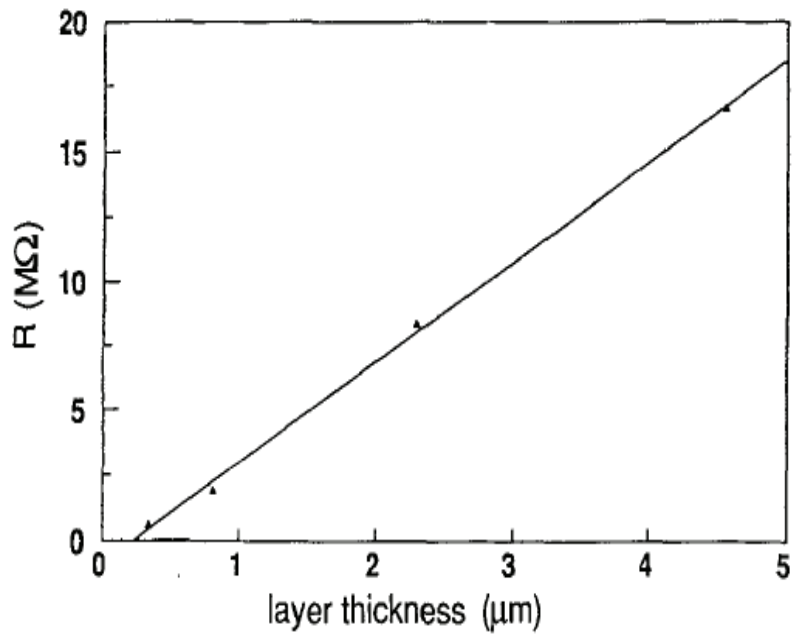

Figure (2-16); The resistance of PSi layer as a function of layer thickness [8].

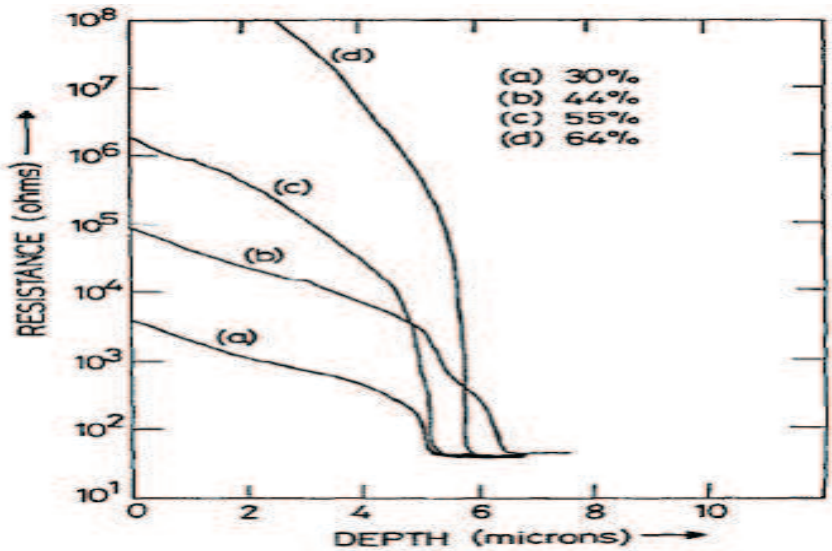

Figure (2-17); The resistance of the PSi layer as a function of its porosity [58].

Generally the resistivity of the PSi layer is very high in comparison to the resistivity of the intrinsic silicon wafer. *Beale et al* [9,49] have reported that the electrical resistivity of PSi layer is five order of magnitude higher than that of intrinsic. But at the same time the calculation of exact value of the resistivity of PSi layer is difficult, since it very sensitive to the ambient atmosphere [46,58]. The most common method to account the resistivity depends on the I-V characteristics and using the following equation [4,8]:

$$R = \rho \frac{d}{A_C} + R_S \quad \quad (2-15)$$

where $R$ ($\Omega$) is the measured resistance, $\rho$ ($\Omega.cm$) is the resistivity of PSi layer, $d$ (μm) is the layer thickness, $A_c$ ($cm^2$) is the contact area and $R_s$ ($\Omega$) is the resistance in series with the PSi layer. The measured resistance of PSi layer ($R$) represents the ratio between the applied voltages $\Delta V$ (V) to the resulted currents $\Delta I$ (A) in the linear portion of the I-V curve for the M/PSi/c-Si/M sandwich structure given as [4]:

$$R = \frac{\Delta V}{\Delta I} \quad \quad (2-16)$$

The series resistance ($R_s$) in equation (2-15) may depend on the resistance to hole transport through the PSi layer due to a depletion of carries in these nanostructures [59] which can be calculated according to *Ray and Chen* [58-60] as:

$$R_S = R_\circ \exp\left(\Delta E / KT\right) \quad\quad\quad\quad\quad\quad (2-17)$$

where $R_\circ$ ($\Omega$) is the resistance at room temperature, $\Delta E$ (eV) is the change in the energy gap due to quantization effects, $T$ (K) is the absolute temperature and $K$ (J/K) is the Boltzmann's constant. The presence of a series resistance in a junction-diode device as shown in figure (2-18) can limit the electrical-to-optical conversion efficiency since the applied voltage will be dropped across the resistance [59].

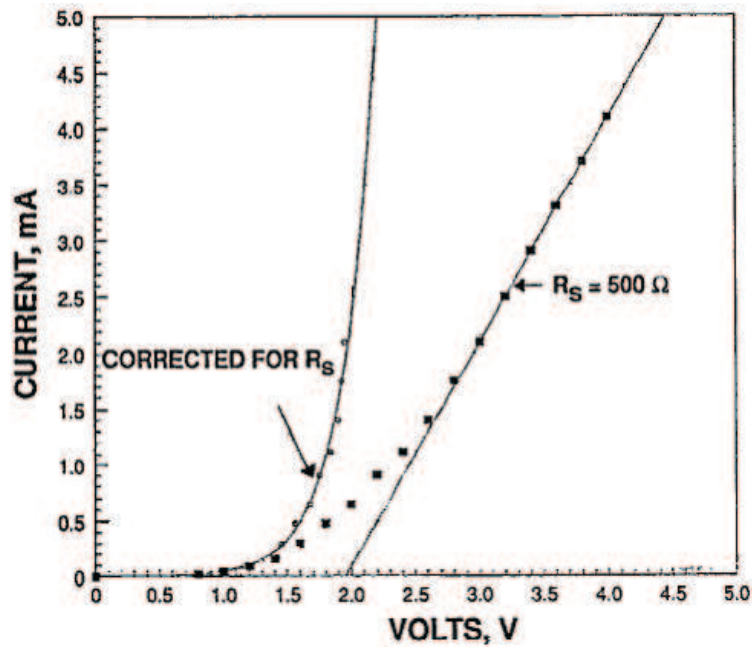

Figure (2-18); The I-V characteristics of PSi-based diode indicating presence of series resistance [59].

The conductivity of the PSi layer is governed by the geometrical effects and structural properties of the layer [22]. *Lee* [42] has demonstrated that the increase in the porosity of PSi layer leads to increase in the activation energy and decrease in the conductivity of PSi layer at room temperature as shown in figure (2-19).

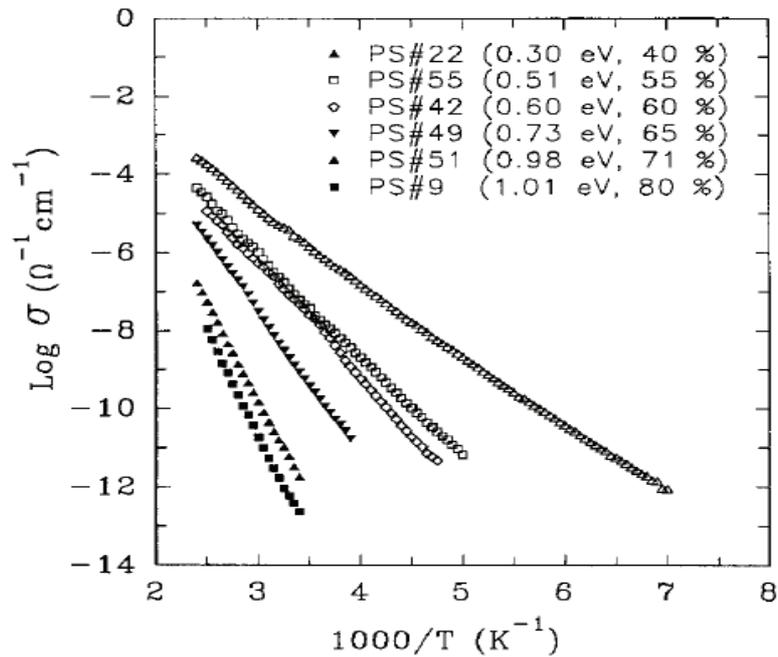

Figure (2-19); Temperature dependence of the dark conductivity for the PSi layer with different porosities [42].

## 2-4-2-3 *Capacitance of Porous Silicon*

Capacitance measurements offer a valuable tool for investigating the electrical properties of PSi layer and devices containing PSi layer [4,8]. *Dimitrov and Pineik et al* [11,40,53] have confirmed that the parameters such as built-in potential, width of the depletion layer and effective carrier density can be calculated from capacitance measurements for the M/PSi/c-Si/M sandwich structure.

The capacitance of PSi layer depends on its morphological properties since it decreases with increasing of layer thickness and porosity as shown in figure (2-20) [8,61].

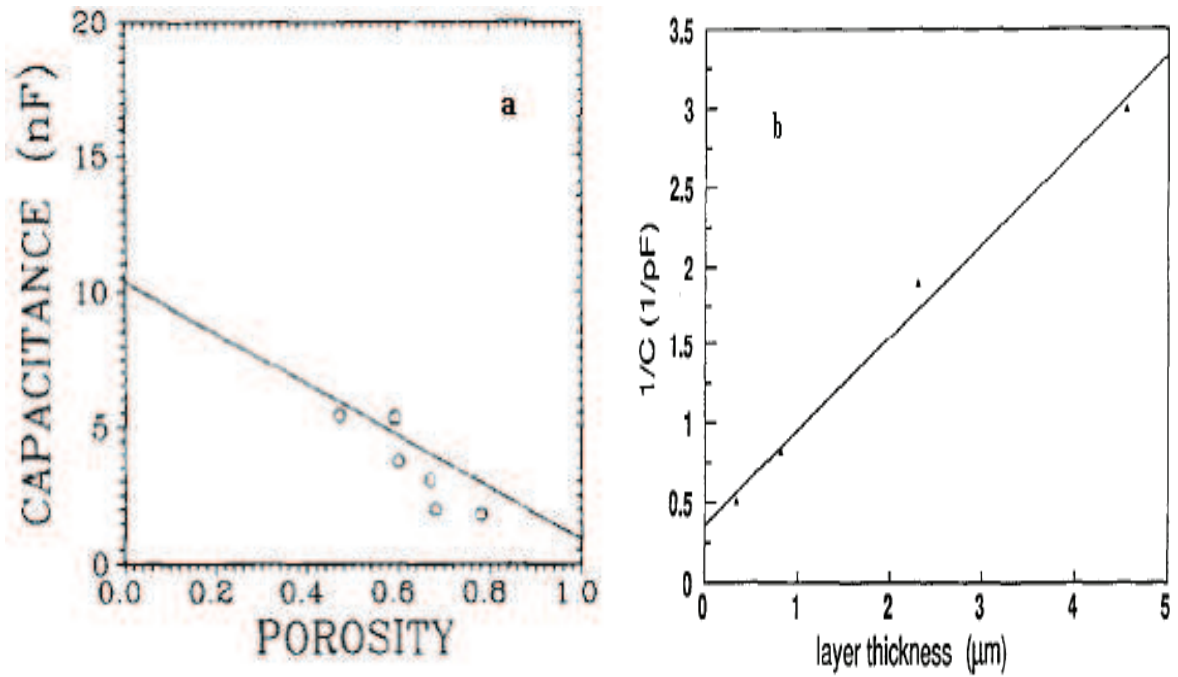

Figure (2-20); The capacitance of PSi layer as a function of the (a) porosity [62] and (b) layer thickness [8].

The capacitance of PSi layer $C_{PSi}$ (F) is given by [40,43]:

$$C_{PSi} = \frac{S \varepsilon_{PSi} \varepsilon_\circ}{d} \quad \quad (2-19)$$

where $\varepsilon_\circ$ (F/cm) and $\varepsilon_{psi}$ (F/cm) are the permittivity of free space and PSi layer respectively, $S$ (cm$^2$) is the PSi area and $d$ (μm) is the thickness of PSi layer.

### 2-4-2-4 *Photocurrent and Sensitivity of PSi-based Structures*

The photocurrent effect is one of the most useful means for finding out the optical and electrical properties of PSi layer, since it can be regarded as a complementary phenomenon of the light emission [4]. *Canham* [62] has reported that the spectroscopic analyses of the photoresponse and dynamic of photocurrent have given some important and selective information about the photoelectronic process in PSi-based sandwich structures.

The structure of the conventional PSi-based photoelectric systems is M/PSi/c-Si/M where PSi layer with the thickness range from few micrometers to ~ 100 μm [45]. The resulted photocurrent is relying on the thickness of PSi layer. *Koshida* [63] has illustrated that the photocurrent of PSi-based sandwich structure reduces with increasing of PSi layer thickness and become almost constant in thickness beyond 30 μm as shown in figure (2-21).

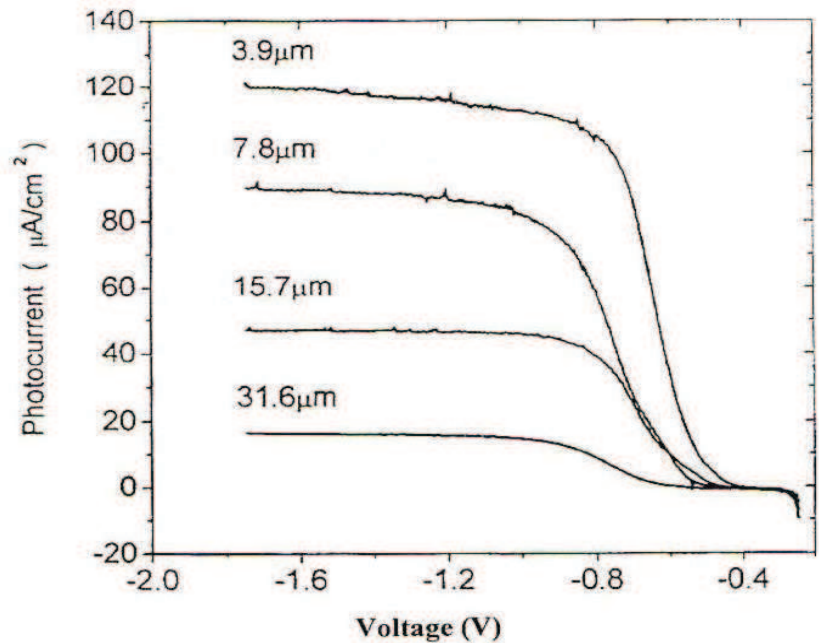

Figure (2-21); Photocurrent as a function of applied voltage for different PSi layer thickness [64].

The origin of photoresponse of the contained devices on PSi layer has been studied by many researchers. *Pulsford* [65] has proved that the sign of the photocurrent at zero-bias voltage regime as shown in figure (2-22) is resulting from transport of photogenerated charge carriers across the PSi/c-Si heterojunction and ascribes this result to the fact that the net current at this interface is determined by the band bending due to the space charge region and by the densities of states on either side of the junction. Space charge region arises from the hole depletion during etching process.

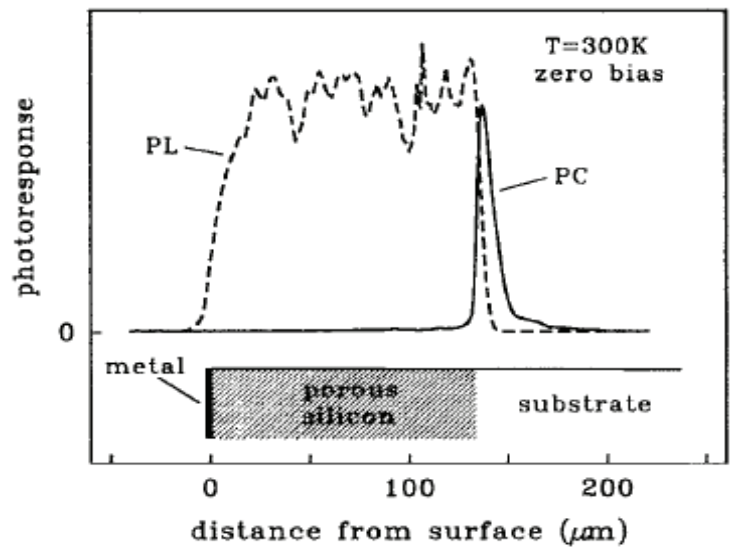

Figure (2-22); The photocurrent (PC) measured across the cleaved edge of the M/PSi/c-Si diode at room temperature under zero applied bias [65].

Furthermore, *Smestad et al* [8,46,48,] have studied I-V characteristics of PSi-based sandwich structures under a wide range of reversely bias voltages. They found that the photocurrent is observed only at reverse bias, and it increases with increasing of illumination intensity as well as the resulted photocurrent is voltage dependent as shown in figure (2-23). At high reverse bias it increases until it arrives to the saturation state while for lower voltage, it tends to diminish. Also they attribute their results to PSi/c-Si heterojunction.

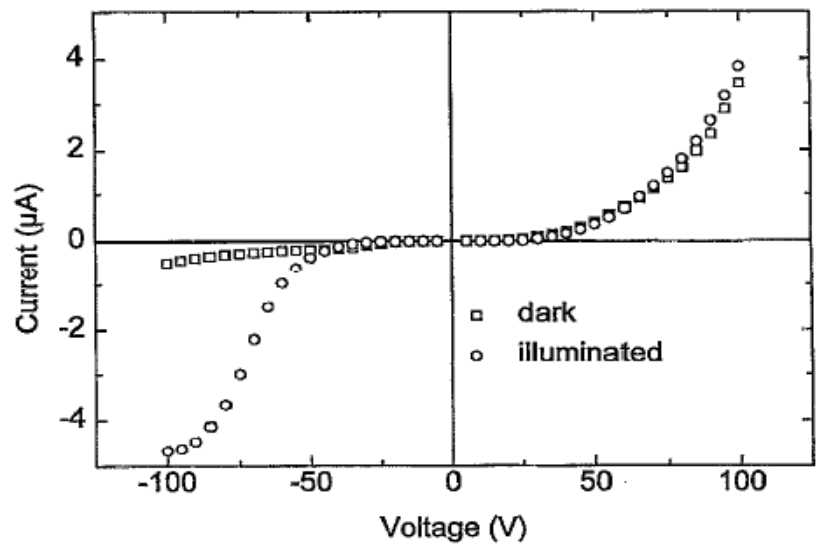

Figure (2-23); The current as a function of applied voltage for PSi-based device, at dark and under illumination [46].

Porous silicon technology provides fast and sensitive photodetection in the visible range [44,66] therefore has attracted large interest by researchers and scientists. *Sveehnikov et al* [45,46] have studied the sensitivity of PSi-based devices with thin and thick PSi layer as shown in figure (2-24).

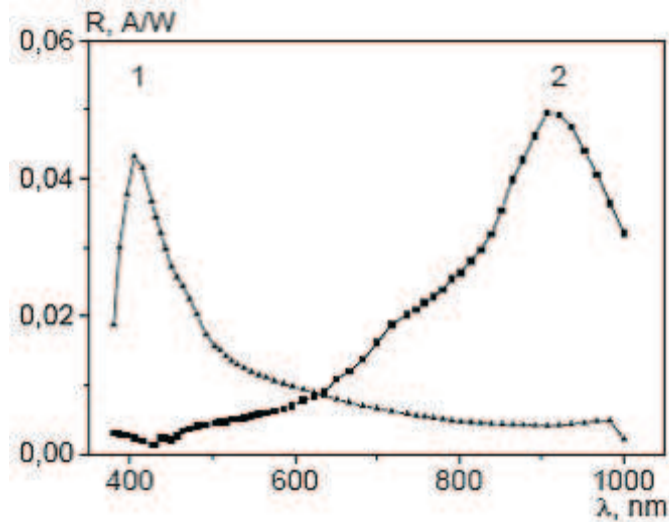

Figure (2-24); Spectral sensitivity of M/PSi/c-Si/M sandwich structure with (1) thick and (2) thin PSi layers [45].

They found that the spectral sensitivity of device with thin PSi layer has the same peak of silicon photodiode while the contrary is device with thick PSi layer.

When the energy of quanta of the incident light exceeds the band gap silicon substrate, but is smaller than that of PSi layer, light is absorbed in bulk silicon and the sign of photoresponse originates from a band bending inside the silicon substrate at its interface to the PSi layer, but when the photon energies of the incident light exceeds the band gap of PSi layer most of incident light would be absorbed by this layer which leads to increasing of photogenerated charge carriers in the PSi layer and then domination of the PSi sensitivity.

The sensitivity according to *Zheng and Diesinger* [64,67] is given as:

$$R_\lambda = \frac{I_{ph}}{P_{in}(\lambda)} \quad \quad (2-20)$$

where $P_{in}$ (W) is the power of incident light and $I_{ph}$ (A) is the photocurrent which is given as [67]:

$$I_{ph} = I_b - I_d \quad \quad (2-21)$$

where $I_b$ (A) is the total reverse current and $I_d$ (A) is the dark current.

The quantum efficiency $Q_E(\lambda)$ is given by [64,66] as:

$$Q_E(\lambda) = \frac{1240}{\lambda} R_\lambda \quad \quad (2-22)$$

where $\lambda$ (nm) is the wavelength of incident light.

## 2-4-3 *Optical Properties*

The optical properties of PSi are currently receiving widespread interest motivated by their potential applications as electroluminescent devices [62]. It has been shown that such structures exhibit dramatic quantum-size effects leading to an increase in the effective band gap [3,19,68] and to efficient radiative-recombination. Photoluminescence (PL) from PSi with energies greater than the band gap of bulk silicon was first reported by *Canham* [18]. But its significance became apparent only when highly efficient visible photo-luminescence was obtained at room temperature by *Canham* [3]. The band-gap luminescence is observed from all types of freshly etched silicon substrate ($P^-$, $P^+$, $n^+$, $n^-$) produced in aqueous or ethanoic HF provided the porosity is high enough to introduce high PL efficiency, but it is not observed from a macroporous layer in an

n⁻-type substrate which has silicon skeleton dimensions too large to expect quantum-size effect [69,70,71].

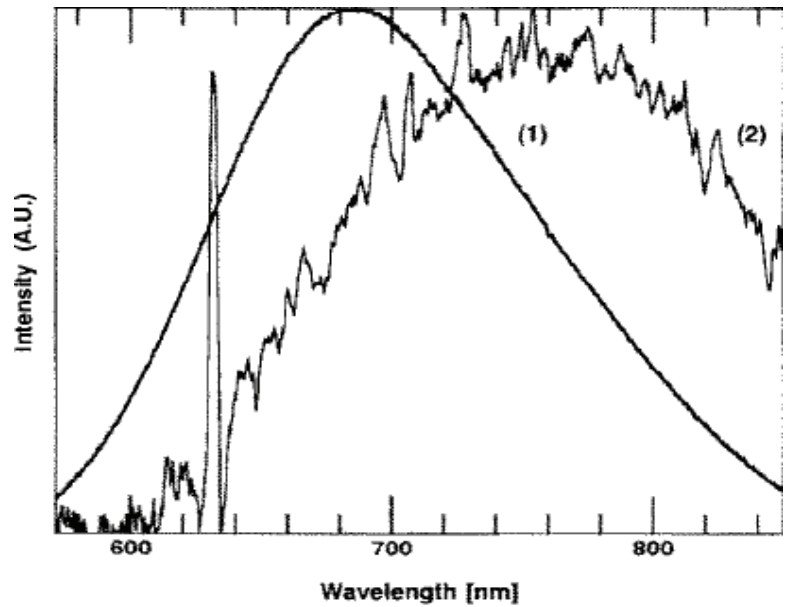

Figure (2-25); Photoluminescence of PSi layer for different excitation energies. Spectrum (1) refers to the 450 nm and (2) to 632 nm [72].

*Bsiesy* [72] has reported that the Photoluminescence could be obtained from PSi layer with high-porosity without the need for subsequent chemical dissolution, but at the same time it depends on the photon energies of incident light as shown in Figure (2-25). The most commonly accepted model for the Photoluminescence is that which arises from electrons and holes localized in quantum crystallites in the PSi layer [3,19].

## 2-4-4 *Chemical Properties*

The surface chemical composition of PSi layer has been studied by X-ray photoelectron spectroscopy and Fourier transform infrared (FTIR) spectroscopy [73,74]. The (FTIR) signal from PSi layer is very strong and easier to measure than that from bulk silicon due to the large specific area [9].

The chemical properties of freshly etched silicon samples are very sensitive to surrounding ambient, where it varies with time during storage in air [18], since the pore surface includes a high density of dangling bonds of silicon for original impurities such as hydrogen and fluorine which are residuals from the electrolyte [4].

*Andrea* [5] has reported that the large specific surface area of PSi layer (200 – 600 m$^2$/cm$^3$) affected chemical properties which will influence the optical, electrical and mechanical properties of PSi layer.

## 2-5 *Applications of Porous Silicon*

Porous Silicon has attracted noticeable attention of many scientists due to the possibility of developing PSi-based devices including light emitting diode, waveguides, solar cells, photodetectors, photomodulators, photoresistor and sensors [7,4,44,65,66]. The high specific surface area of the PSi layer enables applications in chemical/biological sensing and can be exploited in the
future heterogeneous chemical catalysis [75]. A new technology based on PSi layer uses the advantage of the macroporous structures formed on n-type Si wafers (higher specific capacitance) [5].

*Maruska* [59] has reported the fabrication of visible p-n junction light emitting diode (LED) based on PSi material which emits yellow-orange light under forward bias. *Bohn* [66] fabricated photodetector based on PSi layer with responsivity of up to 4 (A/W). *Anderson* [7] manufactured vapor sensor depending on PSi layer. It was found the conductance of the PSi layer increases 600% in the presence of saturated ammonia.

Table (2-1) shows a compilation of relevant applications areas based on the review papers of *Ben-Chorin* [75], *Anderson* [7], *Pulsford* [65], *Steiner* [76], *Seong* [77], *Maruska* [59],………etc.

| Application area | Component based on PS | Utilized property of PS |
|---|---|---|
| Optical applications | Interference filter (Bragg, Fabry-Perot) | Tunable refractive index and layer thickness by manufacturing |
| | Solar cell (antireflection coating) | Tunable refractive index and layer thickness by manufacturing |
| | Waveguide | Tunable refractive index and thickness |
| | | Direct integration with electrical components |
| Optoelectronic applications | Light emitting diode (LED) | Electroluminescence |
| | | Schottky contact with some metals |
| | | p-n junction without ion bombardment |
| | Photodetector (PD) | p-n junction without ion bombardment |
| | | Wavelength dependent refractive index and absorption coefficient |
| | Field emission device (FED) | Free charge carriers generation |
| | Photonic crystal | Designable 2 and 3 D structure by periodically alternated layers with different refractive indices |
| | Optical logic gate | Nonlinear optical absorption |
| Microelectronic applications | Cold cathode | Electron and visible light emission from the surface at room/low temperature |
| | Epitaxial growth of silicon films on the PS surface | Suitable substrate for epitaxial growth |
| | Thermal / electrical insulation (IPOS, SOI, FIPOS) | Low thermal conductivity |
| | | High electrical resistivity |
| | Silicon capacitor (SIKO) | Macroporous structure (n-type Si) |
| | | High specific capacitance |
| Sensors and actuators | Micromechanical structures / MEMS | Selective etching of PS on Si |
| Chemical sensors | Liquid or gas material sensing | Electrical conductivity and capacitance effect |
| | | Band structure change |
| | | Photoluminescence quenching |
| Biological application | Sensing | Compatibility with the living organism |
| | | Change of electrical and optical properties of PS by biomolecules |

Table (2-1); Various applications of porous silicon material [5].

# CHAPTER THREE

# Experimental Procedure

## 3-1 *Introduction*

This chapter includes describing of all instruments and apparatuses that have been employed in the preparation of PSi layers as well as the techniques which have been utilized to study and clarify the characteristics of PSi layers.

## 3-2 *Sample Preparation*

Commercially available mirror-like n-type (111) oriented silicon wafers of (500 μm) thickness with resistivity ($\rho$ = 0.02 and 3.5 Ω.cm) respectively have been used as substrate. Before Photoelectrochemical (PEC) etching process, the silicon wafer has been cut out into small pieces in different dimensions. The interdependent n-type silicon samples in this work have been cut out into small fragments in dimensions of (1.5 × 1 cm). These pieces were rinsed with ethanol to remove dirt followed by etching in dilute (10%) hydrofluoric (HF) acid to remove the native oxide layer.

We deposited aluminum film on the front side of silicon sample in order to creating an ohmic contact. This has been achieved by using thermal evaporation process in vacuum chamber ($10^{-5}$ torr). After (PEC) etching process, the samples were rinsed with dionized water and left in environment for few minutes to dry and then stored in a plastic container filled with ethanol to prevent the formation of oxide layer on the samples as shown in figure (3-1). The PEC etching process was achieved at room temperature.

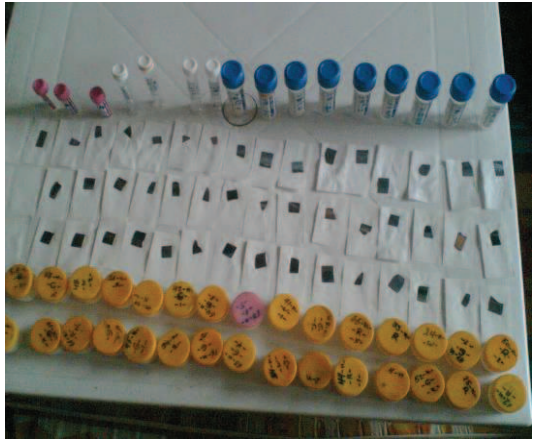 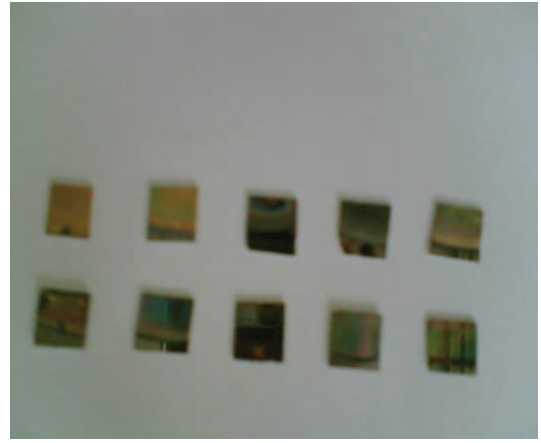

                (a)                 (b)

Figure (3-1); Photographs of samples after PEC etching process;
(a) All prepared samples. (b) The interdependent samples in this work.

### 3-3 *Experimental Set-up of PEC Etching Process*

The simple set-up of PEC etching process consisted of one commercially available CW diode laser with power (2W) and (810 nm) wavelength. The photo-electrochemically etched area for all samples has been 0.6 cm$^2$ and the reached laser power to specimen has been (1.4 W), so the laser power density is 2.33 W/cm$^2$. Dual *Farnel* LT 30/2 power supply as a current source, digital multimeter and an ethanoic solution of 24.5% HF is obtained by 1 volume of ethanol and 1 volume of 49% wt. Hydrofluoric (HF) acid (the HF concentration in the ethanoic solution is given by ((1 × 49%)/(1+1) = 24.5%).

Figure (3-2) depicts a schematic diagram of the PEC set-up. The sample has been mounted on a Teflon cell in such a way that current should not pass from the back surface. At the same time the silicon was mounted as an anode and the electrical circuit was completed by putting a platinum electrode as a cathode in a parallel way to achieve the homogeneous PSi layers. The value of the utilized current density for PEC etching process has been (40 mA/cm$^2$).

Simultaneously, the samples have been irradiated with laser radiation. The distance between the laser source and the sample was (25 cm).

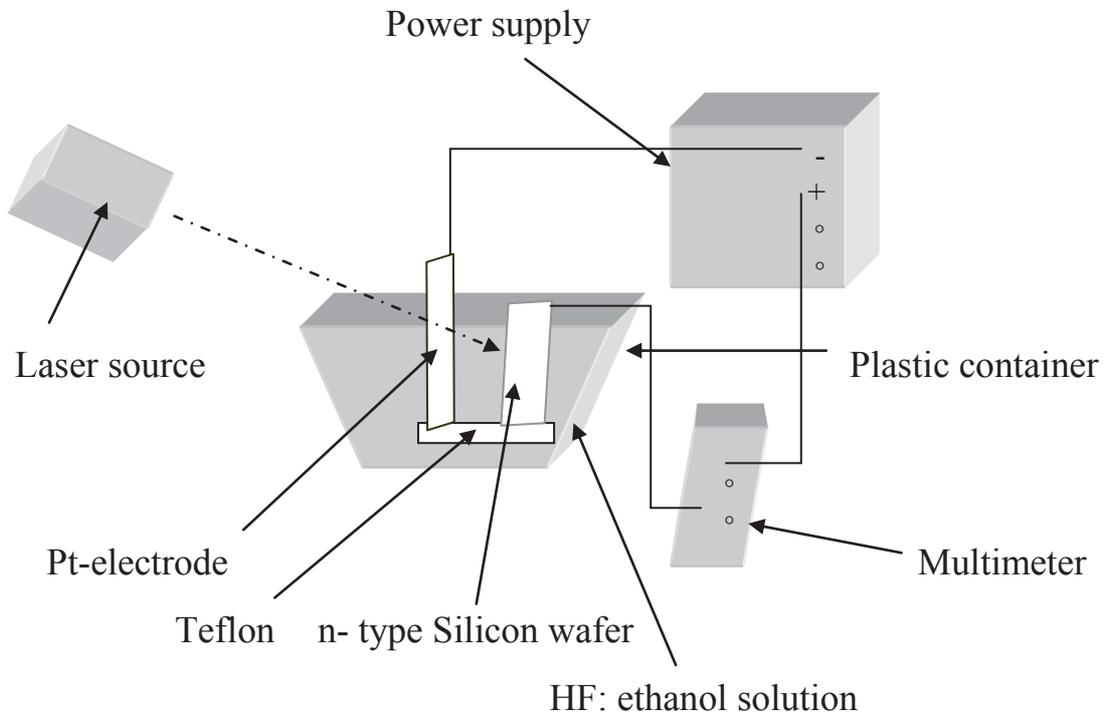

Figure (3-2); Schematic diagram depicts the PEC etching process.

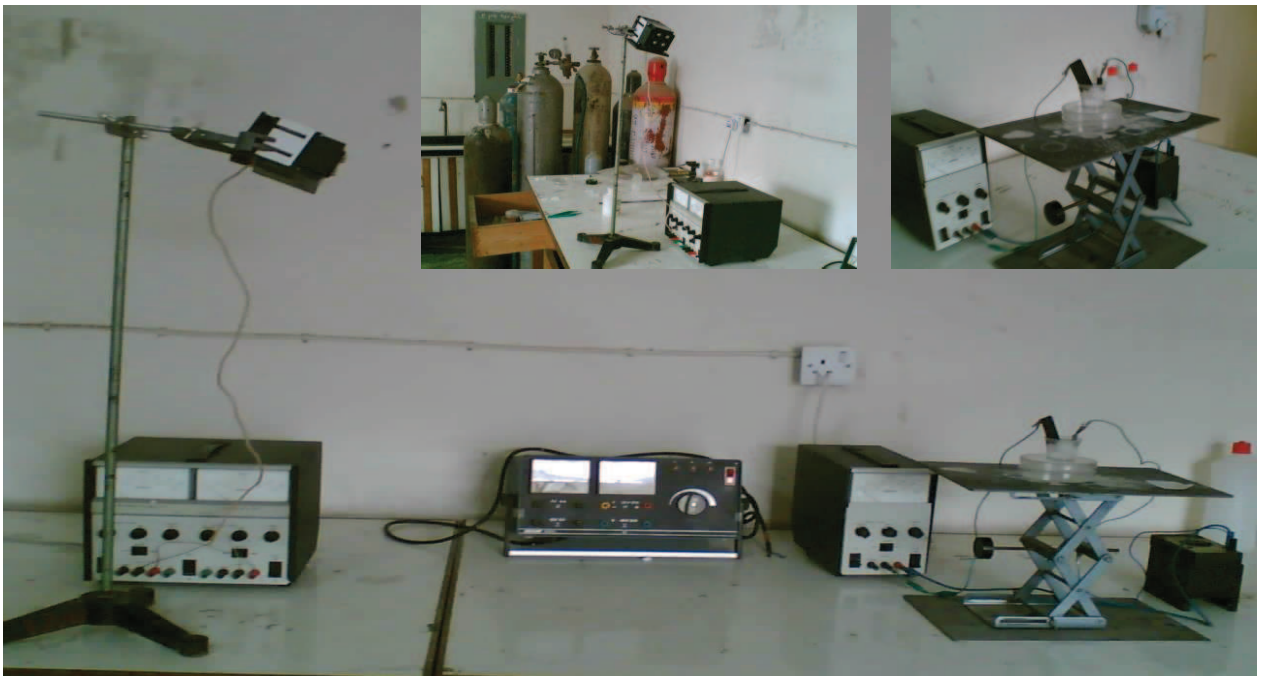

Figure (3-3); Photograph of experimental set-up of the PEC etching process.

### 3-4 *Metallization*

To measure the electrical properties and other related properties of the PSi layer, an ohmic contact with this layer is a necessary point. Aluminum with a high purity (99.9999%) was used to obtain (Al layer with thickness 200 nm) electrodes. Thermal evaporation was achieved using a vacuum evaporation system (*Blazer BAEPVA* 063H), at a vacuum pressure of $10^{-5}$ torr. After the evaporation process, the thickness of evaporated film on a glass substrate was measured using Michelson's device for interference supplied from Phywe Company in Germany. This measurement achieved in department of Physics in Sciences College in Kufa University. Finally, the outward connections were made by connecting copper wires to the electrodes.

### 3-5 Measurements

The structural and electrical properties of PSi layer have been investigated in this work by achieving of some measurements as follows:

### 3-5-1 *Morphological and Structural Measurements*

The structural characteristics of PSi layer have been studied by a range of techniques depending on direct imaging of material such as, high resolution scanning electron microscopy (HRSEM) with a resolution down to 2 nm, transmission electron microscopy with a resolution down to atomic dimensions [77,78], atomic force microscopy (AFM) which gives a direct surface image, scanning tunneling microscopy (STM) that provides imaging information related to the surface electronic structure and X-ray diffraction technique [4].

In this work we have used the following techniques.

### 3-5-1-1 *Scanning Electron Microscopy (SEM) Technique*

The structural properties of PSi layer such as surface morphology, specific surface area, pore width, pore shape, thickness of walls between pores, and layer thickness have been studied in this work by using high resolution scanning electron microscopy (HRSEM) (Leo-1550). The given voltage for each image was (5KV). The magnification of device has been in the range from 2.30 kx to 109.13 kx. The SEM measurements were carried out in the (Institute for Bio-and Nano-systems (IBN2)-Germany).

### 3-5-1-2 *X-Ray Diffraction Technique*

The morphological properties of PSi layer such as nanocrystallite size, the structure aspect of PSi layer (crystalline or amorphous) and lattice constant have been investigated in this work by using X-ray device (Philips-PW 1840) supplied from Philips Company. The source of X-ray radiation has been $CuK_\alpha$ radiation with 0.15405 nm wavelength. The device has been operated at 40 KV and 20 mA emission current. This measurement was achieved in the Company of Geological survey and Mining.

### 3-5-1-3 *Gravimetric Measurements*

The porosity of PSi layer has been calculated by measuring the weight of the samples before and after etching process as well as the weight of sample after removing of PSi layer from sample by immersing it in 1 M KOH solution for 10 s [4,8] and using equation (2-5). *Sartorius* BL210S digital steelyard with accuracy of $10^{-4}$ gm was used to weight the samples. This measurement has been supported by (SEM) images of the samples.

### 3-5-2 *Electrical Measurements*

The electrical and photoelectrical properties of PSi-based device sandwiches (Al/PS/n-Si/Al) have been studied. All measurements were carried out in a dark room and then under daylight illumination. These measurements involve:

### 3-5-2-1 *Current-Voltage Characteristics*

For current-voltage measurements, *Tektronics* CDM 250 multimeter, digital multimeter and a dual *Farnel* LT30/2 power supply were used. The forward current was recorded when a positive voltage was applied to Al metal contact with the PSi layer with respect to the Al electrode on the crystalline silicon substrate as shown in Figure (3-4).

The same measurements were repeated under reverse-bias condition.

The current saturation ($J_s$), the ideality factor (n), the resistivity of the PSi layer ($\rho$), barrier height of junction ($\Phi_{Bn}$) were calculated by using the equations (2-12), (2-13), (2-14) and (2-15) [4,8,36].

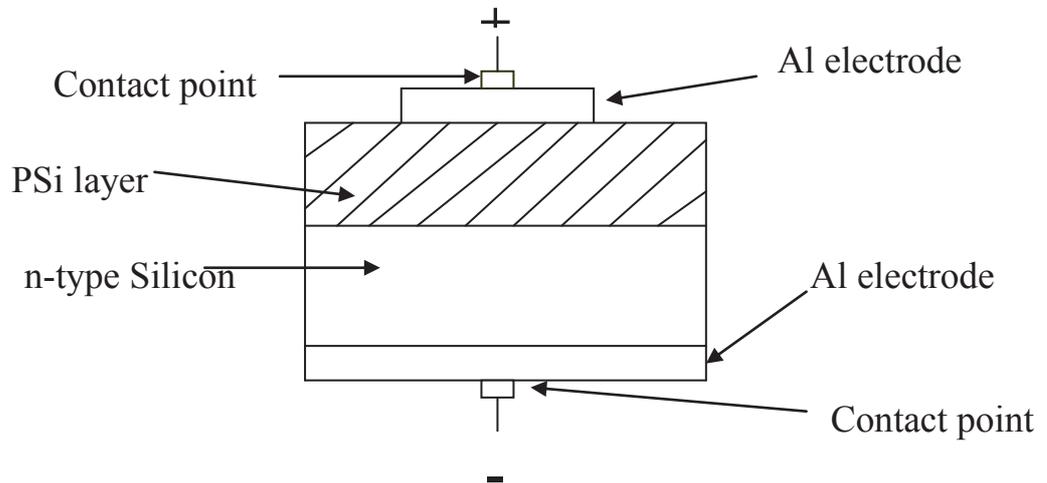

Figure (3-4); Cross-sectional of Al/PSi/n-Si/Al sandwich structure.

### 3-5-2-2 *Spectral Sensitivity Measurements*

The sensitivity of PSi-based devices has been measured using several filters with wavelength range from 320 to 859 nm, in addition to *Tektronics* CDM-250 multimeter and *Farnel* LT30/2 power supply, where the photo-current was measured under 5V reverse biasing condition. The sensitivity $R_\lambda$ and the quantum efficiency $Q_{E(\lambda)}$ are calculated from equations (2-20), (2-21) and (2-22). The power of the incident light at a certain wavelength was measured using silicon power meter.

# CHAPTER FOUR

# Results and Discussion

## 4-1 *Introduction*

This chapter presents results and discussions of the porous silicon properties fabricated by PEC etching process of monocrystalline silicon. The structural and electrical characteristics of PSi layer were investigated depending on the measurements mentioned previously.

## 4-2 *Morphological and Structural Properties*

Porous silicon layer exhibits dramatic and very special structure characterized by the presence of interconnected pores in a single crystal. All morphological properties of the prepared PSi layer such as porosity, layer thickness, pore width, pore shape, the wall thickness between two pores, specific surface area, the morphological aspect of PSi layer (crystalline or amorphous), lattice constant and nanocrystallite size are strongly dependent on the etching conditions. These features of PSi have been studied by direct imaging for its structure by high resolution scanning electron microscopy (HRSEM), X-ray diffraction technique (XRD) and the gravimetric method.

### 4-2-1 *Pore Width and Shape*

In the most general concept, the pore is an etched pit whose depth exceeds its width [4]. The pore initiates as a defect at the crystalline silicon surface and any change in preparation parameters such as silicon substrate resistivity, illumination and etching time leads to PSi layer which has pores with various sizes and shapes [19,37]. Figure (4-1,a,b,c) shows SEM images (top-view) of PSi layer produced at different irradiation times (5-15 min) with constant current density of 40 mA/cm$^2$.

The irradiation has been achieved using diode laser of 2.33 W/cm$^2$ power density and 810 nm wavelength on (3.5 Ω.cm) n-type silicon immersed in 24.5 % HF concentration.

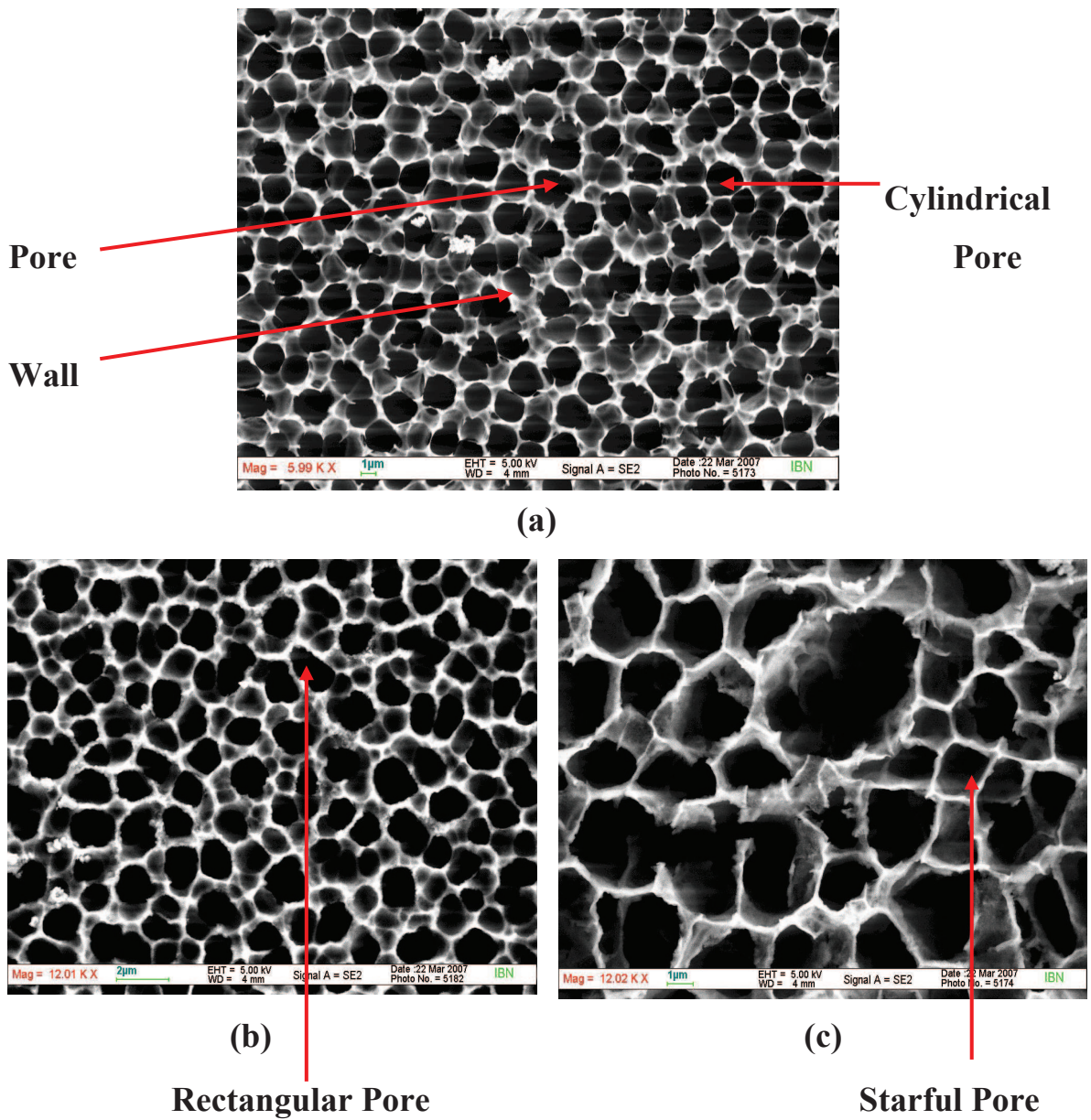

Figure (4-1); SEM images (top-view) of PSi layers prepared at different etching times, (a) 5 min, (b) 10 min and (c) 15 min.

We can note from this figure that the pore width increases with increasing of etching time. This largeness in pore width may be attributed to increasing of holes number on surface of silicon electrode with etching time which leads to preferential dissolution between nearest-neighbor pores, thereby promoting the pore-pore overlap.

However, the etching rates may be different and then leads to nonuniformity in values of the pores width as summarized in table (4-1). The cause of this variation may be due to the nonuniform of the power density distribution of illumination, which leads to nonuniform photocurrent densities consequently, resulting in different pore width. The nonuniform of the power density distribution of illumination is attributed to fact that the laser beam intensity will decrease gradually from its center to its periphery (Gaussian-laser beam). Figure (4-1,a) shows the PSi layer possesses pores with cylindrical shape, while the PSi layer in figure (4-1,b) has nearly structure with rectangular shape. From figure (4-1,c) we can observe PSi layer possessing approximate construction composed of pores with Starful shapes. We can conclude from these results, that the collecting areas of the minority carriers and then the shape for each pore have been determined by the distance between neighboring pores. Our results are in good agreement with those of other investigations [20,28,29,19,78,79]. To avoid ambiguity, all reported values for pore width in table (4-1) were calculated for upper holes of the pores. Figure (4-2,a,b,c) shows SEM images (top-view) of PSi layer produced at different irradiation times (5-15 min) with constant current density of 40 mA/cm$^2$. The irradiation has been achieved using diode laser of 2.33 W/cm$^2$ power density and 810 nm wavelength on (0.02 Ω.cm) n-type silicon immersed in 24.5% HF concentration. Figure (4-2,a,b) shows relationship between etching time and pore width, where the value of pore width increases with increasing of etching time as reported in table (4-1) and also one can note that the pore shapes depend on the separation between them.

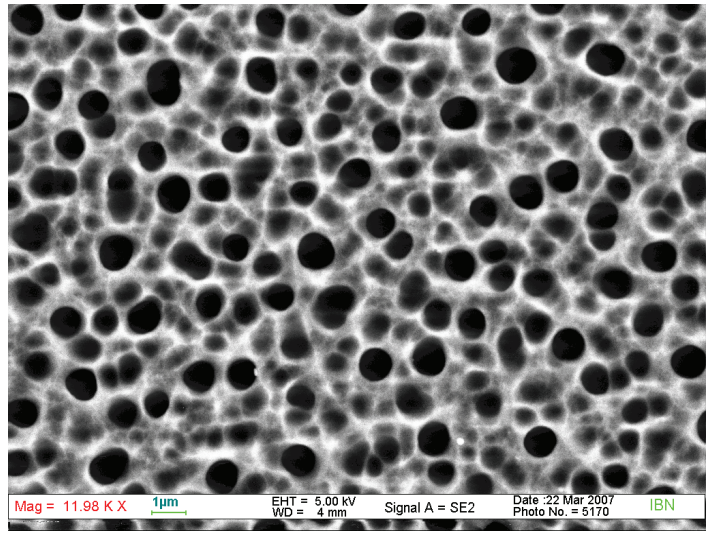

a

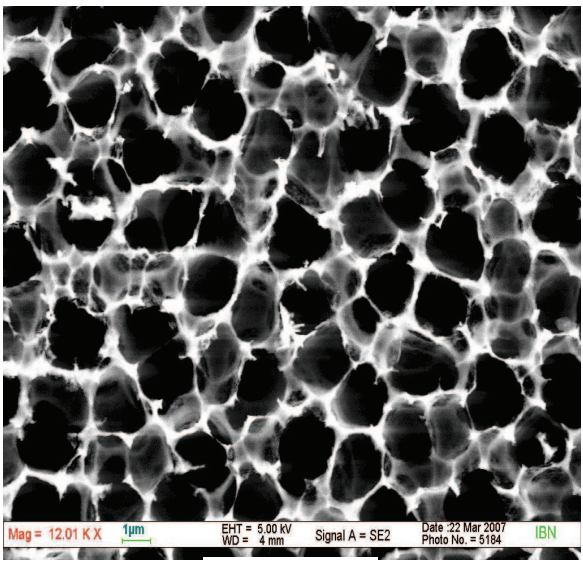

b

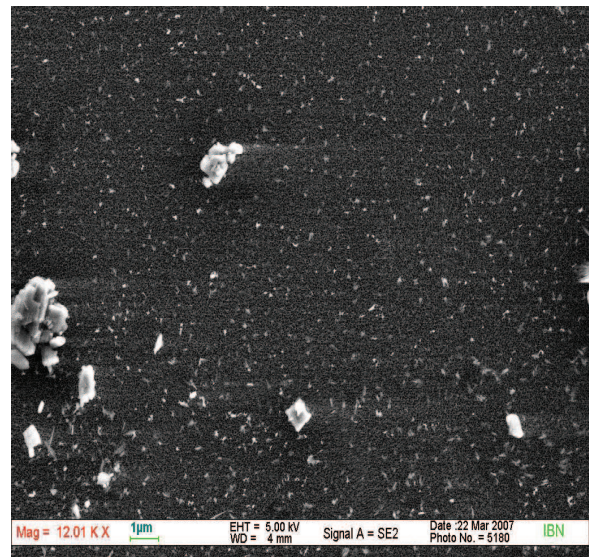

c

Figure (4-2); SEM images (top-view) of PSi layers prepared at different etching times (a) 5 min, (b) 10 min and (c) 15 min.

As etching process proceeds, extra holes reach the surface leading to further dissolving of the silicon and with more time the carriers will be confined to the thin column leading to dissolve these columns and excessive etching takes place until the carriers rearrange again on the whole surface and initiate a new layer as shown in figure (4-2,c).

Table (4-1); Pore width and pore shape of PSi layer at different etching times for high-doped and low-doped substrates resistivities respectively.

| Substrate resistivity (Ω.cm) | Etching time (min) | Pore width (μm) | Pore shape |
|---|---|---|---|
| 3.5 | 5 | 0.58 – 2.16 | Cylindrical |
| 3.5 | 10 | 0.66 – 2.33 | Rectangular |
| 3.5 | 15 | 1.16 – 6.25 | Starful |
| 0.02 | 5 | 0.41 – 1.25 | Cylindrical |
| 0.02 | 10 | 0.5 – 2.08 | Rectangular |
| 0.02 | 15 | 0.04 | Cylindrical |

In summarized framework, we can deduce several facts from figures (4-1), (4-2) and the described data in table (4-1). (1) The pore width of PSi layer increases with increasing of silicon substrate resistivity with respect of etching time and power density distribution of illumination. (2) Nearly the growth of the imaged pores in figure (4-1) is complete, while the opposite is true for pores in figure (4-2). (3) According to pore width value, we can classify photographed PSi layer in figures (4-1) and (4-2) as macroporous silicon layer, except the PSi layer in figure (4-2,c) as mesoporous layer. (4) Only in the case of figure (4-2,c) when the irradiation time is increased the carriers were confined within the thin walls which leads to increased etching for these walls and then complete removing of PSi layer followed by new growth of the pores.

All these results indicate the relationship between pore width and silicon substrate resistivity. For high silicon substrates resistivities, the charge carriers get big chance for initiating and growing of pores. On the other hand, the contrary is right for silicon substrates with low resistivities.

## 4-2-2 *Etching Rate and Layer Thickness*

The etching rate of PEC etching process has attracted a great attention since it describes the etching process speed [78]. The etching rate depends on the formation parameters and is governed by the diffusion rate and drift velocities of holes to the surface [26,35]. We have studied in this direction, the etching rate of PEC etching process to investigate how the etching speed could be governed by effective parameters.

Etching rate is defined as the ratio between PSi layer thickness and the etching time, as shown in equation (2-6).

It was found that the wavelength which produces higher etching rate is 810 nm wavelength. This could be attributed to the fact; the photon energy of 810 nm is 1.53 eV and that is the case of near resonance absorption since it is close to the energy gap of the bulk silicon (1.17 eV), the penetration depth of this wavelength is 22 μm.

Figures (4-3) and (4-4) shows cross sectional SEM images of the prepared PSi layers at various irradiation times (5-15 min) with fixed current density of 40 mA/cm$^2$ on (3.5 and 0.02 Ω.cm) n-type silicon substrates resistivities respectively, immersed in 24.5% HF concentration.

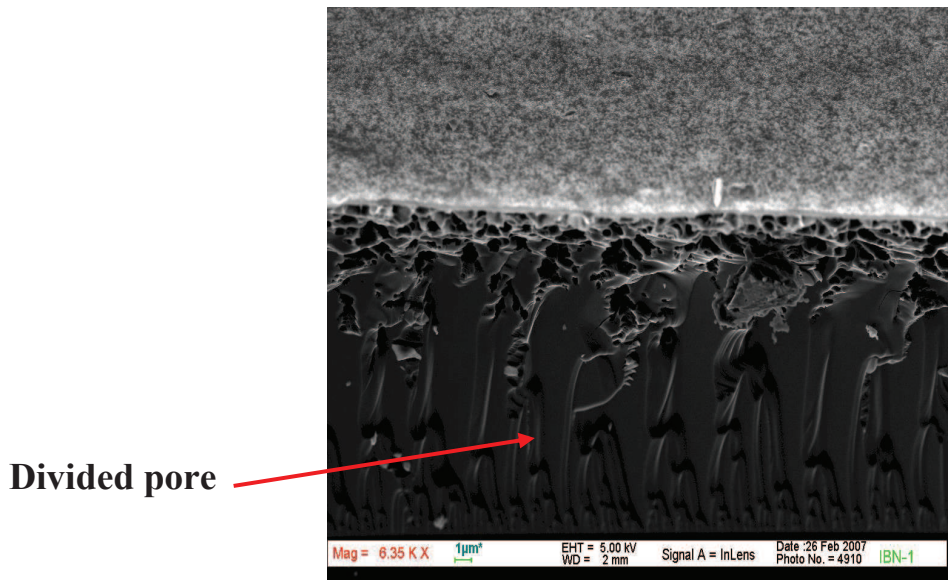

**Divided pore**

a

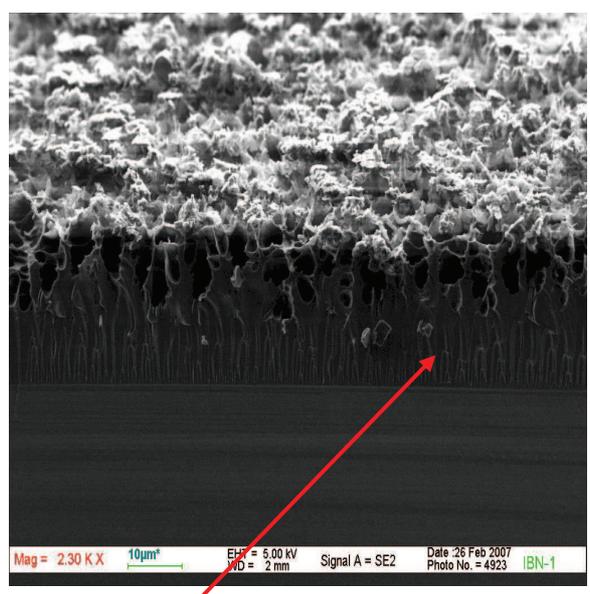

b

**Branching**

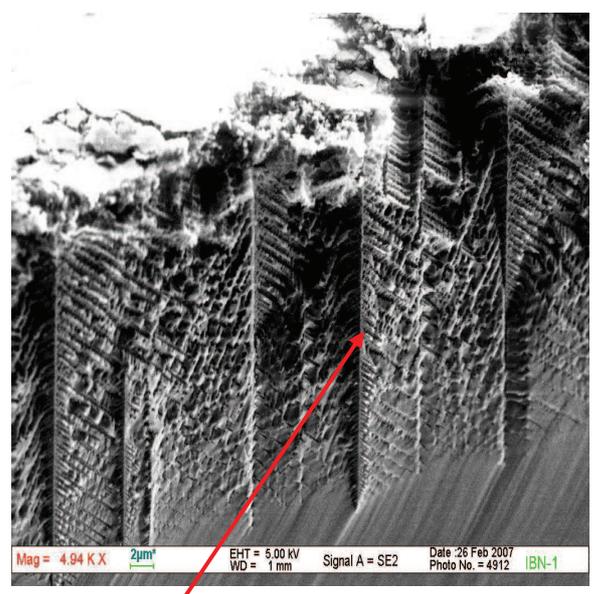

c

**Pores like Fir-Tree**

Figure (4-3); Cross sectional SEM images of the PSi layer, (3.5 Ω.cm) samples; (a) 5 min, (b) 10 min and (c) 15 min.

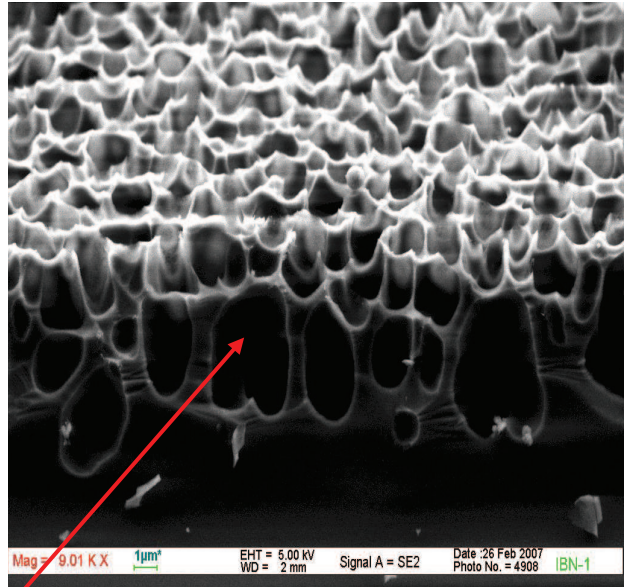

**Overlapped pores**  a

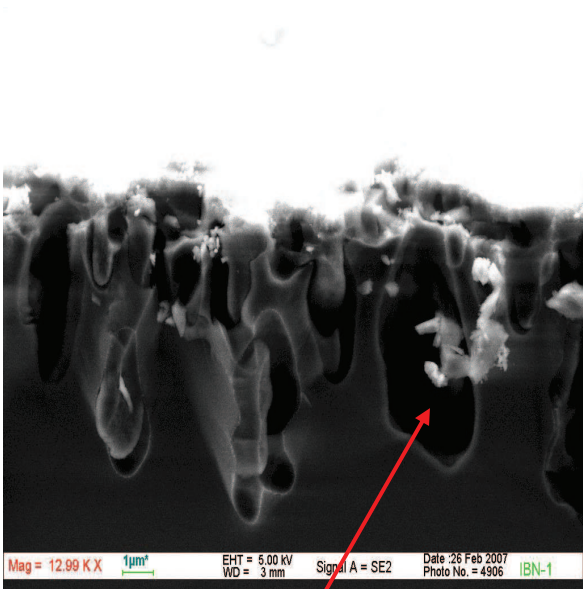

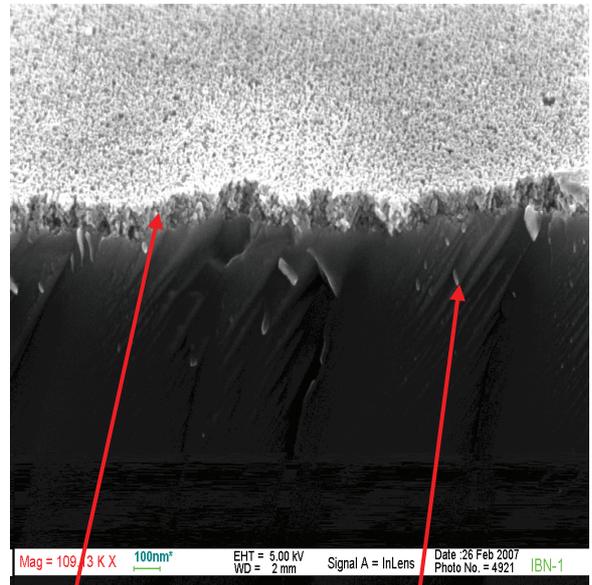

**b**  **c**

**Cone-shaped pore**   **Porous silicon layer**   **Silicon substrate**

Figure (4-4); Cross sectional SEM images of the PSi layer, (0.02 Ω.cm) samples; (a) 5 min, (b) 10 min and (c) 15 min.

From figures (4-3) and (4-4) we can reach to several results as follows:

1- The depth of pores was found to be a linear function of etching time.

2- The local generation of minority carrier determines the pore shape in the following manner: the generation of carriers in deepest layer promote the pore growth at the tips of pores as showed in figure (4-3,a) whereas the generation near surface leads to lateral growth of the pores as showed in figure (4-3,c).

3- The depth of formed pores in PSi layers on low-silicon substrates resistivities was found nonuniform, while the contrary is true for pores in formed PSi layers on high-silicon substrates resistivities.

4- For formed PSi layers on high-silicon substrates resistivities as shown in figure (4-3,a-b,c) it was found that the pore is divided into two or four separate pores. For formed PSi layers on low-silicon substrates resistivities as shown in figure (4-4,a-b,c) the reverse is true; the number of pores has been decreased, while the width of the surviving pores is increasing. In one sentence, one can say the branching depends on the value of silicon substrate resistivity.

5- The strong dependence of the pore shape on its growth which itself relies on crystal orientation might be a result of slightly different etching rates for different crystal planes at the pore tips. This will produce a cone-shaped tip profile (figure (4-4,a) or a fir-tree like structure, figure (4-3,c) with the highest field strength at its center, adjusting the growth direction to <100>.

6- Wide scatter in diameters of the etched pores as shown in figure (4-4,a-b). The width depends on the local efficiency in capturing minority carriers which have been determined by the initial position of the pore in the random pore pattern.

7- Electropolishing may occur for current densities (J) exceeding a critical photocurrent density $J_{PSi}$. Critical photocurrent density ($J_{PSi}$) is a function of etching conditions, especially HF concentration.

Possibly the HF concentration has decreased gradually on length of the pores to the values which lead to critical photocurrent density smaller than current density which produces electropolishing conditions at the bottom of the pores as shown in figure (4-4,c).

For current densities (J) below $J_{PSi}$ the etching process changes and a PSi layer grows as shown in figure (4-3,c).

To avoid ambiguity, we refer to critical current density (J) as a critical photocurrent density here because the photocurrent, which is a parameter of interest of the PEC etching process of n-type silicon, is a nonlinear function of the power density distribution of illumination. Our results are consistent with those of other investigations [18,19,20,21,37,39,79,80].

Figures (4-5) and (4-6) illustrate the correlation between layer thickness and etching rate with etching time for PSi layers prepared on 3.5 and 0.02 Ω.cm silicon substrates resistivities respectively.

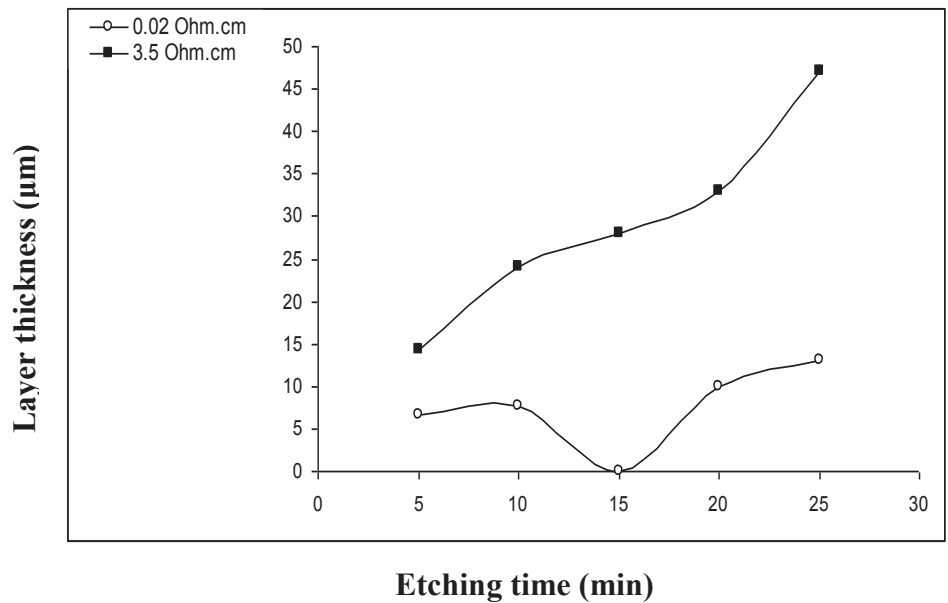

Figure (4-5); The layer thickness as a function of etching time for both substrates resistivities.

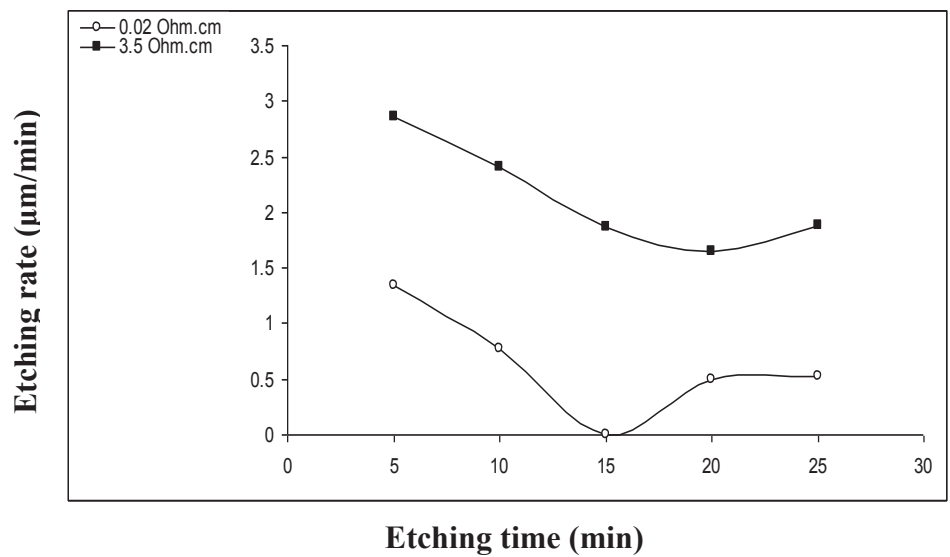

Figure (4-6); The Etching rate as a function of etching time for both substrates resistivities.

We believe that combining between *Noguchi and Beale* [26,49] models is ultra acceptable for clarification of our results which are reported previously and these are sketched in figures (4-5) and (4-6).

The rate-determining step for PEC etching process of silicon is given by the number of holes ($h^+$) available at the electrode surface. If the sample is illuminated during etching, the photocurrent becomes an important parameter besides the breakdown current. If the breakdown mechanism is suppressed the growth of every pore is determined by its success in collecting photogenerated minority carriers.

Since the profile of the power density distribution of laser illumination is assumed to be a Gaussian; the power density distribution will be nonuniform and this results in etched silicon surface initially in three possible ways, two ways are spreading out along the surface from center and one toward the center.

The etching process can start vigorously at various points in the center of the irradiated area because this area has been received most intense of light while the etching process in the other regions would be too slow. In fact the etching rates will diminish continuously from the center of the laser beam to its periphery as shown in figure (4-7). With increasing of etching time, the PSi layer in the center will be deepest, at this stage the light is not completely received at this layer, consequently the number of required holes for etching becomes very small so the etching process would be undoubtedly decreased or until is stopping as shown in figure (4-6) (open dots). In this case, sideways etching at the surface will dominate and lead to a gradual erosion of the columns silicon. At this stage the PSi layer thickness is sufficiently small, light can reach the lower part, resulting in initiating new PSi layer as shown by open dots in figure (4-5).

From the described results in figure (4-5) (black dots), it is concluded that layer thickness is a function of etching time. We can attribute the illustrated behavior of etching process in figure (4-6) in black dots to the fact that the space-charge region (SCR) itself protects the pore walls from being etched. The borderline between space-charge region and bulk is a plane beneath the pore tips. In other words, the whole region between the pores is depleted. The depletion may occur because of trapping of free carriers. Trapping can occur because of the formation of surface states. If all minority carriers are collected by the trench tips there is none left to permit etching of the wall, so the etching rates will decrease.

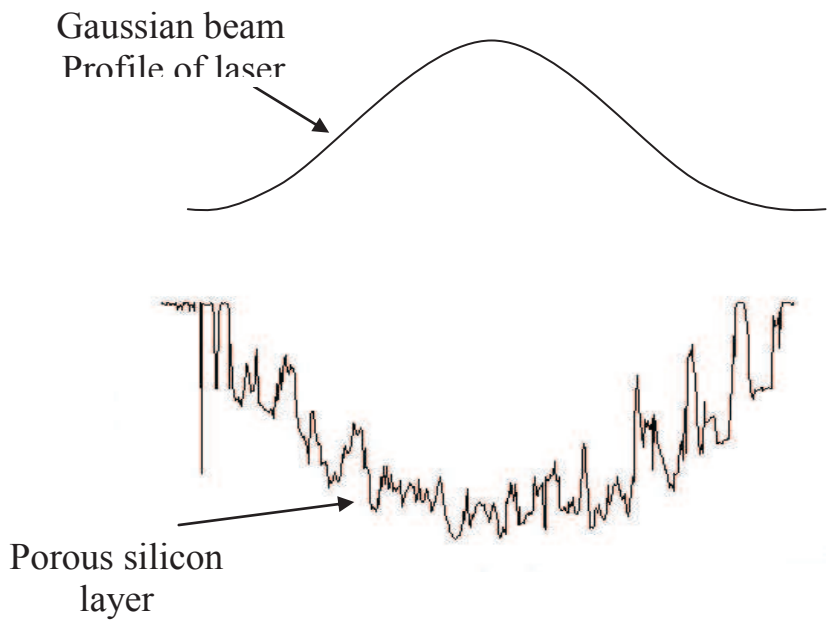

Figure (4-7); A schematic diagram depicts the etching process.

### 4-2-3 *Porosity*

The porosity is defined as the fraction of void within the PSi layer [4]. Porosity is one of important properties of the PSi layer, which can be determined by several methods [19,23,25].

In our work, we have studied the porosity by using gravimetric measurements as shown in equation (2-5). These measurements have been supported by microstructure analysis depending on SEM images.

Figure (4-8) clears the relationship between porosity and etching time of prepared PSi layer at different irradiation times (5-15 min.) with constant current density of 40 mA/cm$^2$. The irradiation has been achieved using diode laser of 2.33 W/cm$^2$ power density and 810 nm wavelength on (3.5 and 0.02 Ω.cm) n-type silicon substrates immersed in 24.5% HF concentration.

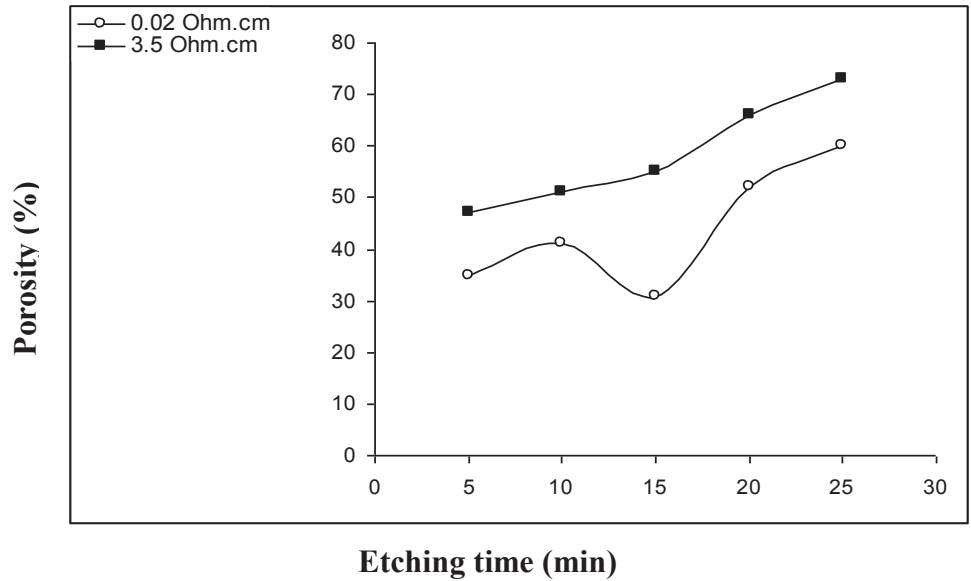

Figure (4-8); The porosity as a function of etching time for both substrates resistivities.

From previous figure, we can see the values of porosity are increasing with increasing of etching time for both two cases, and also easily one can observe that the porosity of formed PSi layers on high-silicon substrates resistivities is larger than that of low-doped sample. These results are ascribed to increasing the number and width of the pores with increasing of etching time. For formed PSi layers on low-silicon substrates resistivities, when the time increases, the nearest pores would be combined (pore dying), leading to fraction amission of void from PSi layer continuously until losing all pores and then initiating new pores with narrow width, consequently the porosity becomes lower as shown in figure (4-8) (open dots), while the branching phenomenon in formed PSi layers on high-silicon substrates resistivities will increase the porosity, the branching density increases with etching time resulting in a large number of voids in PSi layer (figure (4-8) (black dots).

These consequences are consistent with SEM images, where the roughness of surface is a function of etching time and porosity as shown in figures (4-3) and (4-4). Our results agree with outcomes of [3,23,27,29,32,78].

### 4-2-4 *Nanocrystallite Size and Structure Aspect*

Nanocrystallite size is an important feature of the PSi material, and various properties such as conductivity and photoluminescence have been interconnected to this property [4,42]. The morphological phase of PSi layer (crystalline or amorphous) will give good concept for electrical behavior of this material [39,72]. In our case we have calculated the nanocrystallite size by employing Scherer's formula as shown in equation (2-11), and the structure aspect was studied relying on X-ray diffraction measurement.

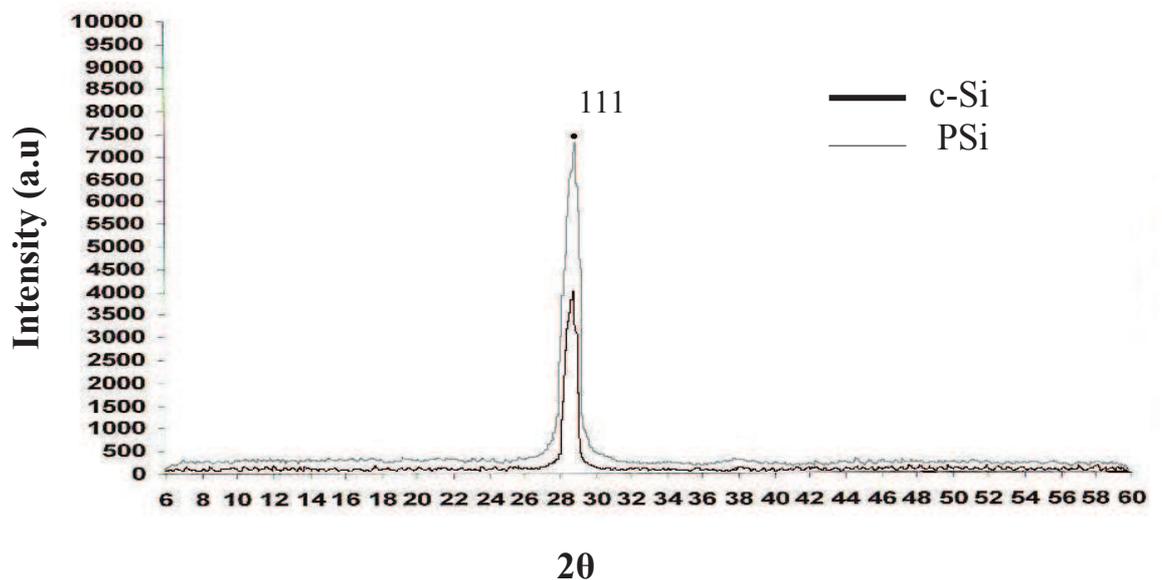

Figure (4-9); X-ray diffraction measurement of silicon and PSi layer.

Figure (4-9) shows X-ray diffraction of crystalline silicon and PSi material respectively. One can note from this figure that the peak of PSi has high value of intensity and expansion, compared with silicon peak. This result is attributed to ray diffraction from crystals with nano-size in the walls between pores. According to this figure, we can confirm that the PSi layer remains crystalline, but it is slightly shifted to a smaller diffraction angle. This result is ascribed to effect of strain which leads to a little expanded lattice parameter and then PSi peak is displaced to small angle diffraction.

Figure (4-10,a,b) shows X-ray diffraction of PSi layers fabricated at various irradiation times (5-15 min.) with fixed current density of 40 mA/cm$^2$ on (3.5 and 0.02 $\Omega$.cm) n-type silicon substrates immersed in 24.5% HF concentration.

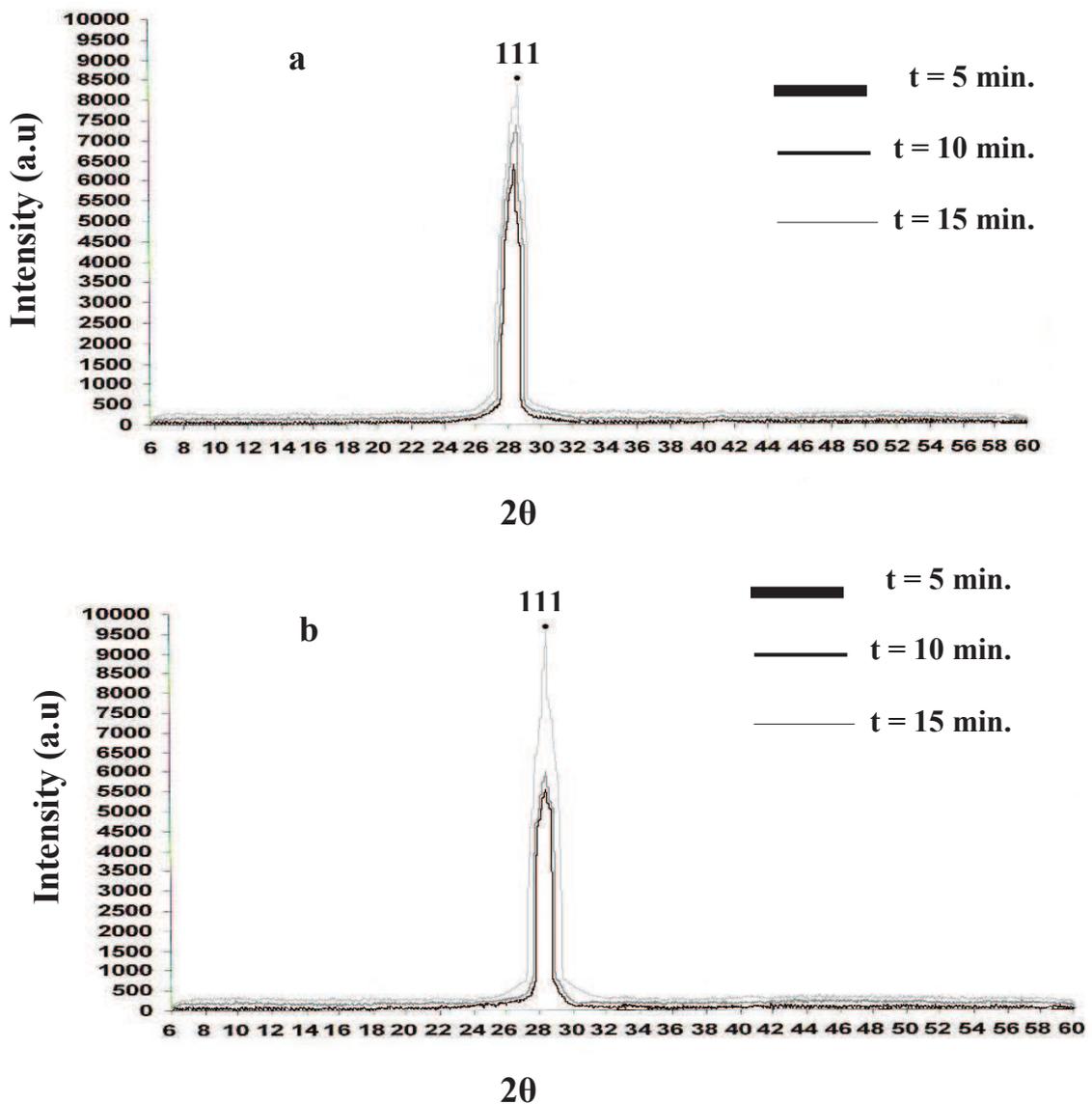

Figure (4-10); X-ray diffraction of PSi layers for (a) 3.5 $\Omega$.cm and (b) 0.02 $\Omega$.cm, silicon substrates respectively.

From figure (4-10,a,b) we could estimate that the full width half maximum (FWHM) and intensity of peak are increasing with increasing of etching time due to littleness of nano-crystals size in PSi layer with the etching time. This result could be observed by plotting nanocrystallite size vs. etching time as shown in figure (4-11).

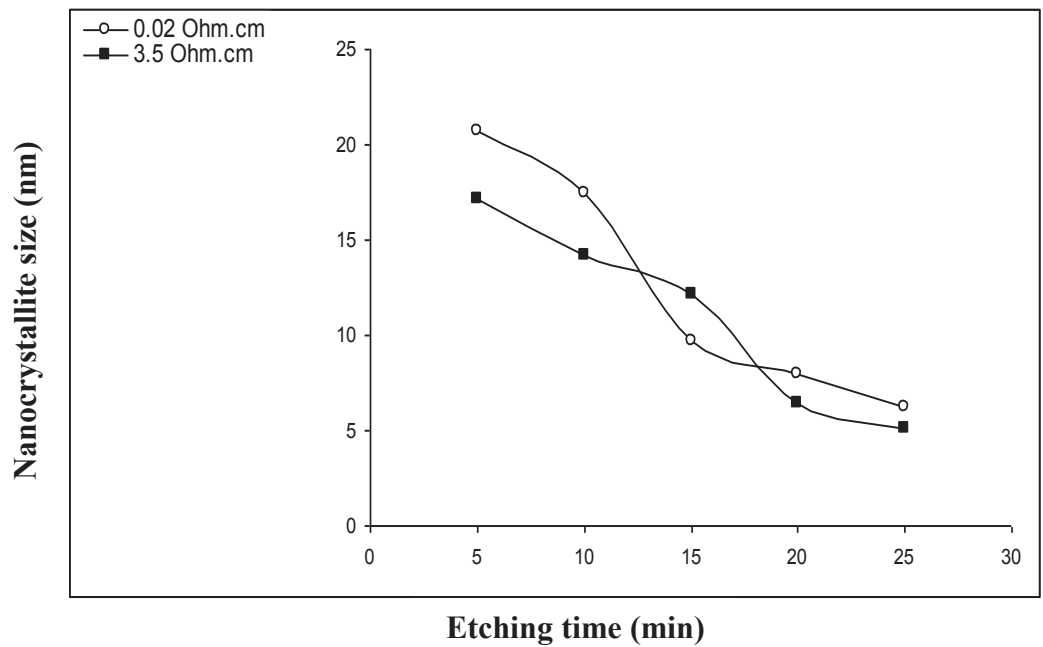

Figure (4-11); The relation between Nanocrystallite size and etching time of PSi layers product at different etching times (5-25 min.).

This figure refers to decreasing of nanocrystals size with increasing of etching time. We could interpret this result as corrosion effect, since this effect increases continuously with increasing of time resulting in very thin silicon fragments which contain crystals of nano-size. Also one can note that the values of crystals size of formed PSi layers on high-silicon substrates resistivities become smaller than that of formed PSi layers on low-silicon substrates resistivities with excess of etching time, except the obtained nanocrystallite size at (15 min.) etching time.

All these results were consistent with SEM measurement, since the walls between pores become thinner with time as shown in table (4-2). Our outcomes are close to results of [32,40,81].

Figure (4-12,a,b) shows the relationship between the nanocrytallite size and energy gap ($E_g$) of PSi layers prepared at different etching times (5-25 min.). The energy gap has been calculated using equation (2-8).

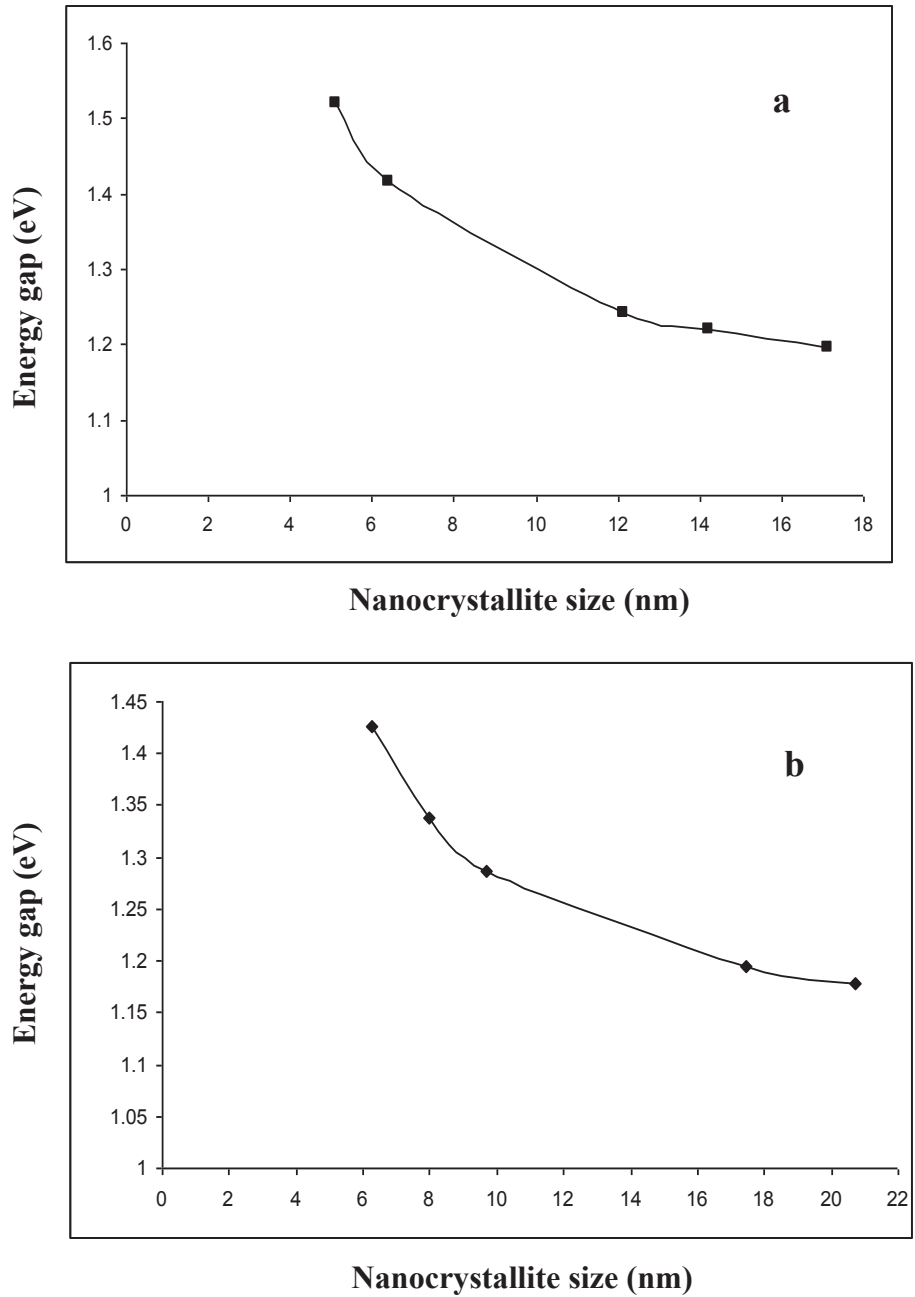

Figure (4-12); The relationship between nanocrystallite size and energy gap; (a) 3.5 $\Omega$.cm and (b) 0.02 $\Omega$.cm, silicon substrates respectively.

From this figure, we can see that the energy gap of the PSi layer has value (1.18 – 1.52 eV) which is bigger than that for silicon substrate (1.12 eV). This result may lead to heterojunction between PSi layer and the silicon substrate. Also one can note that the value of energy gap increases with decreasing of nanocrystallite size for both substrates resistivities. We can explain this result according to *Lehmann* model [19]. The nanocrystallite leads to quantum confinement effect which increases with decreasing of the size of nanocrystals and then leads to widening of effective band-gap. Our results are in a good agreement with results of [2,42,47,72,73,82].

### 4-2-5 *Lattice Mismatch*

Porous silicon exhibits itself as material with special structure. It is recognized by a network of pores affined in a single crystal [4]. *Heuser and Barla* [39,83] have discovered that the diffraction peak of PSi layer was broadened comparing with a silicon single crystal and it is slightly shifted to a smaller diffraction angle.

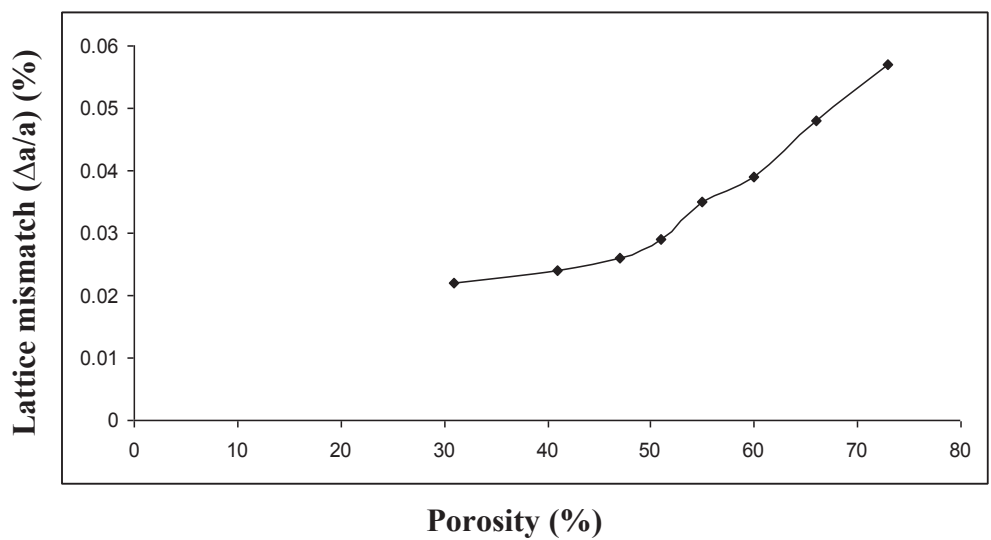

Figure (4-13); The relation between lattice mismatch of PSi layer and its porosity.

Figure (4-13) shows the relation between lattice mismatch and porosity of PSi layers with different porosities. The lattice constant was measured using equations (2-9,2-10). We can see from this figure that the strain effect is roughly a linear function of the porosity which can be changed by using various etching time and substrate resistivity as shown previously. We can attribute this result to expansion in lattice constant which may be as a result of strain effect and the latter may be as a consequence of existence of large internal surface in PSi layer which increases with increasing of porosity.

## 4-2-6 *Specific Surface Area*

Porous silicon consists of a complicated network of pores and columns separated by very thin walls in nanometer sized structures having very large surface area which usually varies from 3-600 $m^2/cm^3$ depending on experimental conditions [4,17,84]. The surface area of PSi layer normally quoted is the specific surface area, defined as the accessible area of solid surface per unit volume of material [84].

We have studied the surface area of samples prepared at various irradiation times (5-15 min.) with fixed current density of 40 mA/cm$^2$ on 3.5 and 0.02 Ω.cm n-type silicon substrates resistivities respectively, immersed in 24.5% HF concentration.

The surface-volume ratio (specific surface area) in $m^2/cm^3$ could be measured by [17]:

$$Surface\ Area\ (m^2/cm^3) = \frac{Area\ of\ one\ pore \times No.\ of\ pores}{Area\ of\ PSi\ structure \times Depth}$$

The pore geometry was considered as cylindrical in shape and thus the area of one pore is [17]:

*The surface area of pore = 2π rh*

where *h*: is the height of pore measured in (m), *r* is the radius of the pore measured in (m).

Table (4-2); Wall thickness and surface area of PSi layer prepared at different etching times for 3.5 Ω.cm and 0.02 Ω.cm substrates resistivities respectively.

| Substrate resistivity (Ω.cm) | Etching time (min) | Wall thickness (μm) | The surface-volume ratio ($m^2/cm^3$) |
|---|---|---|---|
| 3.5 | 5 | 0.25 – 1.08 | 120.29 |
| 3.5 | 10 | 0.20 – 1 | 125.65 |
| 3.5 | 15 | 0.17 – 0.83 | 235.35 |
| 0.02 | 5 | 0.33 – 1.25 | 96.38 |
| 0.02 | 10 | 0.25 – 1.16 | 118.34 |
| 0.02 | 15 | 0.03 | 7.43 |

We can effortlessly see from figure (4-14,a,b,c) and table (4-2) that the specific surface area of PSi layer increases with increasing of etching time. This result is ascribed to interpolation of number of the pores and its area with increment of time, which results in large specific surface area. The important observation is the specific surface area depends on substrate resistivity parameter, since the specific surface area of formed PSi layers on high-silicon substrates resistivities is larger than that of other.

This evection produced from is dependence of etching process on this parameter with considering of etching time and power density distribution of illumination. From the looking to figure (4-14) one can perceive clearly the effect of latter parameters, where the number and height of pores of formed PSi layers on low-silicon substrates have been small compared with other. The reason for this effect is discussed previously in section (4-2-2).

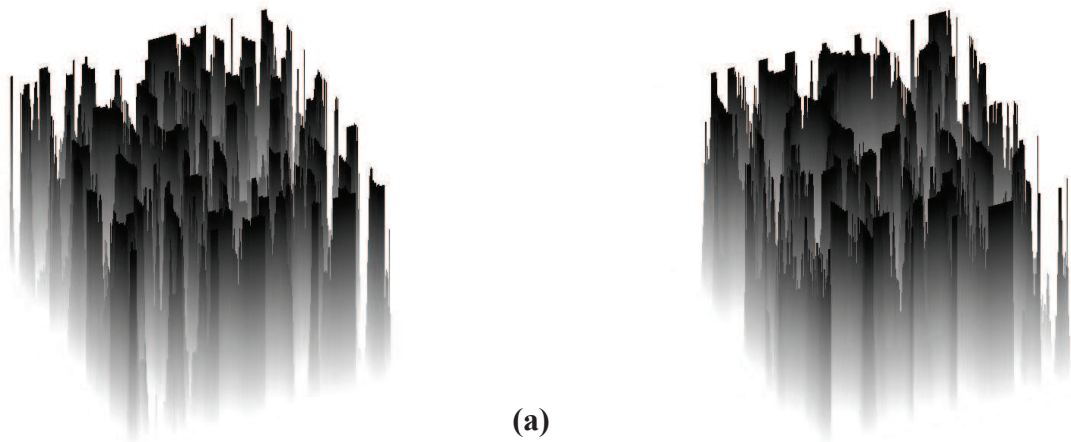

(a)

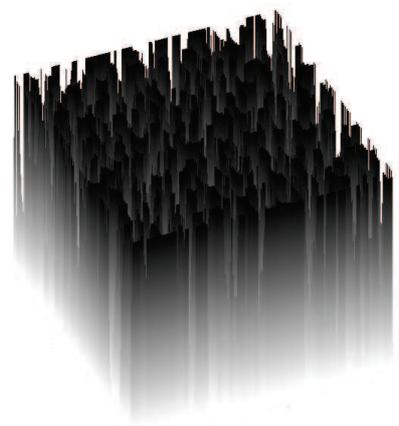
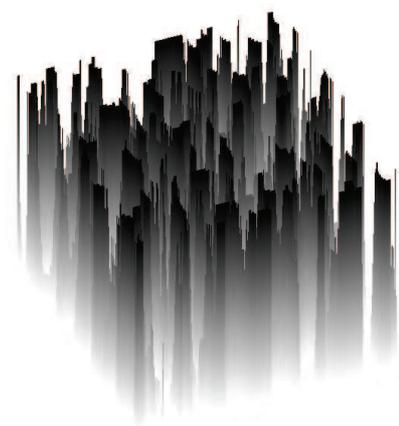

(b)

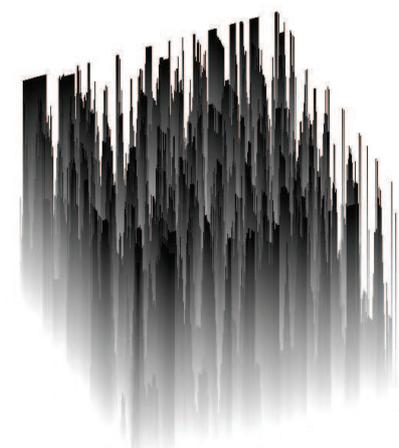
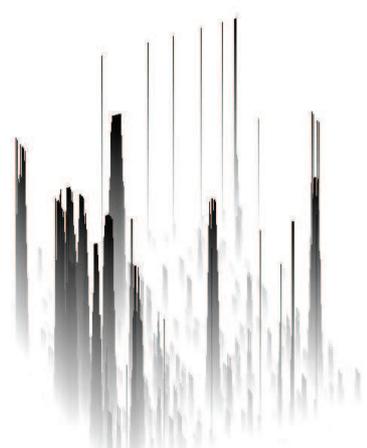

(c)

Figure (4-14); A schematic diagram depicts the surface morphology of samples (a) 5 min., (b) 10 min. and (c) 15 min. for 3.5 Ω.cm (left-hand column) and 0.02 Ω.cm (right-hand column) silicon substrates resistivities.

## 4-3 *Electrical and Photoelectrical Properties*

The interest to study the electrical and photoelectrical properties of PSi layer appears first of all from prospects to developing of many technological applications such as detectors, solar cells, sensors…..etc [17, 45,51]. The current-voltage characteristics, photovoltaic characteristics and the related electrical parameters such as ideality factor, rectification ratio and the charge carrier transport mechanisms in Metal/PSi/c-Si/Metal sandwich structure depend on the structural properties of PSi layer [46,54,78].

In this section we have studied the electrical and photoelectrical properties of Al/PSi/n-Si/Al sandwich structure, and also we have investigated the effect of structural parameters of the PSi layer on these properties. The PSi layers were prepared at different irradiation times (5-25 min.) with constant current density of 40 mA/cm$^2$. The irradiation has been achieved using diode laser of 2.33 W/cm$^2$ power density and 810 nm wavelength on (3.5 and 0.02 Ω.cm) n-type silicon substrates immersed in 24.5% HF concentration.

### 4-3-1 *Current-Voltage Characteristics*

The electrical behavior of metal/PSi/c-Si/metal such as Schottky or heterojunction generally is determined by depending on the characteristics of current-voltage curves.

Figure (4-15,a,b) shows the J-V characteristics of Al/PSi/n-Si/Al sandwich structure which contains PSi layers prepared at (10 min) etching time for (3.5 and 0.02 Ω.cm) n-type silicon substrate respectively. This figure demonstrates that the forward and reverse current densities at room temperature under dark as a function of the applied bias voltage for both resistivities but in same time, one can note the tendency for current saturation at high voltages in forward bias for both two cases.

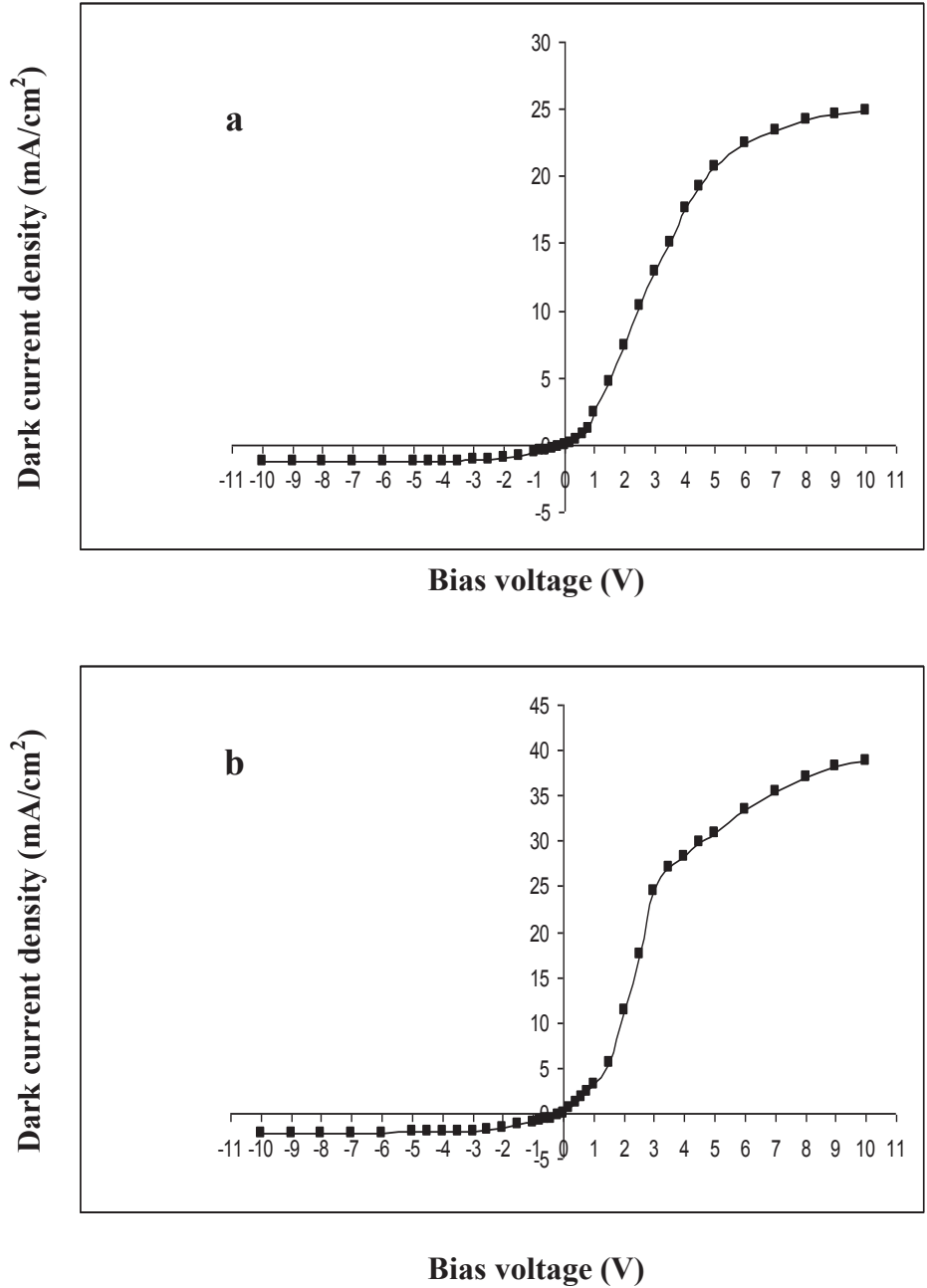

Figure (4-15); The J-V characteristics of Al/PSi/n-Si/Al sandwich structure contain PSi layers prepared on (a) 3.5 $\Omega$.cm and (b) 0.02 $\Omega$.cm Si substrates.

This figure shows J-V characteristics of the junctions formed from the high and low silicon substrates resistivities and shows rectification behavior. The idea that the current through the device is controlled by the Schottky barrier at the Al/PSi interface leads us to interpret the J-V characteristics in terms of a diode equation, but immediately a problem arises:

The curves could fit only using a very high series resistance as well as an unreasonable large value of ideality factor ((a) n = 18 and (b) n = 22 respectively) and also the tendency for current saturation in forward bias. The high resistance was attributed to the PSi layer, but no explanation could be found for the large ideality factor and saturation region in J-V curves. Therefore these results (rectification behavior, large ideality factor as reported in table (4-3) and single saturation region) may be ascribed to depletion layer in the crystalline silicon substrate. In a sense the PSi/n-Si interface behaves like a Schottky diode, where the PSi layer plays the role of a metal. This happens because PSi layer has a high effective density of states at the Fermi level that pins the Fermi level at the PSi/n-Si interface. Inspection of the room temperature J-V characteristics in figure (4-15) indicates a relatively large value of series resistance ($R_s$) for our PSi-based devices. We note that series resistance ($R_s$) may depend on the resistance to hole transport through the PSi layer due to a depletion of carriers in these nanostructures.

The series resistance ($R_s$) is regarded as a function of layer thickness and porosity of PSi layer. In the light of this fact, we can understand the variation in value of $R_s$ for both cases which is considered the main difference between the two devices, where one can easily observe from tendency of J-V curves that the device in first case (PSi layers prepared on high-silicon substrates resistivities) possesses $R_s$ value higher than that in the second case (PSi layers prepared on low-silicon substrates resistivities). This result is congruent with the obtained results in figures (4-5) and (4-6), since the PSi layer has (51%) porosity and (24 μm) layer thickness in the first state while in the second case the PSi layer has (41%) porosity and (7.7 μm) layer thickness.

Figure (4-16,a,b) illustrates the J-V characteristics under dark and at room temperature of the Al/PSi/n-Si/Al sandwich structures, including PSi layers prepared at different etching times (5,15 and 25 min) for both silicon substrate resistivities.

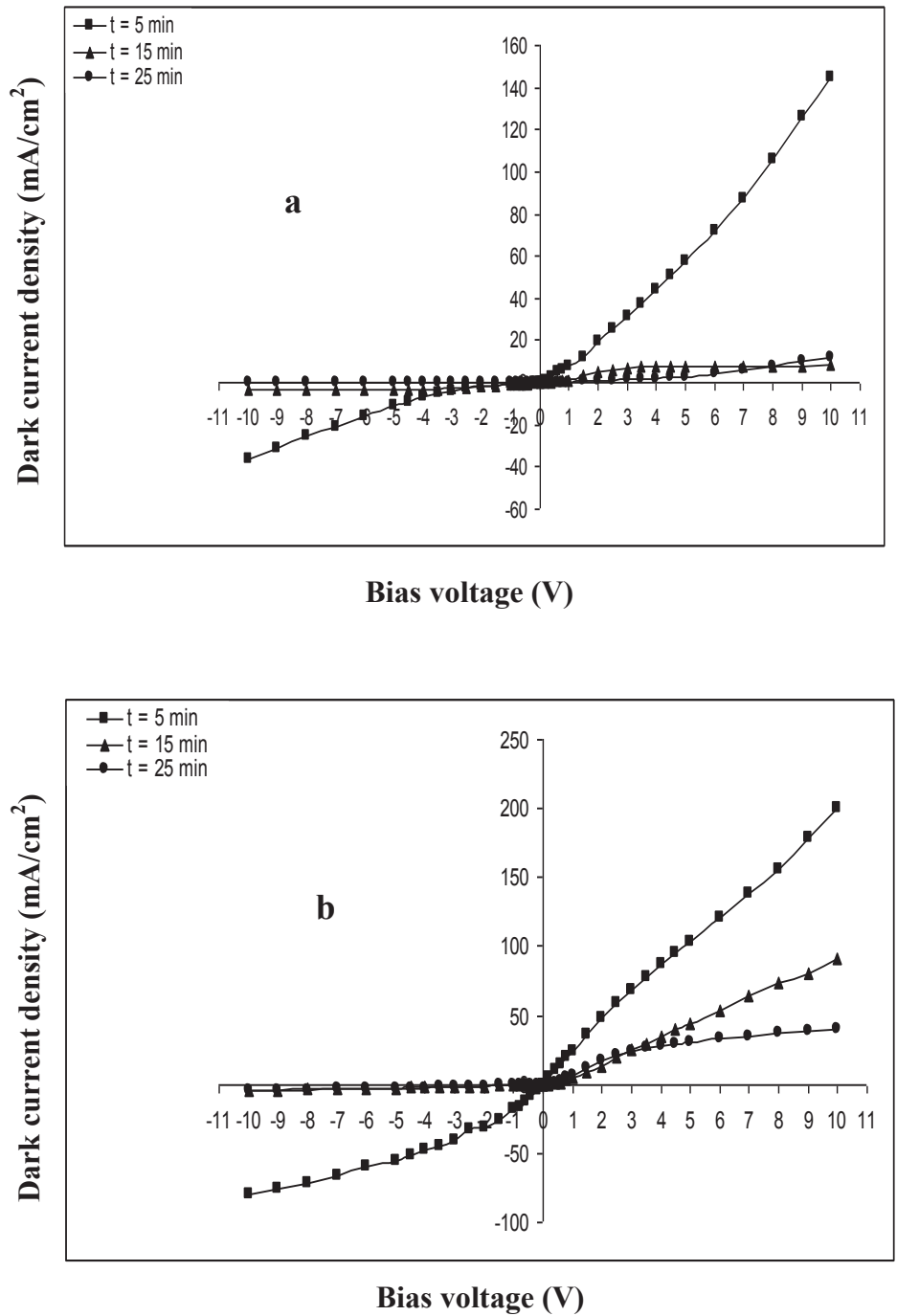

Figure (4-16); The J-V characteristics of Al/PSi/n-Si/Al sandwich structure contains PSi layer prepared on (a) 3.5 $\Omega$.cm and (b) 0.02 $\Omega$.cm Si substrates.

From these curves we can reach several facts as follows:

1- Current-voltage characteristics shows rectification behavior and the rectification ratio ($J_f/J_r$) varied from one sample to another.

2- The current flow in the forward bias is decreasing with increasing of etching time.

3- The value of ideality factor which has been calculated using equation (2-13) is relatively large in the range 10 – 29, and differs from one sample to another.

4- The barrier height ($\Phi_{bn}$) is varying with changing of etching time; the barrier height was calculated using equation (2-14).

In the light of these results, the rectification behavior of the J-V characteristics should not be interpreted in term of Al/PSi interface (Schottky barrier), but rather, the porous silicon/silicon interface (heterojunction). We have taken this concept in account according to the values of ideality factor, which is never unacceptable to be explained by depending on Schottky barrier. These large values of ideality factor demonstrate that the PSi layer has a high density of states, no depletion should occur at the metal/PSi interface. This contact should behave almost in an ohmic manner, on the other hand since the crystalline size is very small as demonstrated in section (4-2-4), the width of the depletion region inside the PSi layer at both interface is small. Furthermore, since the transport in PSi involves hopping, rather than thermally generated carriers the bending of the bands is of no relevance to the injection at the interface. Thus, the metal/PSi interface will be a quasi-ohmic contact. The properties of this interface would be seen only for very low voltages (V < 0.6 V), where we have observed some anomalies.

Again due to the high density of states of the PSi layer which will result in screening of internal field inside the PSi layer, this field would be nearly homogeneously distributed through the PSi layer at higher voltages (V > 1 V).

Therefore, the forward bias characteristics will be controlled by the PSi layer resistance. This result explains the lowering of flow current in forward bias with increasing of etching time, since the porosity of PSi layer increases with etching time and hence the resistance of PSi layer becomes too high which leads to low forward current.

By the way, one can note that the values of forward currents in figure (4-16,a) are quite small, compared with these in figure (4-16,b) this result is attributed to high values of porosity of formed PSi layers on high-silicon substrate resistivities, compared with these on low-silicon substrate resistivities as shown in figure (4-8).

The number of injected charge carriers from the silicon substrate into the PSi layer is equal to the number of charge carriers flowing from the PSi into the substrate (equilibrium case). In forward bias, this latter current will not change, since the Fermi level is fixed at the interface. However, the opposite current from the substrate into the PSi layer would be enhanced due to a slight change of internal voltage drop. This change will be tuned to give the necessary current which may be imposed by the PSi layer resistance. At reverse bias a large part of the applied voltage will drop in the depletion region, the increase in voltage drop continually would lead to eliminating the injected charge carriers from the substrate into the PSi layer. The current then will be equal to the number of injected charge carriers from the PSi into the substrate, which will give the saturation characteristics. For a high reverse bias, additional contributions to the saturation current might appear due to tunneling of electrons from the valence band of the silicon substrate into the PSi or generation-recombination process inside the depletion layer. This would lead to no saturating characteristics as shown in figure (4-16,a).

These results agree with outcomes of [46,54,59,60,71,76].

The calculated values of barrier height ($\Phi_{bn}$) as shown in table (4-3) have been in the range (0.536 – 0.612 eV) and show the slight increase with increasing of etching time. According to Schottky theory, the barrier height ($\Phi_{bn}$) depends completely on the work function of metal ($\Phi_{m}$), but the theoretical value of barrier height of the Al/n-Si equals (0.27 eV) while our results refer to values quite big as shown below.

Therefore we believe that the barrier height resulting from the depletion layer in silicon/porous silicon interface and not be determined by the metal properties, but rather by the PSi layer. This layer covers most of the sample surface, its porosity increases with increasing etching time. Thus its density of states becomes high and as a consequence will affect values of barrier height as small increase. Our results consists with these of [85,86].

Table (4-3); Saturation current density, barrier height and ideality factor of the fabricated silicon/porous silicon junctions.

| Substrate resistivity ($\Omega$.cm) | Etching time (min) | Saturation current density (A/cm$^2$) | Barrier height (eV) | Ideality factor |
|---|---|---|---|---|
| 3.5 | 5 | $5 \times 10^{-3}$ | 0.555 | 29 |
| 3.5 | 10 | $1.6 \times 10^{-4}$ | 0.643 | 18 |
| 3.5 | 15 | $5.5 \times 10^{-4}$ | 0.612 | 24 |
| 3.5 | 20 | $2.5 \times 10^{-3}$ | 0.572 | 29 |
| 3.5 | 25 | $9 \times 10^{-5}$ | 0.658 | 18 |
| 0.02 | 5 | $1 \times 10^{-2}$ | 0.536 | 24 |
| 0.02 | 10 | $5 \times 10^{-4}$ | 0.612 | 22 |
| 0.02 | 15 | $1 \times 10^{-4}$ | 0.596 | 23 |
| 0.02 | 20 | $3 \times 10^{-4}$ | 0.629 | 10 |
| 0.02 | 25 | $2.5 \times 10^{-3}$ | 0.597 | 21 |

Figure (4-17) clears the J-V characteristics under dark and at room temperature of the Al/PSi/n-Si/Al sandwich structure include PSi layer prepared at an etching time (20) min. for high-silicon substrate resistivity. The J-V characteristics of sandwich structure in figure (4-17) show

rectifying behavior and double current saturation. We have attributed these results to silicon/porous silicon heterojunction act as double-Schottky-diode.

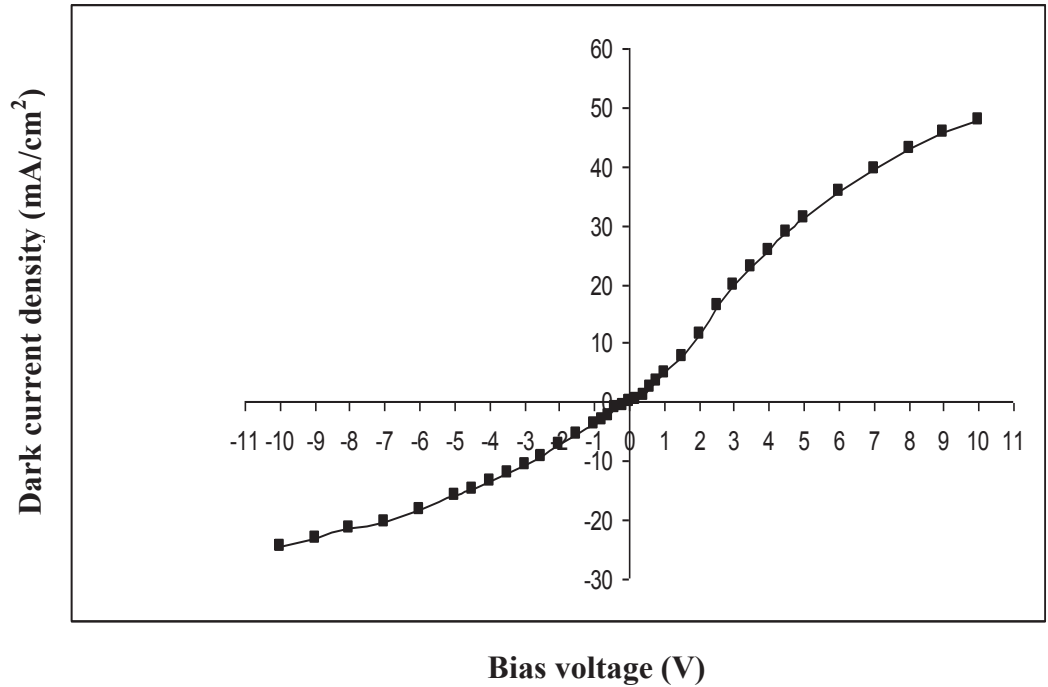

Figure (4-17); The J-V characteristics of Al/PSi/n-Si/Al sandwich structure contains PSi prepared at an etching time 20 min. on 3.5 Ω.cm Si substrate.

The good congruence for our results with *Milnes and Feucht* model [56] prompts us to justify these results in the light of this model. It may be expected that the interface states lie in a thin layer sandwich between two depletion regions having energies lying in a narrow interval and acting as centers with large capture cross-sections for the current carriers.
According to this model the current flows through the heterojunction via two mechanisms i.e. partly by the carriers charge emitted over the tops of potential barriers at the interface without being captured by the interface states (direct transmission) and partly by the charge carriers emitted over one of the tops captured by the interface states and re-emitted over the second top (two-stage transmission).

Figure (4-18) shows the J-V characteristics under dark and at room temperature of the Al/PSi/n-Si/Al sandwich structure include PSi layer prepared at an etching time (20) min. for low-silicon substrate resistivity.

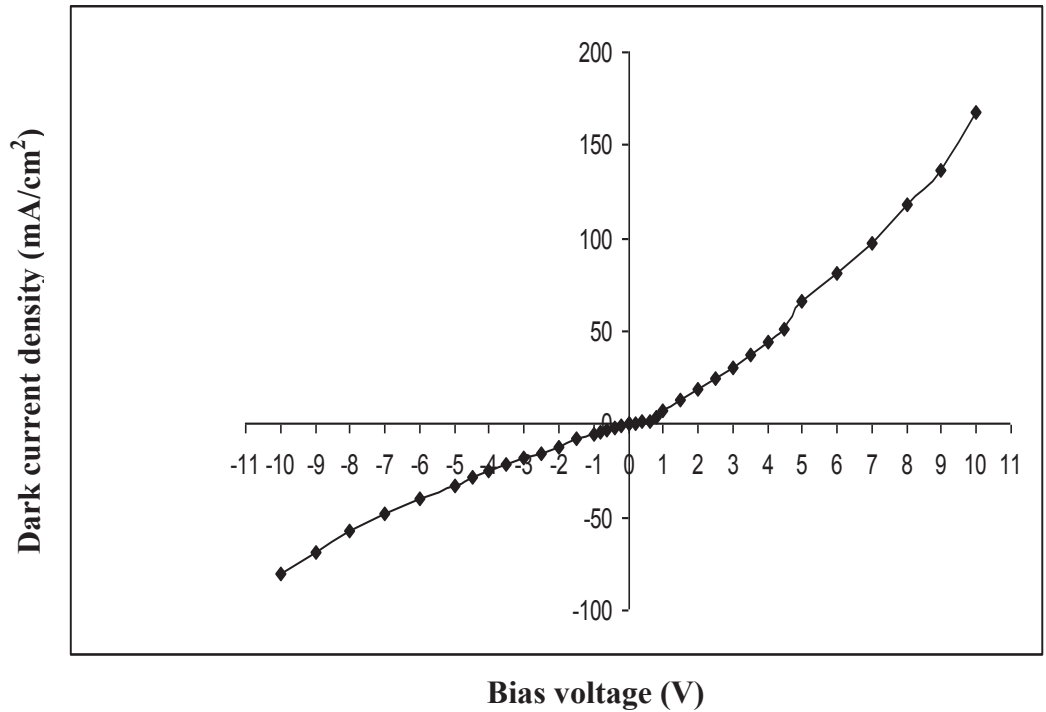

Figure (4-18); The J-V characteristics of Al/PSi/n-Si/Al sandwich structure contains PSi prepared at an etching time 20 min. on 0.02 Ω.cm Si substrate.

The J-V characteristics of sandwich structure in figure (4-18) exhibit symmetrical characteristics for V > 1 V (i.e. the current nearly has the same magnitude for both bias voltages).
*Ben-Chorin* [10] has reported that in this case the transport is controlled by the silicon bulk and that the PSi conductivity has strong field dependence. This behavior has been attributed to electric-field-enhanced hopping conductivity. But we think the current flow at V < 1 V is controlled by Al/PSi (Schottky barrier).

Figure (4-19) presents the rectification factor as a function of etching time and the applied voltage of the Al/PSi/n-Si/Al sandwich structures.

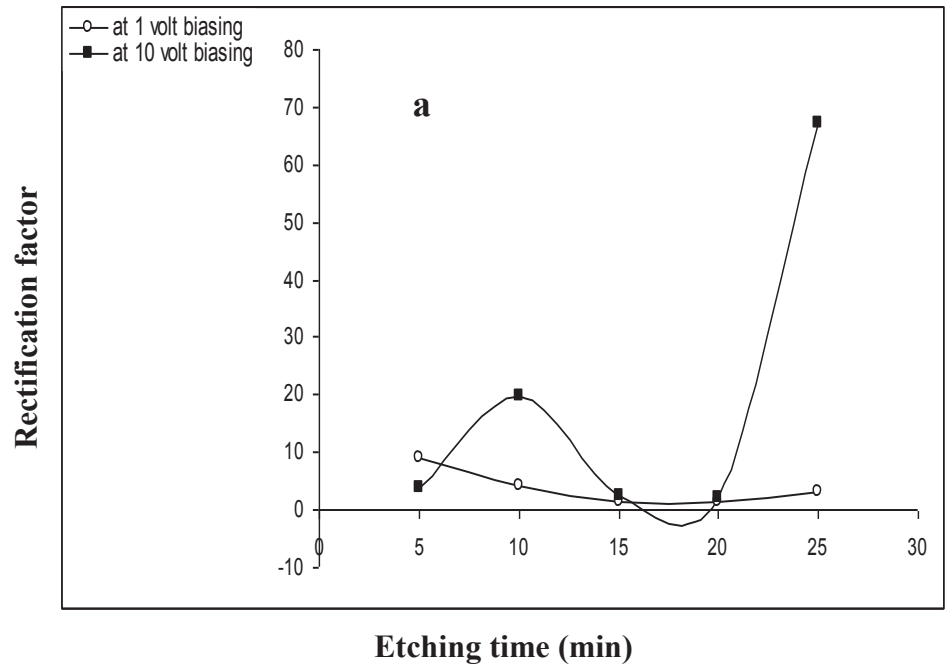

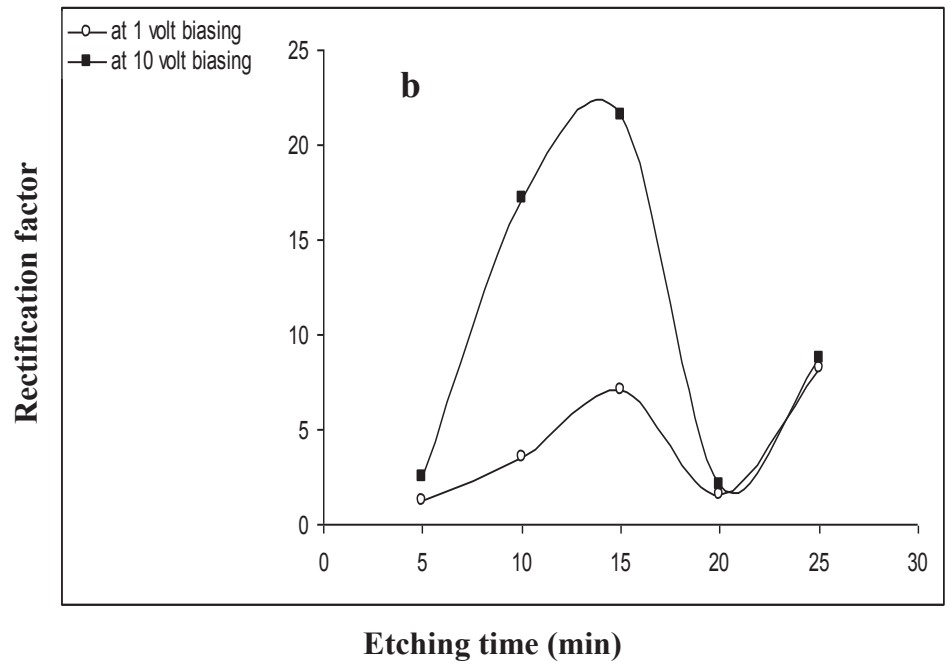

Figure (4-19); The variation in the rectification factor as a function of etching time and applied voltage; (a) 3.5 $\Omega$.cm and (b) 0.02 $\Omega$.cm Si substrates.

Examination of figure (4-19) leads us to the crucial evidence that the rectifying behavior of these devices has been governable by the structural parameters of the entered PSi layers in fabricating these structures since the rectifying factor varies with changing of etching time and the substrate resistivity (i.e. it varies with varying of thickness and porosity of PSi layer) and consequently will determine the transport mechanisms of charge carriers in PSi layer. Therefore, one can note fluctuations in values of the rectification ratio from lowering into rising except the state with low applied voltage (1 V) in figure (4-19,a), where the case is the opposite. On the other hand, we can easily observe that the flow of current in both biases in these devices has been governed by the magnitude of applied voltage. At low applied voltage (1 V) the structure has a small rectification factor. By increasing the applied voltage to (10 V), this factor would be increased to a large magnitude for sandwich structures include PSi layers prepared on both resistivities of the silicon substrates. The origin of rectification characteristics and the effect of applied voltage have been discussed previously.

Figure (4-20,a,b) represents J-V characteristics under dark and under daylight illumination (100 mW/cm$^2$) at room temperature of Al/PSi/n-Si/Al sandwich structure, containing PSi layers prepared on (3.5 Ω.cm) and (0.02 Ω.cm) n-type silicon substrate resistivities respectively. The photocurrent has been observed in reverse bias only. We can see from this figure that the presence of a daylight illumination strongly increases the reverse current. This is very similar to the sensitivity of a photodetector reported in Ref. [88] where it is found that the reverse current depends strongly on illumination. However the shape of the J-V curves refers to photogenerated carriers and associated light absorption takes place in the depletion region of the silicon-porous silicon interface.

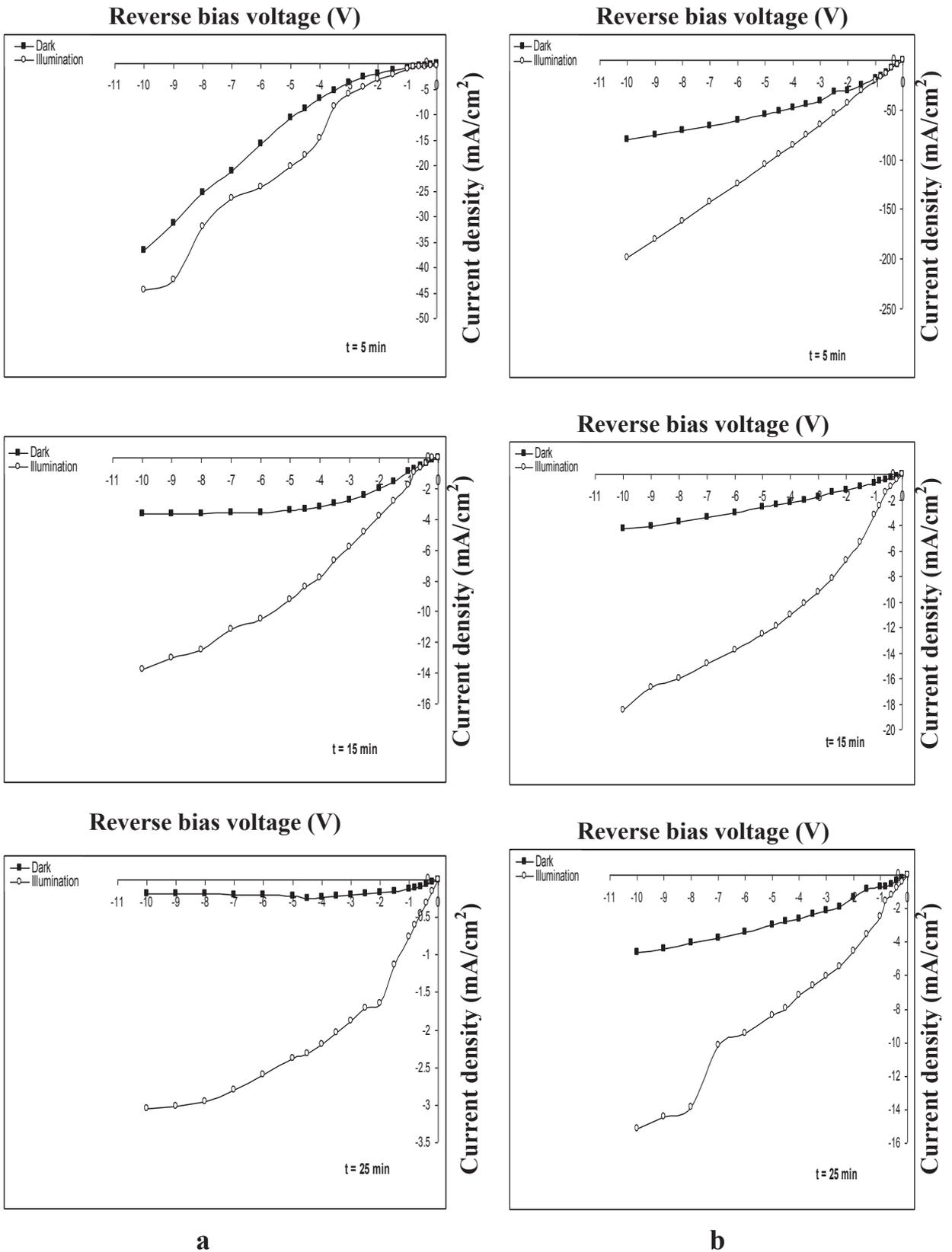

Figure (4-20); The photocurrent as a function of the etching time; (a) 3.5 Ω.cm and (b) 0.02 Ω.cm Si substrates.

The preparation conditions (etching time and substrate resistivity) have very important effect on the photocurrent characteristics of these devices where one can observe that the photocurrent is reduced with increasing of these conditions. We can explain this result in the following manner:

The porosity of PSi layer increases with increasing of substrate resistivity and etching time as showed previously in section (4-2-3). The gradual increasing of porosity leads to increasing resistivity of PSi layer as shown in the next section (figure (4-21)) consequently, the reverse current diminishes gradually.

The photocurrent is voltage dependent: At high reverse bias it saturates but for lower voltage, it tends to diminish. Similar observations were reported by *Maruska et al* [87,89]. We think that this behavior arises from the high resistance of the PSi layer. The influence of a series resistance is well known from solar cell. When the resistance is small, the reverse current is just the sum of the saturation current and a voltage-independent photocurrent; however, high series resistance will limit the current at low voltage, and saturation will be observed only at high reverse bias. In the dark the reverse current is limited by the heterojunction barrier. Under illumination the barrier collapses and the current will be limited by the resistance. Only for high reverse bias will the current be so high that the barrier will control the behavior again.

This is exactly what is seen in figure (4-20).

## 4-3-2 *Resistivity*

One of the most important properties of PSi layer is the electrical resistivity [51]. The resistivity of the macropours silicon exceeded the resistivity of the initial silicon by 1.6 – 15 times, which may be attributed to various effects such as depletion effect [9,22,90].

Figure (4-21) shows the resistivity of prepared PSi layers at different irradiation times (5-25 min.), with constant current density of 40 mA/cm$^2$. The irradiation has been achieved using diode laser of 2.33 W/cm$^2$ power density and 810 nm wavelength on (3.5 and 0.02 Ω.cm) n-type Si substrates immersed in 24.5% HF concentration.

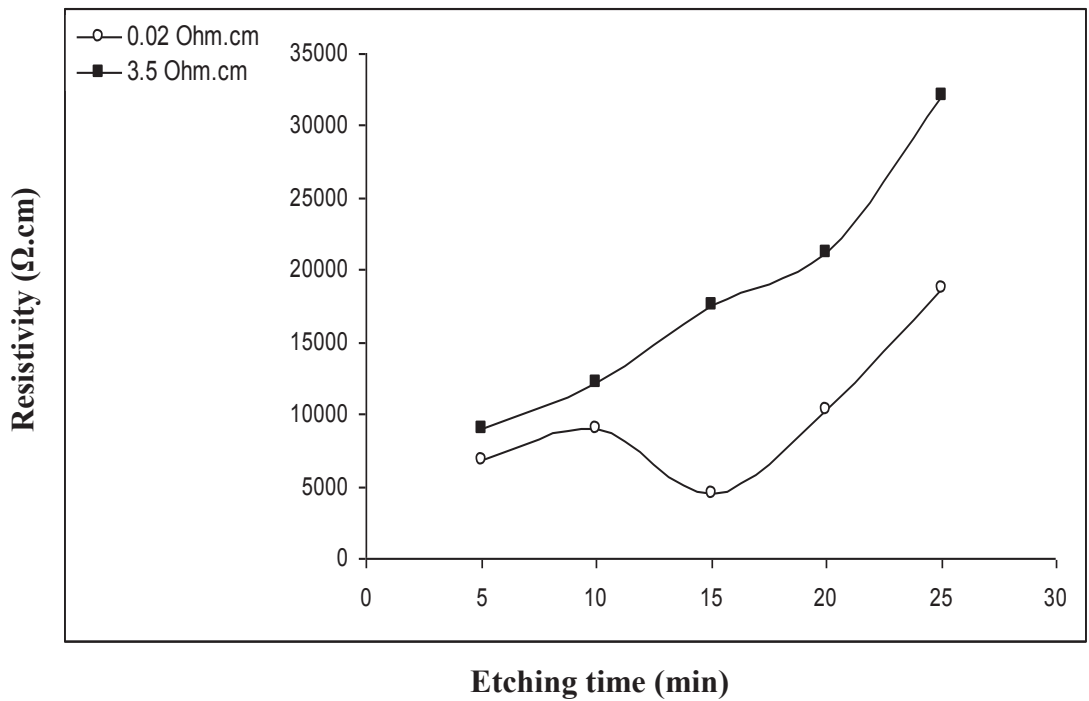

Figure (4-21); The resistivity of the porous silicon layer vs. etching time for both Si substrates resistivities.

The resistivity of PSi layer may determine the J-V characteristics of the PSi-based devices, but at the same time the resistivity of PSi layer strongly depends on preparation parameters as shown in figure (4-21), where the resistivity has been increased with increasing of etching time and substrate resistivity with considering of current density parameter. We are ascribe this fact to the resistance of PSi layer which is very sensitive to its porosity, since it increases by about an order of magnitude for each 10% - 15% rise in porosity and at the same time it is a monotonic function of the resistivity.

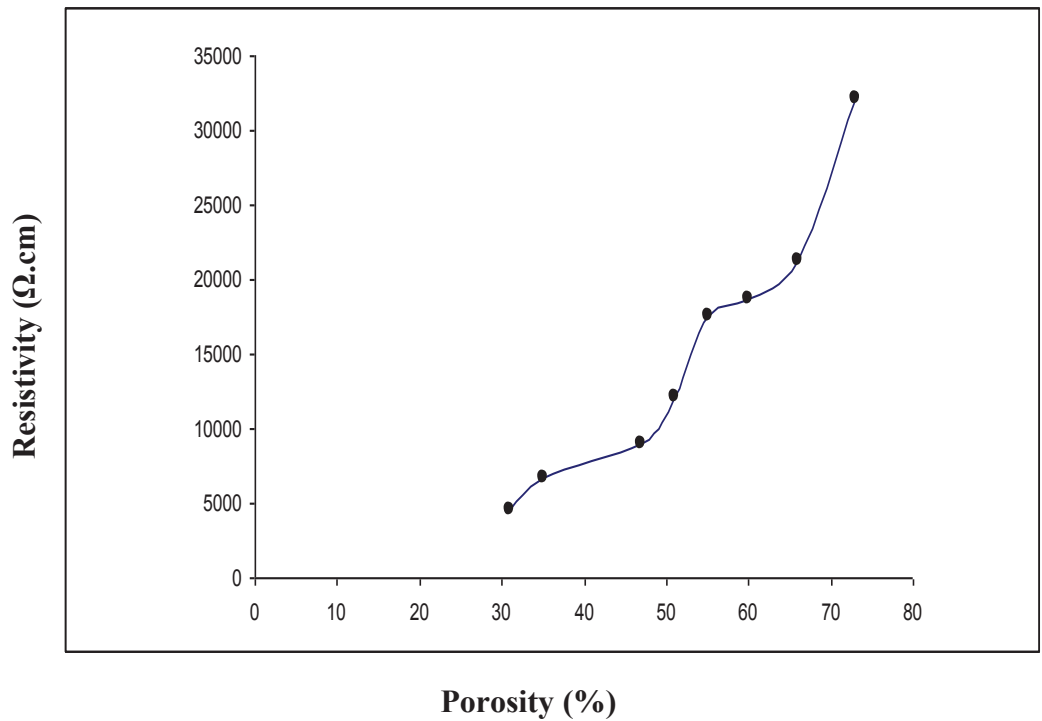

Figure (4-22); The resistivity of PSi layer as a function of its porosity.

Figure (4-22) shows the relationship between the electrical resistivity of PSi layer and its porosity. We can see from this figure that resistivity increases with increasing of porosity. Many assumptions have been reported for clarifying this behavior such as depletion due to quantum confinement or trapping of free carriers…etc.

For our case we believe that depletion effect occurs due to trapping of charge carriers during the preparation of PSi because of the formation of surface states. Therefore, we expect that the depletion effect increases with increasing of porosity consequently the resistivity increases continually.

## 4-3-3 Photocurrent, Spectral Sensitivity and Quantum Efficiency

The photocurrent of PSi-based devices depends on the morphological and structural properties of PSi layer, especialy (layer thickness and porosity). *Diesinger et al* [64] has reported that the relation between photocurrent and layer thickness is inversive relation.

Figure (4-23,a,b) shows the relation between photocurrent density of Al/PSi/n-Si/Al sandwich structures (at different bias voltage), and the formed PSi layers thickness at different irradiation times (5,10,25 min.) with constant current density of 40 mA/cm$^2$. The irradiation has been achieved using diode laser of 2.33 W/cm$^2$ power density and 810 nm wavelength on (3.5 and 0.02 Ω.cm) n-type silicon substrates immersed in 24.5% HF concentration. The sandwich structures have been illuminated by halogen lamp (~ 100 mW/cm$^2$) and the photocurrent was measured by using equation (2-21).

When the structure was illuminated, the electron-hole pairs generated in the depletion layer of PSi/n-Si interface would reduce the barrier for transport of charge carrier. In forward bias this will have no effect since the current is limited by the resistance of PSi layer. In the reverse bias the photogenerated carriers will increase the current strongly, especially at high voltage where the barrier is the main limitation for the current.

*Dimitrov* [11] has reported the high-resistivity PSi layer will affect the photocurrent characteristics. We have obtained similar result as shown in figure (4-23) where the resulting photocurrent is reduced with increasing of layer thickness of the PSi material for both resistivities.

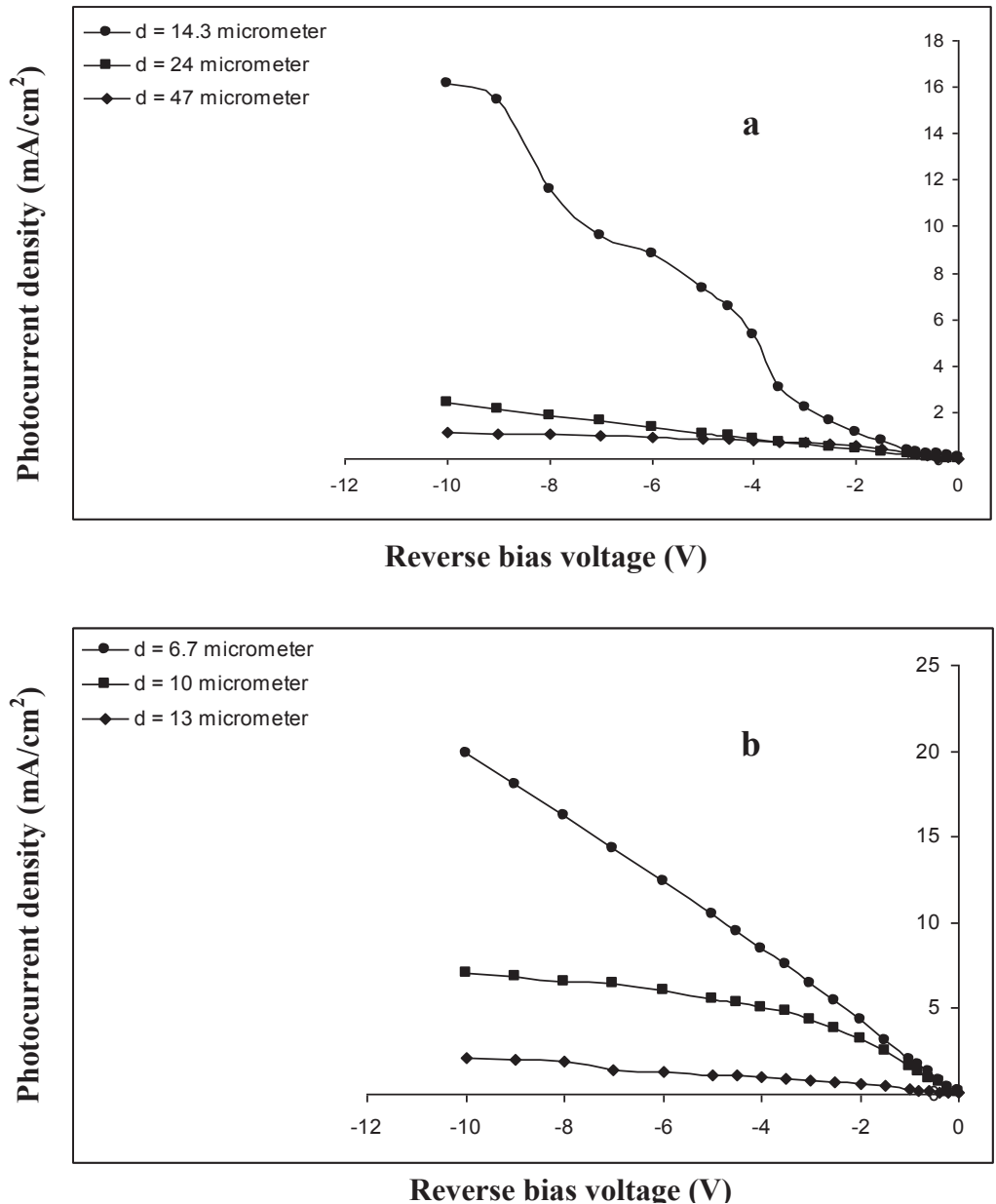

Figure (4-23); Photocurrent against layer thickness at different bias voltages; (a) 3.5 Ω.cm and (b) 0.02 Ω.cm Si substrates.

The spectral sensitivity of PSi-based photoelectric systems depends on the preparation conditions of the PSi layer [45].

Figure (4-24,a,b) illustrates the measured sensitivity ($R_\lambda$) as a function of various incident wavelengths. These measurements were achieved in span of 320 – 859 nm and under 5 V reverse bias voltage of Al/PSi/n-Si/Al sandwich structures. The PSi layers were prepared at different etching times (5 and 25 min.) on 3.5 Ω.cm and 0.02 Ω.cm n-type Si substrates resistivities respectively.

From the curves of figure (4-24,a,b) the sensitivity lies within the visible region, while the value of the sensitivity varies with varying the etching time and substrate resistivity, where the value of sensitivity in the first case increases from 0.26 to 0.32 A/W with decreasing etching time from 25 min. to 5 min. while the sensitivity value in the second case increases from 0.34 to 4.6 A/W with decreasing etching time from 25 to 5 min. also we can note that the spectral sensitivity of both two cases has the same peak at 589 nm. On the other hand, we can note also the existence of weak peak of the sensitivity at long wave length (767 nm).

To explain these results we can divide the sensitivity curve into three regions as follows:

1- The short wavelengths regions (320 – 425 nm), where these wavelengths have large absorption coefficient and then short absorption depth, (i.e. the wavelengths have been absorbed at the surface of PSi layer and the recombination processes at this position is very high, thus the resulted photocurrent is very little and then very small value of sensitivity.

2- The region which include wavelengths in range from 535nm to 589 nm, since these wavelengths may be possess photons energies ($hv$) > PSi band gap it would be absorbed nearly completely in PSi layer which leads to increasing of photogenerated electron-hole pairs in the depletion region of interface c-Si/PSi, and because of high inner electrical field of the depletion region, this does on separating the photogenerated electron-hole pairs and then high value of photocurrent which leads to high sensitivity.

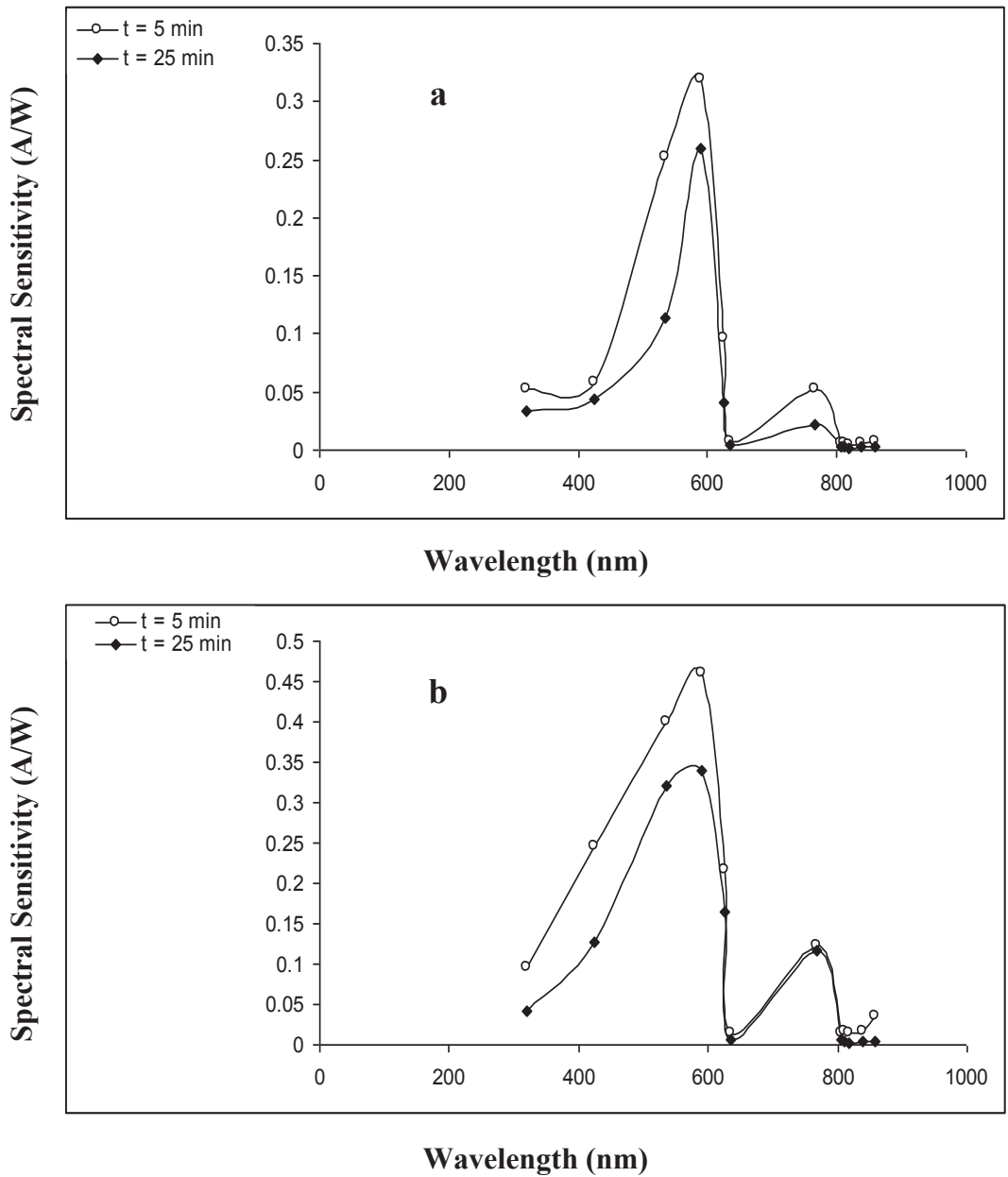

Figure (4-24); Sensitivity as a function of various incident wavelengths at 5 (V) reverse bias voltage; (a) 3.5 Ω.cm and (b) 0.02 Ω.cm Si substrates.

3- The region which include wavelengths in range from 626 nm to 859 nm, because of inhomogeneity of the PSi layer thickness and due to fact that the values of band-gaps of nano-crystals in PSi layer may be larger than the values of incident photons energies, the PSi layer acts as a window for the incident light and then these wavelengths would be absorbed nearly completely in the bulk silicon, and due to fact that the recombination processes in the bulk silicon would be very fast, therefore the life-time of charge carriers become very small and this leads to little photocurrent value and then weak sensitivity.

      Additional to this we can note that value of the sensitivity of sandwich structure include PSi layers prepared at etching times (5 min.) on both silicon substrate resistivity is larger than that of sandwich structure include PSi layers prepared at etching times (25 min.). on the other hand the sensitivity of sandwich structure include PSi layers prepared on 0.02 $\Omega$.cm silicon substrate resistivity is larger than that of sandwich structure include PSi layers prepared on 3.5 $\Omega$.cm silicon substrate resistivity. This result ascribed to decreasing of PSi layer thickness with decreasing of etching time and substrate resistivity (i.e. the small PSi layer thickness leads to little surface resistance and this leads to high photo-conductivity of this PSi layer and the latter is result in high value of the resulted photocurrent which leads to high value of sensitivity.

To avoid ambiguity, the values of sensitivity at the 635 nm wavelength not equal zero for both two cases but it appear such as that because of the transition from high peak to low peak. Where it equal to 0.004 A/W for sandwich structures include PSi layers prepared at etching time (25 min.) on 3.5 $\Omega$.cm silicon substrate resistivities.

      While the sensitivity at the 635 nm equals 0.007 A/W for sandwich structure include PSi layers prepared at etching time (5 min.) on 3.5 $\Omega$.cm silicon substrate resistivity. Also the values of sensitivity at the 635 nm wavelength of the sandwich structure include PSi layers prepared at 25 and 5 min. etching times on 0.02 $\Omega$.cm silicon substrates resistivities respectively equals 0.006 A/W and 0.14 A/W respectively.

By the way the values of quantum efficiency Q.E (λ) at 635 nm of Al/PSi/n-Si/Al sandwich structures include PSi layers prepared at 5 and 25 etching times on (3.5 Ω.cm and 0.02 Ω.cm) n-type silicon substrates resistivities respectively as shown in figure (4-25) also not equal zero, but rather it equals to 1% and 0.7% for sandwich structures include PSi layers prepared at 5 and 25 etching times respectively on 3.5 Ω.cm substrates resistivities. While it equals to 3% and 1% for sandwich structures include PSi layers prepared at 5 and 25 etching times respectively on 0.025 Ω.cm substrates resistivities.

Figure (4-25) represents relationship between the wavelength of the incident light and quantum efficiency Q.E (λ) of Al/PSi/n-Si/Al sandwich structure include PSi layers prepared at 5 and 25 etching times on 3.5 Ω.cm and 0.02 Ω.cm n-type silicon substrates resistivities.

According to the results described in figure (4-25) we can record that the quantum efficiency of PSi-based devices is near the unity as shown in figure (4-25,b) (open dots). As a consequence, we can estimate that the sensitivity of PSi-based devices is better than that of commercial silicon photo-diode in range wavelengths of (320 – 859 nm), and these results are consistent with those of [4,45,60,66]. We believe that the pores can play a role of wave-guides since that the light enters pores after repeated reflections from the pores walls and then will penetrate far deep in PSi layer. Therefore the light will be absorbed strongly in PSi layer compared with bulk silicon.

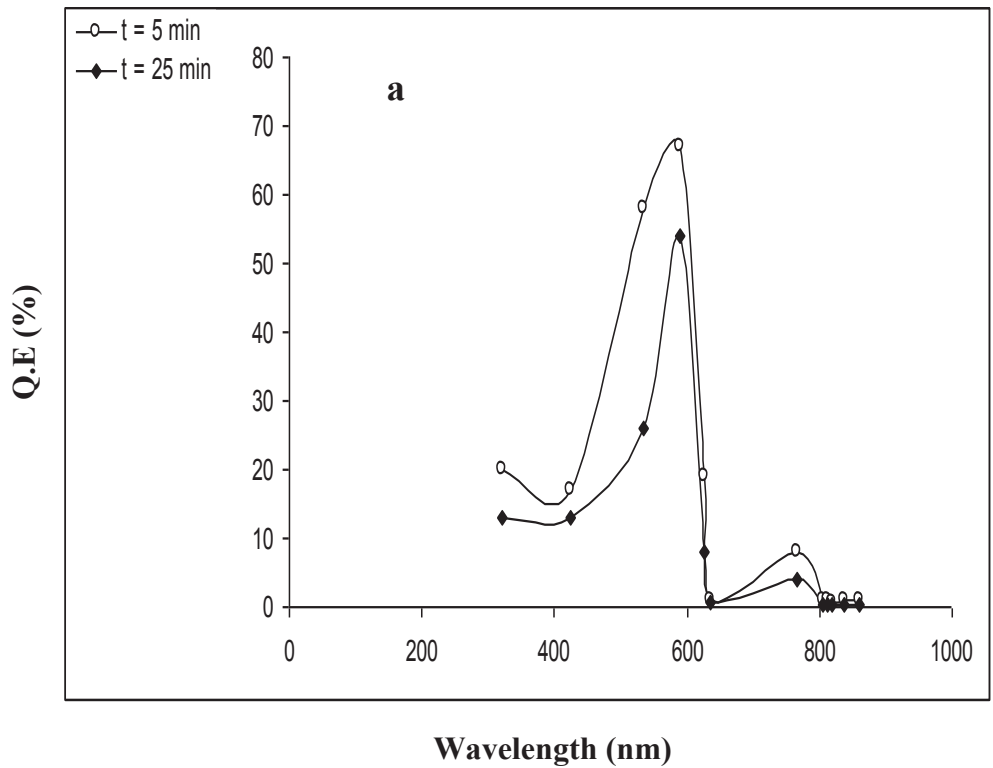

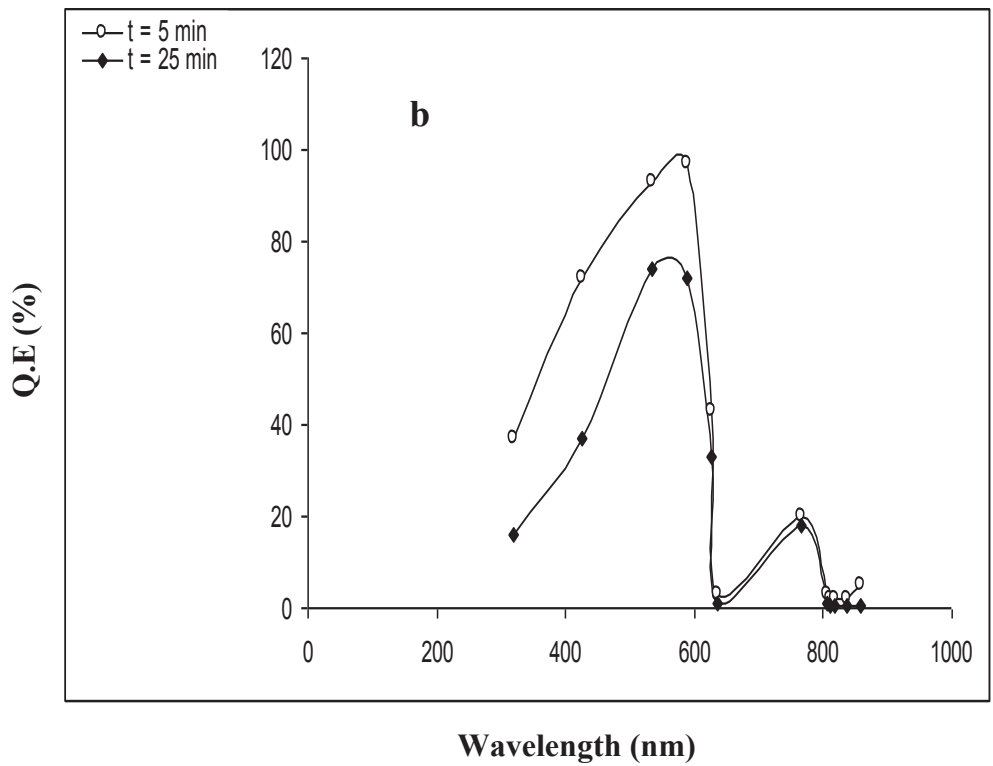

Figure (4-25); Quantum efficiency (Q.E) as a function of various incident wavelengths; (a) 3.5 Ω.cm and (b) 0.02 Ω.cm Si substrates.

# CHAPTER FIVE

# Conclusions and Future Work

## 5-1 *Conclusions*

In light of these results and facts which have been mentioned previously in this work, the photoelectrochemical etching process could consider as a very useful technique in production PSi material with advantageous properties for many applications. We can summarize the main conclusions as follows:

1- Among the etching techniques available today, photoelectrochemical (PEC) etching of n-type silicon is studied particularly. It is a simple, low cost and controllable as well as its characteristics combine between features of electrochemical etching process and the laser-induced etching. Therefore each type of silicon (n- or p-type) could be etched in very short time using this method.

2- Laser diode with 2 W power and 810 nm wavelength as an illumination source in the etching process gives higher etching rates.

3- The morphological properties of the formed PSi material via PEC etching process have been dramatic and very sensitive to preparation conditions. In general, the porosity, pore width, and thickness of PSi layer have increased with increasing of etching time and substrate resistivity, while the wall thickness between pores and nanocrystallite size have decreased with increasing of etching time and substrate resistivity. The pores have nearly cylindrical and rectangular shapes, depending on initiating of pores in random pattern. Additionally the structure aspect of PSi layer is crystalline but we can conclude from shift of its peak to small diffraction angle that the lattice constant has been slightly broadened because of the strain which increases with increasing of etching time and substrate resistivity.

4- The other formation parameters such as illumination and current density have important influence on the etching process and structural properties of resulted PSi layer since the light absorption provides the required electron-hole pairs to initiate the chemical reaction between HF acid and the irradiated area. Therefore distribution of power density of the laser beam in a Gaussian mode affects etching rate and then on the thickness of PSi layer. On the other hand the role of current density in etching process has been obvious from the higher etching rate and also it provides the required holes to etch the deepest layers which do not receive the light.

5- Band gap of each prepared PSi layer via PEC etching process is larger than that of silicon wafer, because of quantum confinement effect which increases with decreasing of nanocrystallite size.

6- The current-voltage characteristics of Al/PSi/n-Si/Al sandwich structures include PSi layers prepared on both substrates resistivities show rectification behavior. We can conclude that the rectifying behavior is due to PSi/n-Si heterojunction, but the rectifying ratio has been varied from one sample to another depending on the structural properties of particular PSi layer and the applied bias voltage, since the higher rectifying ratio has been noticeable at high applied bias voltage for both substrates resistivities.

7- The resulting photocurrent from Al/PSi/n-Si/Al sandwich structure depends strongly on its PSi layer, because it increases with decreasing of PSi layer thickness.

8- The spectral sensitivity of PSi-based device is higher than that of silicon diode, but at the same time it depends on preparation conditions of PSi layer.

9- Porous silicon applications depend on structural and electrical properties of PSi layer and these properties rely on the formation conditions where from what has been mentioned previously we can conclude that the Al/PSi/n-Si/Al sandwich structures which is include PSi layers prepared at etching times (5 and 25 min.) on both (0.02 and 3.5 Ω.cm) n-type silicon substrates resistivities may serve as very sensitive photo-diode in the visible range.

## 5-2 *Future Work*

It would be interesting and important to carry out the following studies in future:

1- Studying of the effect of different concentrations of HF acid on properties of PSi layers prepared by PEC etching process.

2- Studying of the effect of different current densities on PEC etching process.

3- Threshing of the visible electroluminescence from porous silicon prepared by photoelectrochemical etching.

4- Studying of the transport mechanisms of charge carriers under different temperature in porous silicon prepared by PEC etching process.

# *Publications*

Some of results within this thesis have been published in scientific periodicals and conferences.

**1.** Sabah M. Ali Ridha and Oday A. Abbas, ***Morphological Aspects of Porous Silicon Prepared by Photo-Electrochemical Etching***, Journal of Sciences College, Al-Mustansiriyah University, 19 (2007) 1.

**2.** Sabah M. Ali Ridha and Oday A. Abbas, ***Influence of Laser Irradiation Times on Properties of Porous Silicon***, Um-Salama Journal, Girls College of Sciences, University of Baghdad, 4 (2007) 4.

**3.** Sabah M. Ali Ridha and Oday A. Abass, ***Transport Mechanisms of Charge Carriers in Porous Silicon***, Journal of Sciences College, University of Babylon, to be published.